\documentclass[twocolumn]{aastex631}
\usepackage{CJK} % adding Chinese

\usepackage{amsmath}
\usepackage{amssymb}
\usepackage{epsfig}
\usepackage{apjfonts}
\usepackage{natbib}
\usepackage{hyperref}
\usepackage{epstopdf}
\usepackage{verbatim}
\usepackage{enumitem}
\usepackage{color}
\setlist[enumerate]{itemsep=-1mm}
\usepackage{soul, xcolor}
\setstcolor{blue}

\usepackage{mathtools}

\usepackage{bigints}

\newcommand{\dotdeg}{\rlap{.}^\circ}

\submitjournal{AJ (Received: July 10, 2023; Accepted: August 31, 2023)}

\begin{document}

%%%%%%%%%%
%%% TEXT  %%%
%%%%%%%%%%

\begin{CJK*}{UTF8}{gbsn}

\title{ELemental abundances of Planets and brown dwarfs Imaged around Stars (ELPIS): I. Potential Metal Enrichment of the Exoplanet AF Lep b and a Novel Retrieval Approach for Cloudy Self-luminous Atmospheres}

\author[0000-0002-3726-4881]{Zhoujian Zhang (张周健)} \thanks{NASA Sagan Fellow}
\affiliation{Department of Astronomy \& Astrophysics, University of California, Santa Cruz, CA 95064, USA }

\author[0000-0003-4096-7067]{Paul Molli\`{e}re}
\affiliation{Max-Planck-Institut f\"{u}r Astronomie, K\"{o}nigstuhl 17, 69117 Heidelberg, Germany}

\author[0000-0002-1423-2174]{Keith Hawkins}
\affiliation{Department of Astronomy, The University of Texas at Austin, Austin, TX 78712, USA}

\author[0000-0002-0900-6076]{Catherine Manea}
\affiliation{Department of Astronomy, The University of Texas at Austin, Austin, TX 78712, USA}

\author[0000-0002-9843-4354]{Jonathan J. Fortney}
\affiliation{Department of Astronomy \& Astrophysics, University of California, Santa Cruz, CA 95064, USA }

\author[0000-0002-4404-0456]{Caroline V. Morley}
\affiliation{Department of Astronomy, The University of Texas at Austin, Austin, TX 78712, USA}

\author[0000-0001-6098-3924]{Andrew Skemer}
\affiliation{Department of Astronomy \& Astrophysics, University of California, Santa Cruz, CA 95064, USA }

\author[0000-0002-5251-2943]{Mark S. Marley}
\affiliation{Lunar and Planetary Laboratory, University of Arizona, 1629 E. University Boulevard, Tucson, AZ 85721, USA}

\author[0000-0003-2649-2288]{Brendan P. Bowler}
\affiliation{Department of Astronomy, The University of Texas at Austin, Austin, TX 78712, USA}

\author[0000-0001-5365-4815]{Aarynn L. Carter}
\affiliation{Department of Astronomy \& Astrophysics, University of California, Santa Cruz, CA 95064, USA }

\author[0000-0003-4557-414X]{Kyle Franson} \thanks{NSF Graduate Research Fellow}
\affiliation{Department of Astronomy, The University of Texas at Austin, Austin, TX 78712, USA}

\author[0000-0002-0475-3662]{Zachary G. Maas}
\affiliation{Department of Astronomy, Indiana University, Bloomington, IN 47405, USA}

\author[0000-0002-3456-5929]{Christopher Sneden}
\affiliation{Department of Astronomy, The University of Texas at Austin, Austin, TX 78712, USA}

\begin{abstract}
AF~Lep~A+b is a remarkable planetary system hosting a gas-giant planet that has the lowest dynamical mass among directly imaged exoplanets. We present an in-depth analysis of the atmospheric composition of the star and planet to probe the planet's formation pathway. Based on new high-resolution spectroscopy of AF~Lep~A, we measure a uniform set of stellar parameters and elemental abundances (e.g., [Fe/H]$=-0.27 \pm 0.31$~dex). The planet's dynamical mass ($2.8^{+0.6}_{-0.5}$~M$_{\rm Jup}$) and orbit are also refined using published radial velocities, relative astrometry, and absolute astrometry. We use \texttt{petitRADTRANS} to perform chemically-consistent atmospheric retrievals for AF~Lep~b. The radiative-convective equilibrium temperature profiles are incorporated as parameterized priors on the planet's thermal structure, leading to a robust characterization for cloudy self-luminous atmospheres. This novel approach is enabled by constraining the temperature-pressure profiles via the temperature gradient $(d\ln{T}/d\ln{P})$, a departure from previous studies that solely modeled the temperature. Through multiple retrievals performed on different portions of the $0.9-4.2$~$\mu$m spectrophotometry, along with different priors on the planet's mass and radius, we infer that AF~Lep~b likely possesses a metal-enriched atmosphere ([Fe/H]$>1.0$~dex). AF~Lep~b's potential metal enrichment may be due to planetesimal accretion, giant impacts, and/or core erosion. The first process coincides with the debris disk in the system, which could be dynamically excited by AF~Lep~b and lead to planetesimal bombardment. Our analysis also determines $T_{\rm eff} \approx 800$~K, $\log{(g)} \approx 3.7$~dex, and the presence of silicate clouds and dis-equilibrium chemistry in the atmosphere. Straddling the L/T transition, AF~Lep~b is thus far the coldest exoplanet with suggested evidence of silicate clouds.
\end{abstract}

\section{Introduction} 
\label{sec:introduction}

Elemental abundances of exoplanets as measured from spectroscopy provide valuable insights into these planets' origins and formation processes \citep[e.g.,][]{2007prpl.conf..733M}. By comparing the composition of planets to those of their host stars, we can investigate their birth location, relative amounts of gas and dust accreted during their formation, and other phenomena such as late-stage planetesimal bombardment, pebble drift and evaporation, and core erosion \citep[e.g.,][]{2011ApJ...743L..16O, 2014ApJ...794L..12M, 2017MNRAS.469.4102M, 2021Natur.598..580L, 2021A&A...654A..71S, 2021A&A...654A..72S, 2022ApJ...934...74M, 2022arXiv221116877O, 2023ApJ...946...18O}. Our solar system serves as a convenient laboratory for contextualizing the composition of giant planets \citep[e.g.,][]{2004Icar..171..153W, 2005ApJ...622L.145A, 2009Icar..199..351F, 2010SSRv..152..423F, 2014ApJ...789...69H}. Such analysis has been also expanded to extrasolar planets as pioneered by \cite{2011ApJ...743L..16O}, who used the carbon-to-oxygen ratio (C/O) as a metric to probe the planets' formation pathways. 

Measurements of C/O and/or the bulk metallicity have been established for several directly imaged exoplanets, including $\beta$~Pic~b \citep[e.g.,][]{2020A&A...633A.110G}, YSES-1~b \citep[e.g.,][]{2021Natur.595..370Z}, HR~8799~bcde \citep[e.g.,][]{2013Sci...339.1398K, 2017AJ....154...91L, 2020AJ....160..150W, 2023AJ....165....4W, 2020A&A...640A.131M, 2021AJ....162..290R}, GJ~504~b \citep[e.g.,][]{2016ApJ...817..166S}, and 51~Eri~b \citep[e.g.,][]{2017AJ....154...10R, 2017A&A...603A..57S, 2023A&A...673A..98B, 2023MNRAS.tmp.1071W}. Similar measurements have been also made for substellar companions and free-floating brown dwarfs \citep[e.g.,][]{2015ApJ...807..183L, 2017ApJ...848...83L, 2019ApJ...877...24Z, 2022ApJ...936...44Z, 2021MNRAS.506.1944B, 2021A&A...656A..76Z, 2021ApJ...923...19G, 2022ApJ...938...56G, 2022AJ....163..189W, 2022ApJ...937...54X}, as well as irradiated exoplanets \citep[e.g.,][]{2021Natur.598..580L, 2022ApJS..260....3C, 2022ApJ...925L...3F, 2023arXiv230507753A, 2023Natur.614..653A, 2023MNRAS.522.5062B, 2023AJ....165...91B, 2023AJ....166...31F}. C/O is a popular abundance metric since the dominant oxygen and carbon reservoirs, including H$_{2}$O, CO, CO$_{2}$, and CH$_{4}$, are also the main opacity sources in planetary atmospheres. Beyond C/O, other abundance ratios have also been suggested as robust tracers of planet formation, including the nitrogen-to-oxygen ratio (N/O) and the refractory-to-volatile ratio \citep[e.g.,][]{2016ApJ...833..203P, 2020A&A...642A.229C, 2021ApJ...914...12L, 2021A&A...654A..72S, 2022ApJ...934...74M, 2022arXiv221116877O, 2023ApJ...946...18O}. Ultimately, combining all these elemental abundance metrics will provide a more comprehensive understanding of planet formation.

To further our understanding of atmospheric composition and its diversity in the planet and star formation process, we are launching the ELemental abundances of Planets and brown dwarfs Imaged around Stars (ELPIS) program. This program aims to measure the composition of directly imaged planets, brown dwarf companions, and all their host stars through spectroscopy. By exploring the planet-to-star relative abundance as a function of planet mass \citep[e.g.,][]{2011ApJ...736L..29M, 2016ApJ...831...64T, 2019ApJ...874L..31T, 2023AJ....166...85H} and orbital separation, we aim to probe the dominant formation mechanisms in different planet mass regimes and birth locations within protoplanetary disks. The existing census of directly imaged exoplanets, as included in our program, contains about three dozen objects. Looking forward, this list of discoveries is expected to rapidly expand, particularly with the contributions from the Gaia mission \citep{2016AandA...595A...1G}. The astrometric acceleration, or proper-motion anomaly, detected by the long-baseline astrometry from Hipparcos and Gaia has proven to be an efficient method for identifying parent stars of giant planets as compared to blind direct imaging surveys \citep{2018ApJS..239...31B, 2021ApJS..254...42B, 2019A&A...623A..72K, 2022A&A...657A...7K}. Recently, this method has led to new discoveries of imaged exoplanets and brown dwarfs \citep[e.g.,][]{2021ApJ...913L..26B, 2022MNRAS.513.5588B, 2022ApJ...934L..18K, 2023Sci...380..198C, 2023AJ....165...39F}.

One of the most recent exoplanet discoveries driven by astrometric acceleration is AF~Lep~b, which orbits the late-F star AF~Lep~A. This system was independently discovered by three groups \citep{2023AandA...672A..94D, 2023ApJ...950L..19F, 2023AandA...672A..93M}. Using their own astrometry and spectrophotometry observed at different dates, these studies constrained the dynamical mass of the planet to a range of about $3-5$~M$_{\rm Jup}$ and determined an orbital semi-major axis of $8-9$~au. AF~Lep~b is the lowest-mass imaged exoplanet with a dynamical mass measurement to date. The AF~Lep system is part of the $\beta$~Pictoris young moving group with an estimated age of $24 \pm 3$~Myr \citep[e.g.,][]{2015MNRAS.454..593B}. The system also hosts a debris disk located at $40-60$~au \citep{2021MNRAS.502.5390P, 2022A&A...659A.135P}, resembling the Kuiper belt of the solar system. 

AF~Lep~b's dynamical mass and its host star's elemental abundance and age will provide key context for interpreting the emission spectrophotometry of the planet. Therefore, as the first target in the ELPIS program, the AF~Lep system allows for a detailed study of the planet's atmospheric properties and formation history. We first describe our high-resolution spectroscopic observations of the host star AF~Lep~A (Section~\ref{sec:data}), followed by a uniform analysis of the stellar parameters and elemental abundances (Section~\ref{sec:host_param_abund}). Combining published radial velocities, relative astrometry, absolute astrometry, and our newly measured stellar mass, we refine the dynamical mass of AF~Lep~b to be $2.8^{+0.6}_{-0.5}$~M$_{\rm Jup}$ and update its orbital parameters (Section~\ref{sec:orbit}). With atmospheric properties of AF~Lep~b contextualized by evolution models (Section~\ref{sec:evo}), we then perform a retrieval analysis to determine the planet's key properties, including [Fe/H] and C/O (Sections~\ref{sec:retrieval_framework} and \ref{sec:retrieval}). We also introduce a novel retrieval approach that can enable a robust characterization of self-luminous atmospheres, especially those shaped by clouds. Implications of our analysis are discussed in Section~\ref{sec:discussion}, followed by a summary in Section~\ref{sec:summary}.\footnote{Throughout this work, we use subscriptions ``A'' and ``b'' for physical and orbital properties of the host star and the planet, respectively, only in Sections~\ref{sec:host_param_abund}--\ref{sec:orbit}. For the remaining sections, the physical properties refer to AF~Lep~b unless otherwise noted.}

\section{Data}
\label{sec:data}

\subsection{High-resolution Spectroscopy of the Host Star AF~Lep~A}
\label{subsec:tull_host}
We acquired optical (3800~\AA--8800~\AA) spectra of AF~Lep~A on 2023 February 24 UT from the 2.7~m Harlan J. Smith Telescope at McDonald Observatory. The Tull Echelle Spectrograph is utilized in the TS23 mode with the slit plug \#4, leading to a spectral resolution of $R \sim 60,000$. The instrument's encoders are configured to ensure the spectral lines of interest (e.g., atomic lines of H, C, O, Mg, Si, Li) fall within the detector's field of view. Calibration frames, including biases, flats, and Thorium-Argon lamp data, were collected at the beginning of the night. The data reduction follows the standard procedures, including bias subtraction, flat fielding, bad-pixel masking, cosmic-ray removal \citep[via \texttt{DCR} by][]{2004PASP..116..148P}, scattered light subtraction, and optimal spectral extraction. We normalize the continuum of each order assuming a second-order Chebyshev polynomial and then shift the order-stiched spectrum to the stellar restframe by cross-correlating with a solar spectral template using \texttt{iSpec} \citep[][]{2014A&A...569A.111B, 2019MNRAS.486.2075B}.

\subsection{Published Spectrophotometry of the Exoplanet AF~Lep~b}
\label{subsec:data_planet}

The near-infrared spectra of AF~Lep~b were collected from \cite{2023AandA...672A..93M} and \cite{2023AandA...672A..94D}. Both studies used the VLT/SPHERE integral field spectrograph \citep[IFS;][]{2008SPIE.7014E..3EC}. \cite{2023AandA...672A..93M} observed this planet on two different dates on 2022 October 16 UT and 2022 December 20 UT. Their data were reduced through the SPHERE data center \citep[][]{2017sf2a.conf..347D}, leading to an epoch-averaged spectrum spanning 0.94--1.65~$\mu$m ($R\sim30$). \cite{2023AandA...672A..94D} observed AF~Lep~b on 2022 October 20 UT. They reduced data using \texttt{pyKLIP} \citep[][]{2015ascl.soft06001W} and extracted the spectrum over 1.24--1.65~$\mu$m ($R\sim30$). Both studies also obtained the $K1$ (2.11~$\mu$m) and $K2$ (2.25~$\mu$m) photometry using the Infra-Red Dual-beam Imager and Spectrograph \citep[IRDIS;][]{2008SPIE.7014E..3LD} on the same nights as their IFS observations. In addition, \cite{2023ApJ...950L..19F} observed AF~Lep~b using the Keck/NIRC2 camera on 2021 December 21 UT and 2023 February 3 UT. They obtained $L'$-band (3.72~$\mu$m) photometry during these two epochs. 

\begin{figure}[t]
\includegraphics[height=2.35in]{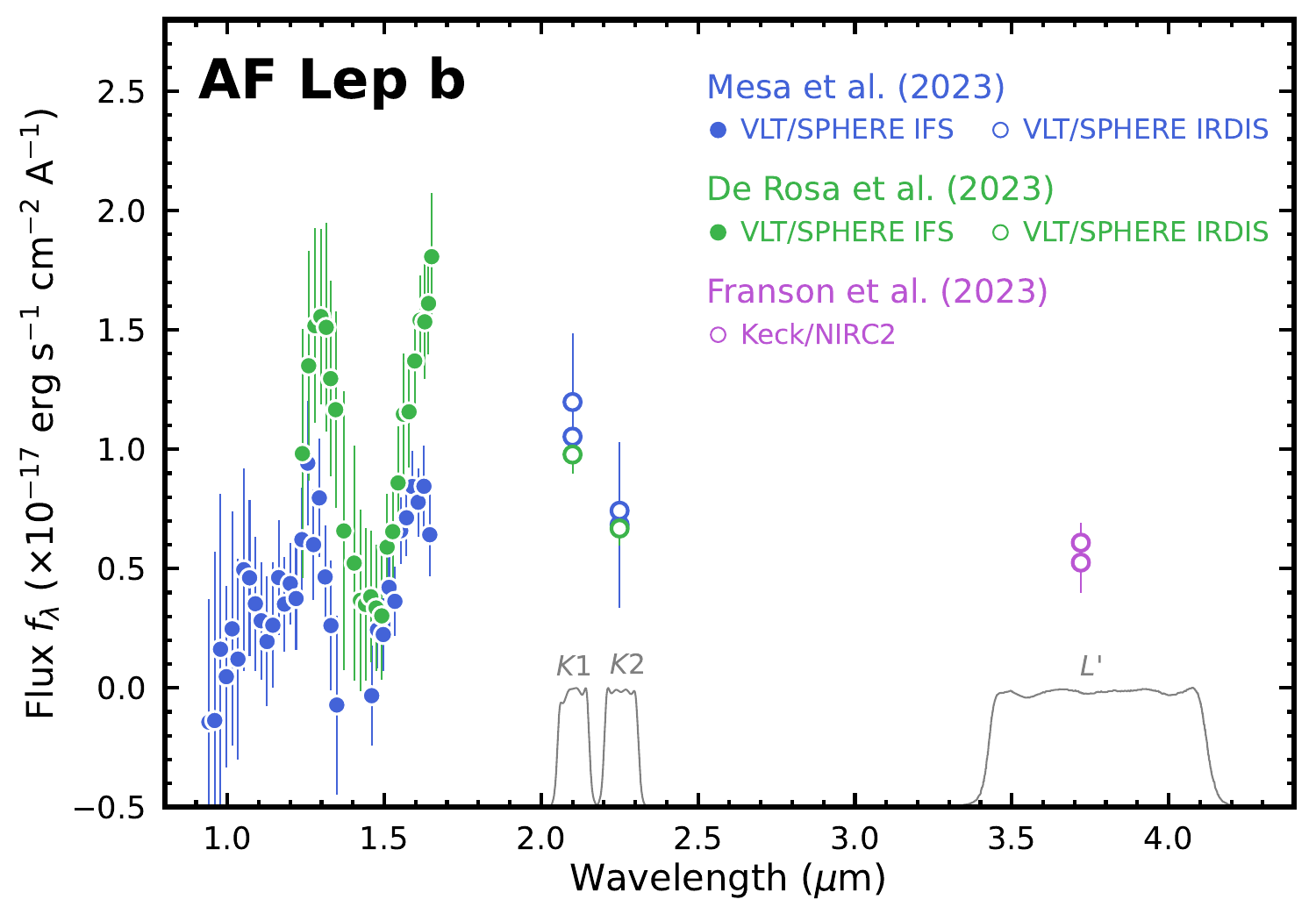}
\caption{Spectrophotometry of AF~Lep~b from \citeauthor{2023AandA...672A..93M} (\citeyear{2023AandA...672A..93M}; blue), \citeauthor{2023AandA...672A..94D} (\citeyear{2023AandA...672A..94D}; green), and \citeauthor{2023ApJ...950L..19F} (\citeyear{2023ApJ...950L..19F}; purple). Photometry is converted from magnitudes into fluxes (unless reported in the literature) based on zero points provided by \cite{2017AJ....154..218N}. Response curves of $K1/K2/L'$ bands (grey) are obtained from the VLT/SPHERE and Keck/NIRC2 websites. }
\label{fig:specphot}
\end{figure}

Figure~\ref{fig:specphot} summarizes all the published spectrophotometry of AF~Lep~b, including two spectra, three sets of $K1/K2$ photometry, and two $L'$ photometry. It is notable that the fluxes of two SPHERE/IFS spectra differ, with a reduced $\chi^{2} = 3.2$ if their flux difference is assumed to be zero. In other words, the \cite{2023AandA...672A..94D} spectrum is approximately $1.9$ times brighter than the \cite{2023AandA...672A..93M} spectrum in overlapping wavelengths.\footnote{This scaling factor $k = 1.9$ is calcuated by minimizing the $\chi^{2}$ metric $\sum_{i=1}^{N_{\rm pix}}{ (f_{i, \rm D} - k f_{i, \rm M})^{2} / (\sigma_{i, \rm D}^{2} + k^{2} \sigma_{i, \rm M}^{2}) }$. The $f_{i}$ and $\sigma_{i}$ represent the spectral flux and uncertainty in a given pixel $i$ ($N_{\rm pix}$ in total in overlapping wavelengths), with subscriptions ``D'' and ``M'' for the dataset of \cite{2023AandA...672A..94D} and \cite{2023AandA...672A..93M}, respectively. } In contrast, the planet's $K1/K2/L'$ photometry from SPHERE/IRDIS and Keck/NIRC2, observed on different dates and processed by different pipelines, is consistent with each other within uncertainties. 

The discrepant spectral fluxes of AF~Lep~b could potentially be attributed to atmospheric variability, which is common for young, low-gravity imaged planets and brown dwarfs \citep[e.g.,][]{2016ApJ...818..176Z, 2022AJ....164..239Z, 2019MNRAS.483..480V, 2022ApJ...924...68V}. Variability tends to have a stronger impact on fluxes at shorter wavelengths. However, \cite{2023AandA...672A..93M} measured the planet's photometry in $J$ and $H$ bands by using their IFS data collected over two epochs (see their Table~4) and found that these photometric data are consistent within the uncertainties. In addition, the photometric measurements of AF~Lep~b in $K1$, $K2$, and $L'$ bands over multiple epochs also show consistency (Figure~\ref{fig:specphot}). Therefore, the variability scenario cannot be confirmed based on the currently available data. Dedicated spectrophotometric monitoring of AF~Lep~b is warranted to investigate its top-of-atmosphere inhomogeneity.

An alternative explanation of the discrepant spectral fluxes could be attributed to the systematic differences in data reduction procedures between \cite{2023AandA...672A..93M} and \cite{2023AandA...672A..94D}. Speckle subtraction and flux calibration are both key sources of the systematics in the resulting emission spectra of imaged planets. Negative or positive speckle residuals near the location of planet detection can contribute an additive offset to the spectrum, while uncertainties in the calibration of the planet's flux relative to the host star's flux can contribute a multiplicative scaling factor. The scaling factor of 1.9 between the two IFS spectra of AF~Lep~b suggests a large calibration systematics of $90\%$, indicating that flux calibration might not be the primary source of the discrepancy. In addition, \cite{2023AandA...672A..94D} mentioned the presence of strong negative speckle residuals near AF~Lep~b in their reduced data and they suspected that these speckle-noise artifacts are responsible for the discrepant planet astrometry measured from their own IFS and IRDIS data. The negative speckle residuals might also lead to over-estimated spectral fluxes. Moreover, it is worth noting that \cite{2023AandA...672A..93M} performed spectral differential imaging (SDI), while \cite{2023AandA...672A..94D} deliberately skipped this procedure. SDI can introduce striping patterns that affect the extracted emission spectrum \citep[e.g., Figure~1 of][]{2023AandA...672A..93M}. 

In our work, we assume that the discrepant IFS spectral fluxes between \cite{2023AandA...672A..93M} and \cite{2023AandA...672A..94D} are impacted by the speckle residuals and SDI systematics. When combining both spectra for the subsequent atmospheric retrievals of AF~Lep~b, we incorporate an additive flux offset as a free parameter for each spectrum. We also perform retrievals for individual spectra, without incorporating any flux offsets. As discussed in Section~\ref{subsubsec:metal_enrichment}, the retrievals on different sets of spectra (the \citealt{2023AandA...672A..93M} spectrum, the \citealt{2023AandA...672A..94D} spectrum, or both spectra combined by offsets) consistently predict a metal-enriched atmosphere of AF~Lep~b.

\begin{deluxetable}{lll}
\setlength{\tabcolsep}{10pt} 
\tablecaption{Properties of AF Lep A} \label{tab:stellar_params} 
\tablehead{ \multicolumn{1}{l}{Parameter} &  \multicolumn{1}{l}{Value} &  \multicolumn{1}{l}{Reference} } 
\startdata 
Spectral Type  &  F8  &  Gray06  \\
Age (Myr)  &  $24 \pm 3$  &  Bell15  \\
\hline
\multicolumn{3}{c}{Astrometric Properties} \\
\hline
RA  &  05:27:04.78  &  Gaia16, Gaia22  \\
Dec  &  $-$11:54:04.26  &  Gaia16, Gaia22  \\
Parallax\tablenotemark{\scriptsize a} (mas)  &  $37.254 \pm 0.020$  &  Gaia16, Gaia22  \\
Distance (pc)  &  $26.825 \pm 0.014$  &  Bail21  \\
\hline 
\multicolumn{3}{c}{Photometric Properties} \\
\hline 
Tycho $B$ (mag)  &  $6.944 \pm 0.015$  &  H{\o}g00  \\
Tycho $V$ (mag)  &  $6.358 \pm 0.010$  &  H{\o}g00  \\
Hipparcos $H_{\rm p}$ (mag)  &  $6.421 \pm 0.002$  &  Ande12  \\
DR2 $G$ (mag)  &  $6.1803 \pm 0.0008$  &  Gaia16, Gaia18  \\
DR2 $BP$ (mag)  &  $6.501 \pm 0.003$  &  Gaia16, Gaia18  \\
DR2 $RP$ (mag)  &  $5.755 \pm 0.002$  &  Gaia16, Gaia18  \\
2MASS $J$ (mag)  &  $5.268 \pm 0.027$  &  Cutr03  \\
2MASS $H$ (mag)  &  $5.087 \pm 0.026$  &  Cutr03  \\
2MASS $K_{\rm s}$ (mag)  &  $4.926 \pm 0.021$  &  Cutr03  \\
$W1$ (mag)  &  $4.915 \pm 0.179$  &  Cutr14  \\
$W2$ (mag)  &  $4.783 \pm 0.060$  &  Cutr14  \\
\hline 
\multicolumn{3}{c}{Physical Properties} \\
\hline 
$T_{\rm eff}$ (K)  &  $5997 \pm 147$  &  This Work  \\
$\log{(g)}$ (dex)  &  $4.30 \pm 0.05$  &  This Work  \\
$M$ ($M_{\odot}$)  &  $1.09 \pm 0.06$  &  This Work  \\
$R$ ($R_{\odot}$)  &  $1.21 \pm 0.06$  &  This Work  \\
$\log{(L_{\rm bol}/L_{\odot})}$ (dex)  &  $0.235 \pm 0.010$  &  This Work  \\
$\xi$ (km s$^{-1}$)  &  $2.2 \pm 0.3$  &  This Work  \\
$v\sin{(i_{\star})}$ (km s$^{-1}$)  &  $<61.5$ &  This Work  \\
\hline 
\multicolumn{3}{c}{Elemental Abundances} \\
\hline 
{\rm [Fe/H]} (dex)  &  $-0.27 \pm 0.31$  &  This Work  \\
{\rm [Mg/H]} (dex)  &  $-0.11 \pm 0.21$  &  This Work  \\
{\rm [Ca/H]} (dex)  &  $-0.32 \pm 0.26$  &  This Work  \\
\enddata
\tablenotetext{a}{This parallax is the reported value in Gaia DR3. In our isochrone analysis (Section~\ref{subsec:stellar_params_isochrones}), we apply a zero-point of $-0.024$~mas and inflate its uncertainty by $30\%$.}  
\tablerefs{H{\o}g00: \cite{2000AandA...355L..27H}, Cutr03: \cite{2003yCat.2246....0C}, Gray06: \cite{2006AJ....132..161G}, Ande12: \cite{2012AstL...38..331A},  Cutr14: \cite{2014yCat.2328....0C}, Bell15: \cite{2015MNRAS.454..593B}, Gaia16: \cite{2016AandA...595A...1G}, Gaia18: \cite{2018AandA...616A...1G}, Bail21: \cite{2021AJ....161..147B}, Gaia22: \cite{2022arXiv220800211G},} 
\end{deluxetable}

\section{Stellar Parameters and Elemental Abundances of AF Lep A} 
\label{sec:host_param_abund}

\subsection{Initial Spectroscopic Analysis}
\label{subsec:init_stellar_params_spec}

We measure the stellar parameters of AF~Lep~A, including its effective temperature $T_{\rm eff, A}$, surface gravity $\log{(g_{A})}$,\footnote{Throughout this manuscript, we use ``$\log{()}$'' and ``$\ln{()}$'' for 10-based and natural logarithm, respectively.} iron abundance [Fe/H]$_{A}$, microturbulent velocity $\xi_{A}$, and spectral broadening (induced by the projected rotational velocity $v\sin{i_{\star}}$, macroturblent velocity, and the instrumental broadening). This measurement is established by analyzing the Tull spectrum of the host star using the Brussels Automatic Code for Characterizing High accUracy Spectra \citep[\texttt{BACCHUS};][]{2016ascl.soft05004M}. The setup of \texttt{BACCHUS} and our spectral analysis follow \cite{2020MNRAS.496.2422H}. 

The \texttt{BACCHUS} code derives stellar atmospheric parameters using the standard excitation/ionization balance technique. This technique determines the effective temperature by ensuring that there is no correlation between the excitation potential of absorption features and their measured abundances. In addition, the surface gravity is constrained by balancing the abundances of Fe~\textsc{i} and Fe~\textsc{ii} (i.e., ionization balance). The microturbulence velocity is derived by verifying there is no correlation between the abundance of Fe~\textsc{i} and its reduced equivalent width (i.e., the equivalent width divided by the wavelength). The spectral broadening is constrained by ensuring that the Fe abundances derived by the equivalent widths (which are insensitive to the broadening effect) are consistent with those derived using the line core (which is sensitive to the broadening). The abundances of individual Fe lines are derived using both equivalent widths and $\chi^2$ minimization between the observed spectrum and spectral synthesis. \texttt{BACCHUS} employs the \texttt{TURBOSPECTRUM} \citep{2012ascl.soft05004P} code for spectral synthesis, assuming the Local Thermodynamic Equilibrium (LTE) and adopting the MARCS model atmosphere \citep[][]{2008A&A...486..951G}. 

The fifth version of the Gaia-ESO atomic line list \citep{2021A&A...645A.106H} is used in \texttt{BACCHUS}. Hyperfine structure splitting is included for Sc~\textsc{i}, V~\textsc{i}, Mn~\textsc{i}, Co~\textsc{i}, Cu~\textsc{i}, Ba~\textsc{ii}, Eu~\textsc{ii}, La~\textsc{ii}, Pr~\textsc{ii}, Nd~\textsc{ii}, Sm~\textsc{ii} \citep[see more details in][]{2021A&A...645A.106H}. We also include the molecular line lists for CH \citep{2014A&A...571A..47M}, SiH (from the Kurucz line lists.\footnote{\url{http://kurucz.harvard.edu/linelists/linesmol/}}), CN, NH, OH, MgH and C$_{2}$ (T. Masseron, private communication).

AF~Lep~A has a high rotational velocity \citep[$v\sin{i_{\star}}$ of $50-55$~km~s$^{-1}$; e.g.,][]{2005ApJS..159..141V, 2005yCat.3244....0G, 2007AJ....133.2524W, 2009A&A...493.1099S, 2014MNRAS.444.3517M, 2021A&A...645A..30Z}. The stellar rotation period is $P_{\rm rot} = 1.007 \pm 0.009$~day (\citealt{2023ApJ...950L..19F}; also see \citealt{2015A&A...574A..25J, 2023AandA...672A..94D}), which falls on the short-period end of the distribution of late-F stars \citep[e.g.,][]{2014ApJS..211...24M}. The fast rotation leads to line broadening and blending of spectral features, particularly for Fe~\textsc{i}, Fe~\textsc{ii}, and other species of interest. This effect reduces the number of high-quality spectral lines available in our analysis. Therefore, our initial spectral analysis based on \texttt{BACCHUS} leads to stellar parameters with compromised precision, including an effective temperature of $5917 \pm 259$~K, a logarithmic surface gravity of $4.1 \pm 0.7$~dex, and an iron abundance of $-0.25 \pm 0.20$~dex. To further improve the precision of these stellar parameters, we feed these initial spectroscopic $T_{\rm eff, A}$, $\log{(g_{A})}$, and [Fe/H]$_{A}$ into the subsequent isochrone analysis (Section~\ref{subsec:stellar_params_isochrones}) to derive the adopted stellar properties. The results from the isochrone analysis are then used to refine and constrain the abundances of Fe and other elements (Section~\ref{subsec:stellar_abund}).

\subsection{Isochrone Analysis}
\label{subsec:stellar_params_isochrones}
We combine the spectroscopic $T_{\rm eff,A}$, $\log{(g_{A})}$, and [Fe/H]$_{A}$ (from Section~\ref{subsec:init_stellar_params_spec}) with broad-band photometry and parallax of AF~Lep~A and model them using \texttt{isochrones} \citep[][]{2015ascl.soft03010M}. The MESA Isochrones and Stellar Tracks \citep[MIST;][]{2016ApJS..222....8D, 2016ApJ...823..102C} are used. To construct the spectral energy distribution (SED) of AF~Lep~A, we collect its optical and infrared photometry from Tycho-2 \citep[][]{2000AandA...355L..27H}, Hipparcos \citep[][]{2012AstL...38..331A}, Gaia~DR2 \citep[][]{2016AandA...595A...1G, 2018AandA...616A...1G}, 2MASS \citep{2003yCat.2246....0C}, and AllWISE \citep{2014yCat.2328....0C}. The $W3$ and $W4$ from AllWISE are excluded to avoid the contaminating flux from the debris disk in the same planet system \citep[see Figure~1 of][]{2021MNRAS.502.5390P}. We adopt a photometric uncertainty floor of 0.03~mag if the reported magnitude in a given band is more precise, in order to account for any external calibration uncertainties of observed photometry, as well as systematic errors of synthetic photometry by \texttt{isochrones} \citep[also see][]{2019A&A...628A..94A, 2022A&A...662A.125F}. Filter response curves of $G/BP/RP$ photometry and the $G$-band magnitude calibration are all from \cite{2018A&A...619A.180M}. The parallax is taken from Gaia DR3 \citep[][]{2016AandA...595A...1G, 2022arXiv220800211G}, with its uncertainty inflated by $30\%$ \citep[e.g.,][]{2021ApJS..254...42B, 2021MNRAS.506.2269E, 2021A&A...649A...5F, 2021AJ....161..214Z} and the zero point computed via \texttt{gaiadr3-zeropoint} \citep{2021A&A...649A...4L}.

We feed the spectroscopic $T_{\rm eff, A}$, $\log{(g_{A})}$, and [Fe/H]$_{A}$, photometry, and parallax of AF~Lep~A into the \texttt{isochrones}. This analysis infers the age, distance, equivalent evolutionary point, stellar mass ($M_{A}$), radius ($R_{A}$), and bolometric luminosity ($L_{\rm bol, A}$), and also refines $T_{\rm eff, A}$, $\log{(g_{A})}$, and [Fe/H]$_{A}$ obtained from the initial spectroscopic analysis. We fix the $V$-band extinction at zero. Beyond default parameter priors set in \texttt{isochrones}, we adopt a log-uniform prior for the age parameter, over the $3\sigma$ confidence interval of the $\beta$~Pictoris moving group's age of $24 \pm 3$~Myr \citep[][]{2015MNRAS.454..593B}. The \texttt{isochrones} code employs \texttt{PyMultiNest} \citep{2008MNRAS.384..449F, 2009MNRAS.398.1601F, 2019OJAp....2E..10F, 2014A&A...564A.125B} and we set $10^{4}$ live points to sample the parameter posteriors. Systematic uncertainties of $2.4\%$ in $T_{\rm eff, A}$, $5\%$ in $M_{A}$, and $2\%$ in $L_{\rm bol, A}$ are incorporated as additional Gaussian noise into the derived stellar parameters, following suggestions by \cite{2022ApJ...927...31T}. We then re-compute $R_{A}$ and $\log{(g_{A})}$ from the modified $(T_{\rm eff,A}, L_{\rm bol,A})$ and $(M_{A}, R_{A})$ posteriors, respectively. 

This isochrone analysis provides a uniform set of stellar parameters for AF~Lep~A, with the adopted values and uncertainties summarized in Table~\ref{tab:stellar_params}. As discussed in Appendix~\ref{app:stellar_params_compare_lit}, our estimated stellar properties (particularly [Fe/H]$_{A}$) are consistent with those derived by previous work.

\subsection{Elemental Abundances}
\label{subsec:stellar_abund}

To measure the elemental abundances of AF~Lep~A, the isochrone-based $T_{\rm eff, A}$ and $\log{(g_{A})}$ (from Section~\ref{subsec:stellar_params_isochrones}) are used as input for \texttt{BACCHUS} to re-analyze the Tull spectrum. This analysis refines [Fe/H]$_{A}$, microturbulent velocity, and the spectral broadening (providing an upper limit for $v\sin{i_{\star}}$); also, the abundances of individual species, including C, O, Mg, Si, and Ca, and measured. For each spectral absorption feature of each element, we create a set of synthetic spectra corresponding to various [X/Fe] abundances spanning from $-0.6$~dex to $+0.6$~dex. A $\chi^2$ minimization is then performed between the observed and synthetic spectra. Our reported stellar [X/H] values are the median of derived [X/H] across all lines for a given species. The uncertainty of [X/H] is taken as the dispersion in this ratio across all lines. If only one absorption line is used, we conservatively assume an [X/Fe] uncertainty of 0.10~dex. 

We also determine the propagated uncertainty in [X/H] due to the uncertainties of the stellar effective temperature, surface gravity, and microturbulent velocity. Specifically, we perturb the $T_{\rm eff, A}$, $\log{(g_{A})}$, and the $\xi_{A}$ one at a time by their 1$\sigma$ uncertainties listed in Table~\ref{tab:stellar_params} and re-determine [X/H]. Changes in abundances due to these perturbations allow us to determine the uncertainty in [X/H] due to the uncertainties of stellar parameters. We find that the uncertainty in [Fe/H] is $\pm$0.08~dex, $\pm$0.04~dex, and $\pm$0.05~dex for perturbations of $\Delta$T$_{\rm eff, A}$= 150~K, $\Delta\log{(g_{A})}$ = 0.05~dex, and $\Delta\xi$ = 0.30~km s$^{-1}$, respectively. Furthermore, for perturbations in $T_{\rm eff, A}$, $\log{(g_{A})}$, and the $\xi$ at the same level as listed above, we find that the uncertainty in [Mg/H] is $\pm$0.20~dex, $\pm$0.02~dex, and $\pm$0.06~dex, respectively; the uncertainty in [Ca/H] is $\pm$0.16~dex, $\pm$0.02~dex, and $\pm$0.07~dex, respectively. Thus, we incorporate in quadrature an additional uncertainty of $0.08$~dex in [Fe/H], $0.20$~dex in [Mg/H], and $0.16$~dex in [Ca/H]. 

Due to the rotational broadening in the stellar spectrum, the abundances of C, O, and Si cannot be reliably measured. Therefore, the stellar C/O ratio is not determined. We are able to constrain the abundances of Fe, Mg, and Ca, as listed in Table~\ref{tab:stellar_params}.

\begin{figure}[t]
\includegraphics[height=3.in]{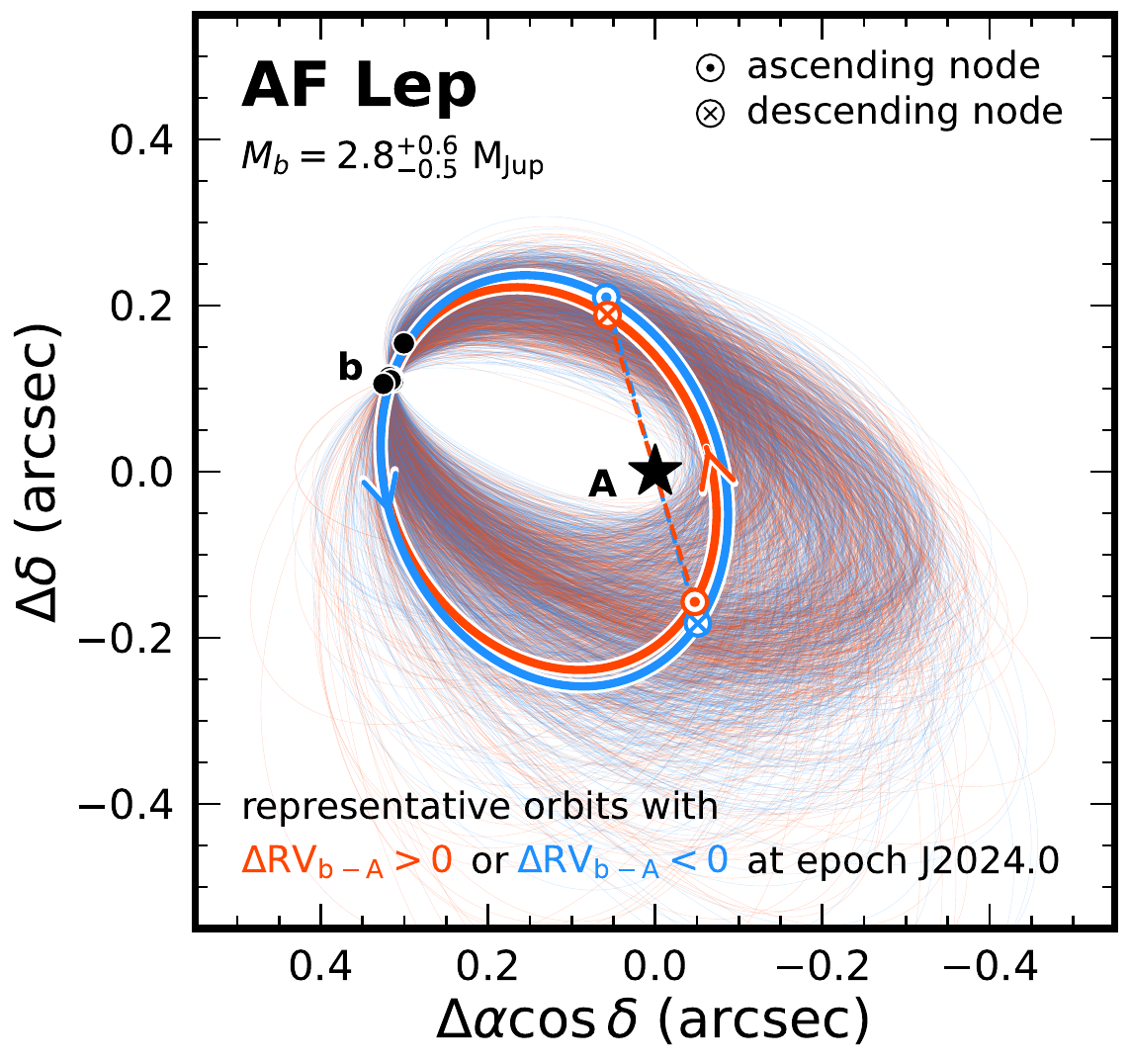}
\caption{Fitted orbits of AF~Lep~b, with the red and blue color denoting positive and negative $\Delta$RV$_{\rm b-A}$ at epoch J2024.0, respectively. A positive (or negative) $\Delta$RV$_{\rm b-A}$ value means the planet moves, relative to its host star, away from (or toward) the observer at a given epoch. Specific sign and value of $\Delta$RV$_{\rm b-A}$ has not been observed to date. Thin lines are random orbits drawn from the posteriors, and each thick line corresponds to the maximum-likelihood orbital solution for a given sign of $\Delta$RV$_{\rm b-A}$. The orbital solution with positive and negative $\Delta$RV$_{\rm b-A}$ values have similar shapes but the locations of their ascending and descending nodes are nearly swapped. We use the black star to show AF~Lep~A and black circles to trace the observed relative astrometry of AF~Lep~b. }
\label{fig:orbit}
\end{figure}

\section{Refined Dynamical Mass and Orbit\\of AF Lep \lowercase{b}} 
\label{sec:orbit}

The dynamical mass of AF~Lep~b provides key constraints on this planet's atmospheric properties (see Section~\ref{sec:retrieval}). However, previous orbit analyses of this planetary system led to different mass estimates, including $4.3^{+2.9}_{-1.2}$~M$_{\rm Jup}$ by \cite{2023AandA...672A..94D}, $3.2^{+0.7}_{-0.6}$~M$_{\rm Jup}$ by \cite{2023ApJ...950L..19F}, and $5.24^{+0.09}_{-0.10}$~M$_{\rm Jup}$ by \cite{2023AandA...672A..93M}. This discrepancy occurs mainly because the relative astrometry used in these studies was measured at different epochs over different baselines. Here we combine all published relative radial velocities (RVs) of the host star and the relative and absolute astrometry of the system, as well as our newly measured stellar mass (Section~\ref{sec:host_param_abund}), to provide the latest updates to the dynamical mass and the orbit of AF~Lep~b.

\subsection{RVs, Relative Astrometry, and Absolute Astrometry}
\label{subsec:rv_astro_data}

We obtain all 20 epochs of RVs of AF~Lep~A measured by \cite{2017AJ....153..208B} using Keck/HIRES. Among these RVs, 6 and 14 epochs were observed before and after the HIRES CCD upgrade on 2004 August 18 UT, respectively. These two sets of RV measurements are thus treated as separate instruments.\footnote{These RVs were treated as the same instrument in previous studies of AF~Lep, although our orbit analysis implies that relative RVs alone are not providing tight constraints on the orbital architecture of this system.} The pre-upgrade RVs span 1.1~years, with a linear trend of $149 \pm 50$~m~s$^{-1}$~yr$^{-1}$ and an RMS of 100~m~s$^{-1}$. The post-upgrade RVs span 9.1~years, with a linear trend of $-14 \pm 5$~m~s$^{-1}$~yr$^{-1}$ and an RMS of 162~m~s$^{-1}$. \cite{2023AandA...672A..94D} also measured RVs of AF~Lep~A using the ARC Echelle Spectrograph at Apache Point Observatory over 5 epochs in late~2022. These latest RVs have a typical uncertainty ($\approx 1.2$~km~s$^{-1}$) which is about 20 times larger than that of Keck/HIRES measurements ($\approx 62$~m~s$^{-1}$), and are thus excluded in our analysis.  

For the relative astrometry between A and b components, we collect all individual measurements by \cite{2023AandA...672A..94D}, \cite{2023ApJ...950L..19F}, and \cite{2023AandA...672A..93M} based on VLT/SPHERE and Keck/NIRC2, spanning a baseline of 1.1~years. The orbital motion of AF~Lep~b is demonstrated by its increasing position angle with a rate of $6\dotdeg8 \pm 1\dotdeg2$~yr$^{-1}$, although this planet's angular separation from its host star remains nearly constant during the monitoring (with a slope of $-2 \pm 10$~mas~yr$^{-1}$). There is also a significant difference between the Gaia and the joint Hipparcos-Gaia long-term proper motions of AF~Lep~A \citep[reduced $\chi^{2}$ is 77 for a constant proper-motion model;][]{2021ApJS..254...42B}, suggesting this star has an astrometric acceleration of $2.5 \pm 0.3$~m~s$^{-1}$~yr$^{-1}$ caused by the planet's gravitational perturbation.

{ 
\begin{deluxetable*}{llccl}
\setlength{\tabcolsep}{4pt} 
\tablecaption{Orbit analysis of AF Lep} \label{tab:orbparams} 
\tablehead{ \multicolumn{1}{l}{Parameter\tablenotemark{\scriptsize a}} &  \multicolumn{1}{l}{Unit} &  \multicolumn{1}{c}{Median$\pm1\sigma$} &  \multicolumn{1}{c}{$2\sigma$ Confidence Interval} &  \multicolumn{1}{l}{Adopted Prior} } 
\startdata 
\multicolumn{5}{c}{Fitted Parameters} \\ 
\hline 
Mass of AF Lep A $M_{A}$ &  $M_{\odot}$ & $1.09^{+0.07}_{-0.07}$  &  $(0.94, 1.23)$  &  $\mathcal{N}(\mu=1.07, \sigma^{2}=0.06^{2})$ \\  
Mass of AF Lep b $M_{b}$ &  $M_{\rm Jup}$ & $2.8^{+0.6}_{-0.5}$  &  $(1.8, 4.1)$  &  $1/M$ (log-flat) \\  
Semi-major axis $a$ &  au & $8.2^{+1.3}_{-1.7}$  &  $(5.8, 11.6)$  &  $1/a$ (log-flat)  \\  
$\sqrt{e_{b}}\sin{\omega_{\star}}$ &  -- & $0.04^{+0.55}_{-0.62}$  &  $(-0.82, 0.82)$  &  Uniform  \\  
$\sqrt{e_{b}}\cos{\omega_{\star}}$ &  -- & $0.03^{+0.38}_{-0.38}$  &  $(-0.68, 0.67)$  &  Uniform  \\  
Inclination $i_{b}$ &  degree & $55^{+8}_{-13}$  &  $(26, 69)$  &  $\sin{(i)}$ with $i \in [0, 180^{\circ}]$  \\  
PA of the ascending node\tablenotemark{\scriptsize b} $\Omega_{b}$ &   &   &    &    \\  
$\Omega_{b}$  component 1 ($\Delta$RV$_{\rm b-A,ref}>0$) & degree  & $243^{+18}_{-26}$  &  $(182, 333)$  &  Uniform  \\  
$\Omega_{b}$  component 2 ($\Delta$RV$_{\rm b-A,ref}<0$) & degree  & $63^{+16}_{-28}$  &  $(-1, 168)$  &  Uniform  \\  
Mean longitude at J2010.0\tablenotemark{\scriptsize b} $\lambda_{\rm ref,\star}$ &   &   &    &    \\  
$\lambda_{\rm ref,\star}$ component 1 ($\Delta$RV$_{\rm b-A,ref}>0$) & degree  & $340^{+41}_{-100}$  &  $(177, 416)$  &  Uniform  \\  
$\lambda_{\rm ref,\star}$ component 2 ($\Delta$RV$_{\rm b-A,ref}<0$) & degree  & $165^{+36}_{-104}$  &  $(-35, 231)$  &  Uniform  \\  
Parallax $\varpi$ &  mas & $37.25^{+0.02}_{-0.02}$  &  $(37.21, 37.29)$  &  $\mathcal{N}(\mu=37.254, \sigma^{2}=0.019^{2})$  \\  
System Barycentric Proper Motion in RA $\mu_{\alpha}\cos{(\delta)}$ &  mas~yr$^{-1}$ & $17.11^{+0.03}_{-0.03}$  &  $(17.05, 17.17)$  &  Uniform \\  
System Barycentric Proper Motion in DEC $\mu_{\delta}$ &  mas~yr$^{-1}$ & $-49.19^{+0.03}_{-0.03}$  &  $(-49.24, -49.13)$  &  Uniform  \\  
RV Jitter for pre-upgrade HIRES $\sigma_{\rm jit,pre-HIRES}$ &  m s$^{-1}$ & $124^{+67}_{-43}$  &  $(38, 320)$  &  $1/\sigma_{\rm jit,pre-HIRES}$ (log-flat)  \\  
RV zero point for pre-upgrade HIRES ZP$_{\rm pre-HIRES}$ &  m s$^{-1}$ & $-143^{+11}_{-11}$  &  $(-164, -121)$  &  Uniform  \\  
RV Jitter for post-upgrade HIRES $\sigma_{\rm jit,post-HIRES}$ &  m s$^{-1}$ & $168^{+46}_{-33}$  &  $(109, 280)$  &  $1/\sigma_{\rm jit,pre-HIRES}$ (log-flat)  \\  
RV zero point for post-upgrade HIRES ZP$_{\rm post-HIRES}$ &  m s$^{-1}$ & $64^{+12}_{-15}$  &  $(45, 80)$  &  Uniform  \\  
\hline 
\multicolumn{5}{c}{Derived Parameters} \\ 
\hline 
Eccentricity $e_{b}$  &  --  &  $0.4^{+0.3}_{-0.2}$  &  $(0.0, 0.8)$  &  --  \\  
Period $P_{b}$  &  year  &  $22.3^{+5.6}_{-6.7}$  &  $(13.4, 38.3)$  &  --  \\  
Argument of periastron\tablenotemark{\scriptsize b} $\omega_{\star}$  &    &    &    &    \\  
$\omega_{\star}$ component 1 ($\Delta$RV$_{\rm b-A,ref}>0$)  &  degree  &  $50^{+45}_{-85}$  &  $(-92, 131)$  &  --  \\  
$\omega_{\star}$ component 2 ($\Delta$RV$_{\rm b-A,ref}<0$)  &  degree  &  $231^{+44}_{-85}$  &  $(65, 306)$  &  --  \\  
Time of periastron\tablenotemark{\scriptsize c} $T_{0}$  &  JD  &  $2456906^{+720}_{-358}$  &  $(2455810, 2462691)$  &  --  \\  
Periastron separation $a_{b}(1-e_{b})$  &  au  &  $5.3^{+2.3}_{-2.7}$  &  $(1.4, 9.0)$  &  --  \\  
\enddata 
\tablenotetext{a}{Orbital parameters all correspond to the orbit of AF~Lep~b except for $a$, $\omega_{\star}$, and $\lambda_{\rm ref,\star}$. The first parameter $a$ corresponds to the system's (instead of individual components') semi-major axis, and the latter two parameters correspond to the orbit of the host star AF Lep A.}  
\tablenotetext{b}{Posteriors of $\Omega$, $\lambda_{\rm ref,\star}$, and $\omega_{\star}$ are bimodal. We divide each parameter posterior into two components, with each corresponding to a positive and a negative $\Delta$RV$_{\rm b-A, ref}$, respectively. Here $\Delta$RV$_{\rm b-A, ref}$ denotes the relative RV between the exoplanet and its host star at epoch 2024.0. The parameter confidence interval of each component is reported separately.}  
\tablenotetext{c}{$T_{0}$ is computed as $t_{\rm ref} - P \times (\lambda_{\rm ref,\star} - \omega_{\star})/360^{\circ}$, where $t_{\rm ref} = 2455197.5$~JD (i.e., epoch J2010.0).} 
\end{deluxetable*} 
}

\subsection{Orbit Analysis}
\label{subsec:orvara}
We use \texttt{orvara} \citep[][]{2021AJ....162..186B} to constrain the dynamical mass and orbit of AF~Lep~b by fitting all available RVs, relative astrometry, and absolute astrometry (Section~\ref{subsec:rv_astro_data}). There are 15 free parameters in our orbit analysis, including the mass of the host star ($M_{A}$), the dynamical mass of the planet ($M_{b}$), the semi-major axis of the planetary system ($a$), eccentricity ($e_{b}$), inclination ($i_{b}$), position angle of the ascending node of the planet's orbit ($\Omega_{b}$), the argument of the periastron of the host star's orbit ($\omega_{\star}$), mean longitude of the host star's orbit at epoch J2010.0 ($\lambda_{\rm ref,\star}$), marginalized parallax ($\varpi$) and proper motion ($\mu_{\alpha}\cos{(\delta)}$ and $\mu_{\delta}$) of the system, as well as the zero points (ZP) and jitter terms ($\sigma_{\rm jit}$) for pre- and post-upgrade Keck/HIRES RV measurements. Adopted priors for these parameters are summarized in Table~\ref{tab:orbparams}. In particular, we assume a Gaussian prior for M$_{A}$, with the mean and standard deviation as $1.09 \pm 0.06$~M$_{\odot}$ based on the stellar analysis (Section~\ref{subsec:stellar_params_isochrones}).

This analysis employs the parallel-tempering Markov Chain Monte Carlo (MCMC) sampler \citep[][]{2013PASP..125..306F, 2016MNRAS.455.1919V}. We run the MCMC with 50 temperatures and 100 walkers over $5\times10^{5}$ steps. Chains are saved every 50 steps, and the first 5000 samples from each walker of the thinned chains are removed as burn-in. 

The resulting orbital solution for AF~Lep~b is bimodal due to the lack of information about the orbital RV of the planet, i.e., the relative RV between the planet and the host star $\Delta$RV$_{\rm b-A} \equiv $ RV$_{b}-$RV$_{A}$ \citep[e.g.,][]{2020ApJ...894..115P, 2023AJ....165...73Z, 2023AJ....166...48D}. We compute $\Delta$RV$_{\rm b-A}$ at a reference epoch J2024.0 based on the MCMC chains and the following equation,
\begin{equation} 
\begin{aligned} 
\Delta{{\rm RV}_{\rm b-A,ref}} =& - \frac{2\pi\sin{i} \sqrt{M_{A} + M_{b}}}{\sqrt{a(1-e^{2})}}  \\
& \times \left[\cos{(\nu_{\rm ref} + \omega_{\star})} + e \cos{(\omega_{\star})} \right] 
\end{aligned} 
\end{equation} 
where $\nu_{\rm ref}$ is the true anomaly of the planet at epoch J2024.0. The inferred orbital parameters can be divided into two subsets corresponding to positive and negative $\Delta$RV$_{\rm b-A, ref}$ values. The two solution modes for the orbit of AF~Lep~b produce nearly identical confidence intervals for most parameters, except for $\Omega$, $\omega_{\star}$, and $\lambda_{\rm ref,\star}$, which have different median values by approximately 180$^{\circ}$ between the two modes. Representative orbits for these two solution modes are shown in Figure~\ref{fig:orbit}. 

We have refined the dynamical mass of AF~Lep~b to be $M_{b} = 2.8^{+0.6}_{-0.5}$~M$_{\rm Jup}$. Several orbital parameters of the planet have been also updated, with median values and confidence intervals summarized in Table~\ref{tab:orbparams}. Parameter posteriors and the comparison between the observed data and fitted orbits are shown in Appendix~\ref{app:orbit_posterior}.

We also re-assess the spin-orbit alignment of the AF~Lep system by using the updated planet's orbital parameters and stellar properties. Adopting a $v\sin{(i_{\star})} = 50 \pm 5$~km~s$^{-1}$ \citep{2005yCat.3244....0G}, a stellar radius of $1.21 \pm 0.06$~R$_{\odot}$ (Table~\ref{tab:stellar_params}), and a stellar rotation period of $1.007 \pm 0.009$~day \citep{2023ApJ...950L..19F}, we follow \cite{2023AJ....165..164B} and infer that the inclination of the stellar spin axis is $i_{\star} = {60^{\circ}}^{+14^{\circ}}_{-10^{\circ}}$. If we switch to a different $v\sin{(i_{\star})}$ of $54.7 \pm 0.5$~km~s$^{-1}$ \citep{2014MNRAS.444.3517M}, which has better precision and is among the highest value in the literature, then the inferred $i_{\star} = {67^{\circ}}^{+10^{\circ}}_{-6^{\circ}}$. These estimated inclinations of the stellar spin axis are consistent with the planet's orbital inclination ($i_{b} = {55^{\circ}}^{+8^{\circ}}_{-13^{\circ}}$) within $1\sigma$. Thus, the minimum misalignment angle between the spin axis of AF~Lep~A and the orbit of AF~Lep~b, $|i_{\star} - i_{b}|$, is consistent with zero \citep[also see Section~2 of][]{2023AJ....165..164B}. Given that the orientation of the stellar spin axis is unknown, the true spin-orbit misalignment angle can be potentially larger. Therefore, as previously suggested by \cite{2023ApJ...950L..19F}, the architecture of the AF~Lep system could be consistent with either spin-orbit alignment or misalignment.

\begin{figure*}[t]
\begin{center}
\includegraphics[height=6.in]{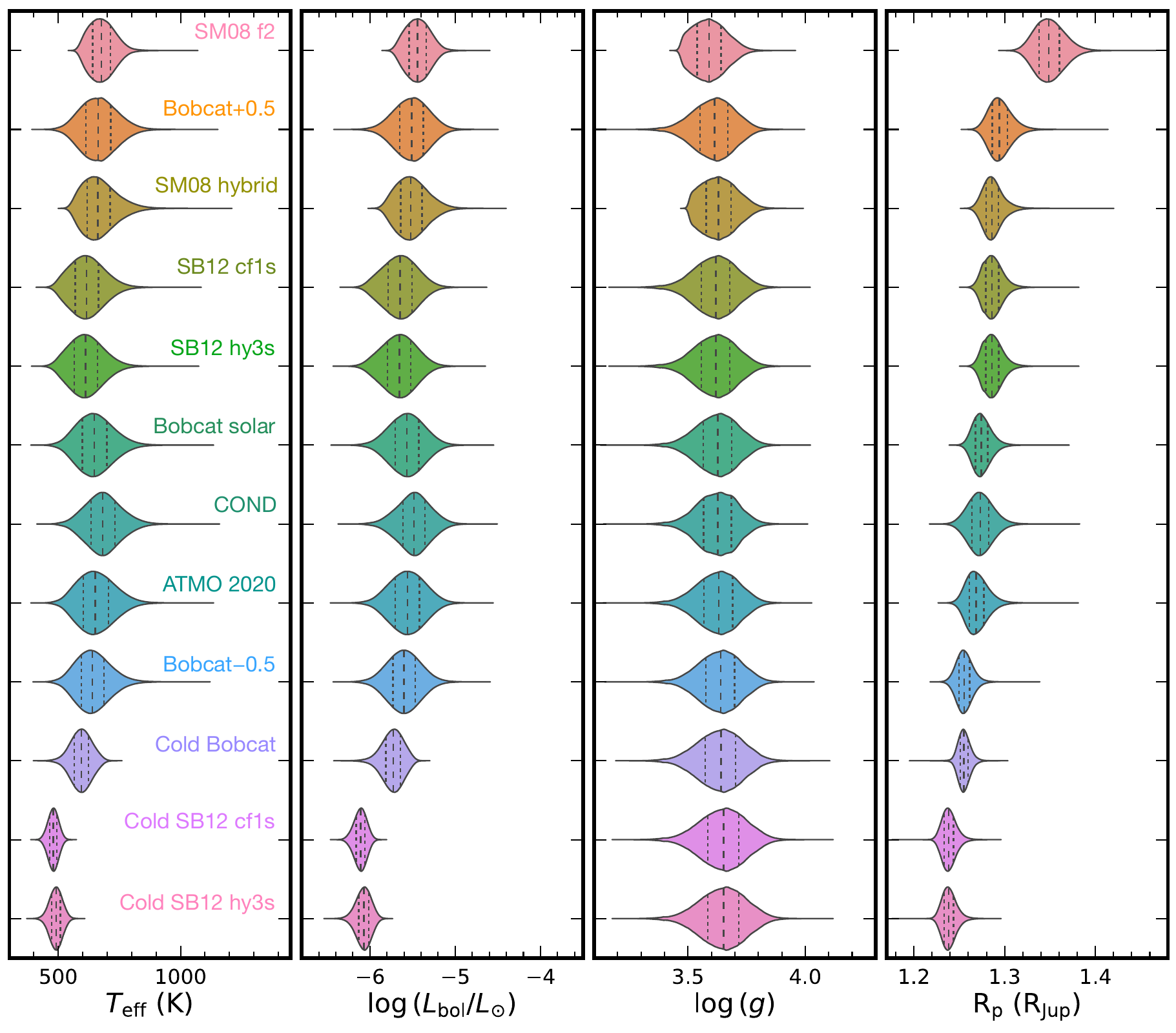}
\caption{Violin plot for the physical properties of AF~Lep~b inferred by various evolution models (sorted by radii), based on the planet's dynamical mass and age. The vertical dashed line inside each component highlights the median value of the parameter, while the two dotted lines mark the first and third quartiles of the distribution. The labels and the parameter confidence intervals are shown in Table~\ref{tab:evoparams}.   }
\label{fig:evo_violin}
\end{center}
\end{figure*}

{ 
\begin{deluxetable*}{llcccc}
\setlength{\tabcolsep}{8pt} 
\tablecaption{Properties of AF Lep b based on evolution models (assuming planet and host star are coeval)} \label{tab:evoparams} 
\tablehead{ \multicolumn{1}{l}{Label\tablenotemark{\scriptsize a}}  & \multicolumn{1}{l}{Evolution Model} &  \multicolumn{1}{c}{$T_{\rm eff}$} &  \multicolumn{1}{c}{$\log{(L_{\rm bol}/L_{\odot})}$} &  \multicolumn{1}{c}{$\log{g}$} &  \multicolumn{1}{c}{$R$} \\ 
\multicolumn{1}{l}{} &  \multicolumn{1}{l}{} &  \multicolumn{1}{c}{(K)} &  \multicolumn{1}{c}{(dex)} &  \multicolumn{1}{c}{(dex)} &  \multicolumn{1}{c}{($R_{\rm Jup}$)} } 
\startdata 
\multicolumn{6}{c}{Hot-Start Models} \\ 
\hline 
SM08 f2  & \cite{2008ApJ...689.1327S}: cloudy ($f_{\rm sed} = 2$) and [Fe/H]$=0$  &  $675^{+54}_{-50}$  &  $-5.44^{+0.15}_{-0.14}$  &  $3.59^{+0.08}_{-0.07}$  &  $1.349^{+0.017}_{-0.015}$  \\ 
SM08 hybrid  & \cite{2008ApJ...689.1327S}: hybrid and [Fe/H]$=0$  &  $661^{+76}_{-60}$  &  $-5.52^{+0.20}_{-0.17}$  &  $3.63^{+0.08}_{-0.07}$  &  $1.286^{+0.011}_{-0.009}$  \\ 
Bobcat$-$0.5  & \cite{2021ApJ...920...85M}: cloudless and [Fe/H]$=-0.5$  &  $639^{+72}_{-65}$  &  $-5.60^{+0.19}_{-0.19}$  &  $3.64^{+0.09}_{-0.09}$  &  $1.255^{+0.009}_{-0.008}$  \\ 
Bobcat solar  & \cite{2021ApJ...920...85M}: cloudless and [Fe/H]$=0$  &  $647^{+76}_{-70}$  &  $-5.57^{+0.20}_{-0.20}$  &  $3.63^{+0.08}_{-0.09}$  &  $1.274^{+0.011}_{-0.009}$  \\ 
Bobcat$+0.5$  & \cite{2021ApJ...920...85M}: cloudless and [Fe/H]$=+0.5$  &  $663^{+76}_{-71}$  &  $-5.51^{+0.20}_{-0.20}$  &  $3.61^{+0.08}_{-0.09}$  &  $1.294^{+0.013}_{-0.011}$  \\ 
SB12 cf1s  & \cite{2012ApJ...745..174S}: cloudless and [Fe/H]$=0$  &  $615^{+73}_{-67}$  &  $-5.65^{+0.20}_{-0.21}$  &  $3.62^{+0.09}_{-0.09}$  &  $1.286^{+0.011}_{-0.010}$  \\ 
SB12 hy3s  & \cite{2012ApJ...745..174S}: hybrid and [Fe/H]$=+0.5$  &  $612^{+71}_{-66}$  &  $-5.66^{+0.20}_{-0.21}$  &  $3.62^{+0.09}_{-0.09}$  &  $1.286^{+0.011}_{-0.010}$  \\ 
ATMO 2020  & \cite{2020AandA...637A..38P}: cloudless and [Fe/H]$=0$  &  $651^{+79}_{-71}$  &  $-5.56^{+0.21}_{-0.21}$  &  $3.63^{+0.09}_{-0.09}$  &  $1.269^{+0.013}_{-0.011}$  \\ 
COND  & \cite{2003AandA...402..701B}: cloudless and [Fe/H]$=0$  &  $682^{+73}_{-70}$  &  $-5.48^{+0.19}_{-0.20}$  &  $3.63^{+0.08}_{-0.09}$  &  $1.273^{+0.014}_{-0.013}$  \\ 
\hline 
\multicolumn{6}{c}{Cold-Start Models} \\ 
\hline 
Cold Bobcat & \cite{2021ApJ...920...85M}: cloudless and [Fe/H]$=-0.5$  &  $594^{+42}_{-42}$  &  $-5.73^{+0.12}_{-0.13}$  &  $3.64^{+0.09}_{-0.10}$  &  $1.255^{+0.007}_{-0.006}$  \\ 
Cold SB12 cf1s & \cite{2012ApJ...745..174S}: cloudless and [Fe/H]$=0$  &  $480^{+21}_{-21}$  &  $-6.11^{+0.07}_{-0.08}$  &  $3.65^{+0.09}_{-0.10}$  &  $1.238^{+0.008}_{-0.007}$  \\ 
Cold SB12 hy3s & \cite{2012ApJ...745..174S}: hybrid and [Fe/H]$=+0.5$  &  $491^{+25}_{-26}$  &  $-6.07^{+0.08}_{-0.09}$  &  $3.65^{+0.09}_{-0.10}$  &  $1.238^{+0.008}_{-0.007}$  \\ 
\enddata 
\tablenotetext{a}{The label of each evolution model shown in Figure~\ref{fig:evo_violin}.}  
\end{deluxetable*} 
}

\section{Contextualizing the Properties of \\AF Lep \lowercase{b} via Evolution Models}
\label{sec:evo}

In preparation for the atmospheric retrievals of AF~Lep~b, an analysis of this planet's properties predicted by evolution models is performed. This analysis serves several purposes, including providing important context for the retrieval analysis and helping to avoid inferring unphysical solutions that often occur in atmospheric studies, such as an excessively small radius and/or large surface gravity of gas-giant exoplanets. The detailed motivation for this evolution model analysis is described in Section~\ref{sec:evo_motivation}, and the results of the analysis are presented in Section~\ref{sec:evo_params}.

\subsection{Motivation: On the Discrepancies between Atmospheric vs. Evolution Model Predictions}
\label{sec:evo_motivation}
Characterizing properties of directly imaged planets and brown dwarfs usually relies on two types of models: atmospheric models and thermal evolution models. Atmospheric models involve retrievals (a.k.a. inverse-modeling; e.g., \citealt{2015ApJ...807..183L, 2017MNRAS.470.1177B, 2019A&A...627A..67M}; also see \citealt{2023RNAAS...7...54M}) or pre-computed temperature-pressure-composition profiles in radiative-convective equilibrium (RCE), and corresponding synthetic spectra over a grid of parameters \citep[a.k.a. forward-modeling; e.g.,][]{1997ApJ...491..856B, 2008ApJ...689.1327S, 2012RSPTA.370.2765A, 2012ApJ...756..172M, 2018ApJ...854..172C, 2020AandA...637A..38P, 2021ApJ...920...85M, 2021ApJ...923..269K, 2023ApJ...942...71M}. These models are fitted to observed spectrophotometry to constrain atmospheric physical properties, such as effective temperature, surface gravity, and radius.

Thermal evolution models adopt the upper boundary condition set by RCE and are typically provided as tables, which calculate an object's effective temperature, bolometric luminosity, and radius, as a function of age, for a given set of modeled masses. \citep[e.g.,][]{2003AandA...402..701B, 2007ApJ...655..541M, 2008ApJ...683.1104F, 2008ApJ...689.1327S, 2012ApJ...745..174S, 2020AandA...637A..38P, 2021ApJ...920...85M}. Measurements of any two variables from the set of $\{$mass, age, bolometric luminosity$\}$ can provide estimates of the evolution-based properties, including $T_{\rm eff}$, $\log{(g)}$, and $R$. 

While ideally, atmospheric and evolution models would yield consistent predictions, discrepancies often arise in practice. \cite{2020ApJ...891..171Z, 2021ApJ...916...53Z} found differences between atmospheric properties, inferred from spectra and atmospheric models, and evolution model predictions, inferred from the objects' bolometric luminosities and their host stars' ages. These differences can be significant, with variations of up to $120$~K in $T_{\rm eff}$ and $0.5$~dex in $\log{(g)}$. Also, studies on large samples of free-floating late-T and Y dwarfs have noted discrepancies between spectroscopic $\log{(g)}$ and the values based on evolution models \citep[e.g.,][]{2019ApJ...877...24Z, 2022ApJ...936...44Z, 2021ApJ...921...95Z}. Some brown dwarfs have atmospheric $\log{(g)}$ that would imply unphysical ages (e.g., older than the Universe) according to evolution models. In addition, the ``small radius problem'' has also been encountered in retrieval analyses of brown dwarfs, where retrieved radii from spectra are much smaller than the radii based on evolution given these objects' ages \citep[e.g.,][]{2020ApJ...905...46G, 2021MNRAS.506.1944B, 2022ApJ...930..136L, 2022ApJ...937...54X, 2023arXiv230304885H}. 

These discrepancies highlight the systematics of atmospheric models, including uncertainties in opacities (e.g., alkali and CH$_{4}$ line lists) and the assumptions about the chemical (dis-)equilibrium, thermal structure, and clouds. Evolution models, although not entirely free of systematics \citep[e.g.,][]{2009ApJ...692..729D, 2018AJ....156..168B, 2020AJ....160..196B, 2023AJ....165...39F}, provide important context for atmospheric model predictions. %, including information about elemental compositions and thermal structures that are typically not directly constrained by evolution models. 
Considering the predictions of evolution models is crucial when interpreting the parameters inferred from retrieval or forward modeling analyses, as it helps to account for these discrepancies and provide additional insights into the objects' properties.

\subsection{Evolution Model Analysis}
\label{sec:evo_params}

To contextualize our subsequent retrieval analysis of AF~Lep~b, we derive the properties of this planet using the following evolution models.\footnote{The AMES-DUSTY \citep[][]{2000ApJ...542..464C, 2002A&A...382..563B} and the BHAC15 evolution models \citep{2015A&A...577A..42B} are commonly used but are not applicable for AF~Lep~b, given that more than half of the posteriors of this planet's dynamical mass and age are outside the parameter space of these two sets of evolution models.} 
\begin{enumerate}
\item[$\bullet$] The hot-start \cite{2008ApJ...689.1327S} evolution models with two versions: $f_{\rm sed} = 2$ and the hybrid version. Both versions assume solar metallicity.
\item[$\bullet$] The hot-start and cold-start Sonora-Bobcat by \cite{2021ApJ...920...85M}, with [Fe/H] $= -0.5$, $0$, and $+0.5$. The cold-start models are available only at solar metallicity \citep[also see Section~5.5 of][]{2019AJ....158...13N}. 
\item[$\bullet$] The hot-start and cold-start models by \cite{2012ApJ...745..174S}. Cloudless atmospheres are assumed for the solar metallicity models, while cloudy atmospheres are assumed for the $3\times$ solar metallicity models. 
\item[$\bullet$] The hot-start ATMO2020 models by \cite{2020AandA...637A..38P} with solar metallicity.
\item[$\bullet$] The hot-start AMES-COND models by \cite{2003AandA...402..701B} with solar metallicity.
\end{enumerate}

Assuming AF~Lep~b is coeval with its host star \citep[$t = 24 \pm 3$~Myr;][]{2015MNRAS.454..593B}, the planet's dynamical mass ($M = 2.8^{+0.6}_{-0.5}$~M$_{\rm Jup}$) is combined with this assumed age to determine the $T_{\rm eff}$, $\log{(L_{\rm bol} / L_{\odot})}$, $\log{(g)}$, and $R$.\footnote{As explained in Section~\ref{sec:evo_motivation}, evolution-based properties of objects can be derived by any two variables from the set of $\{M, t, L_{\rm bol}\}$. Here we use mass and age since AF~Lep~b's bolometric luminosity might not be reliably measured from the spectrophotometry observed to date, which has short wavelength coverage. Also, the bolometric correction for this young planet requires assumptions built upon large uncertainties (e.g., spectral type), compared to those of the dynamical mass and age for this planet.} We use the MCMC chain of dynamical mass from the orbit analysis (Section~\ref{sec:orbit}) and generate an equal-size distribution of $1.5 \times 10^{6}$ random ages, sampled from a normal distribution $\mathcal{N}(\mu=24\ {\rm Myr}, \sigma=3\ {\rm Myr})$ truncated at zero. The evolution models are interpolated in linear scales for $\log{g}$ and $\log{(L_{\rm bol} / L_{\odot})}$ and in logarithmic scales for $T_{\rm eff}$, $R$, $M$, and $t$. No extrapolation is conducted outside the convex hull of each model grid.

Figure~\ref{fig:evo_violin} and Table~\ref{tab:evoparams} summarize the inferred physical properties of AF~Lep~b using various evolution models. The effective temperature and bolometric luminosity estimates are consistent among all hot-start models. The cold-start models predict slightly lower values since these models assume lower initial entropies at a given planet mass. Surface gravities are similar across all models and the radii are consistent.  

Considering that the formation and accretion process of the planet can occur over a few Myrs along with the dispersal of the protoplanetary disk \citep[e.g.,][]{2014prpl.conf..475A}, it is likely that AF~Lep~b is younger than its host star. While the typical disk lifetime is about $3$~Myr (e.g., \citealt{2009AIPC.1158....3M}; also see reviews by \citealt{2011ARA&A..49...67W} and \citealt{2022arXiv220309759D}), here we explore a slightly more extreme case where AF~Lep~b formed 10~Myr after its host star. We thus derive another set of physical properties by assuming a planet age of $t = 14 \pm 3$~Myr. Compared to the results assuming coevality with the host star, all evolution models predict $30-120$~K hotter $T_{\rm eff}$, $0.01-0.03$~dex lower $\log{(g)}$, $0.01-0.05$~R$_{\rm Jup}$ larger $R$, and $0.1-0.3$~dex brighter $\log{(L_{\rm bol}/L_{\odot})}$. These inferred planet parameters are summarized in Appendix~\ref{app:evo_params_young}. 

Regardless of the assumed age of AF~Lep~b, the $3\sigma$ confidence intervals of the planet's radius inferred from the evolution models fall within the range of $1.20-1.55$~R$_{\rm Jup}$. This range will be implemented as a uniform prior for $R$ in the subsequent retrieval analysis to address the ``small radius problem'' encountered in other recent retrieval studies.\footnote{Our subsequent retrieval analysis suggests that AF~Lep~b likely has a metal-enriched atmosphere with [Fe/H] above 1.0~dex (see Table~\ref{tab:retrieval_params_key_runs} and Section~\ref{subsubsec:metal_enrichment}). This high metallicity value is beyond the [Fe/H] grid of the existing evolution models of exoplanets and brown dwarfs (see Table~\ref{tab:evoparams}). Future modeling efforts that self-consistently combine the planet interior models and atmospheric models with significant metal enrichment are warranted to provide context for the radius of planets such as AF~Lep~b.  }

\section{Atmospheric Retrieval Framework} 
\label{sec:retrieval_framework}

To characterize the atmospheric composition of AF~Lep~b, we use the \texttt{petitRADTRANS} code \citep[][]{2019A&A...627A..67M} to perform chemically-consistent retrievals for the planet's spectrophotometry (Section~\ref{subsec:data_planet}). A novel parameterization approach for the temperature-pressure (T-P) profile is also introduced (Sections~\ref{subsec:tp_profile}--\ref{subsec:rce_prior}) that can lead to a robust characterization of cloudy self-luminous atmospheres for giant planets and brown dwarfs.

\subsection{Temperature Model}
\label{subsec:tp_profile}

\begin{figure}[t]
\begin{center}
\includegraphics[height=2.5in]{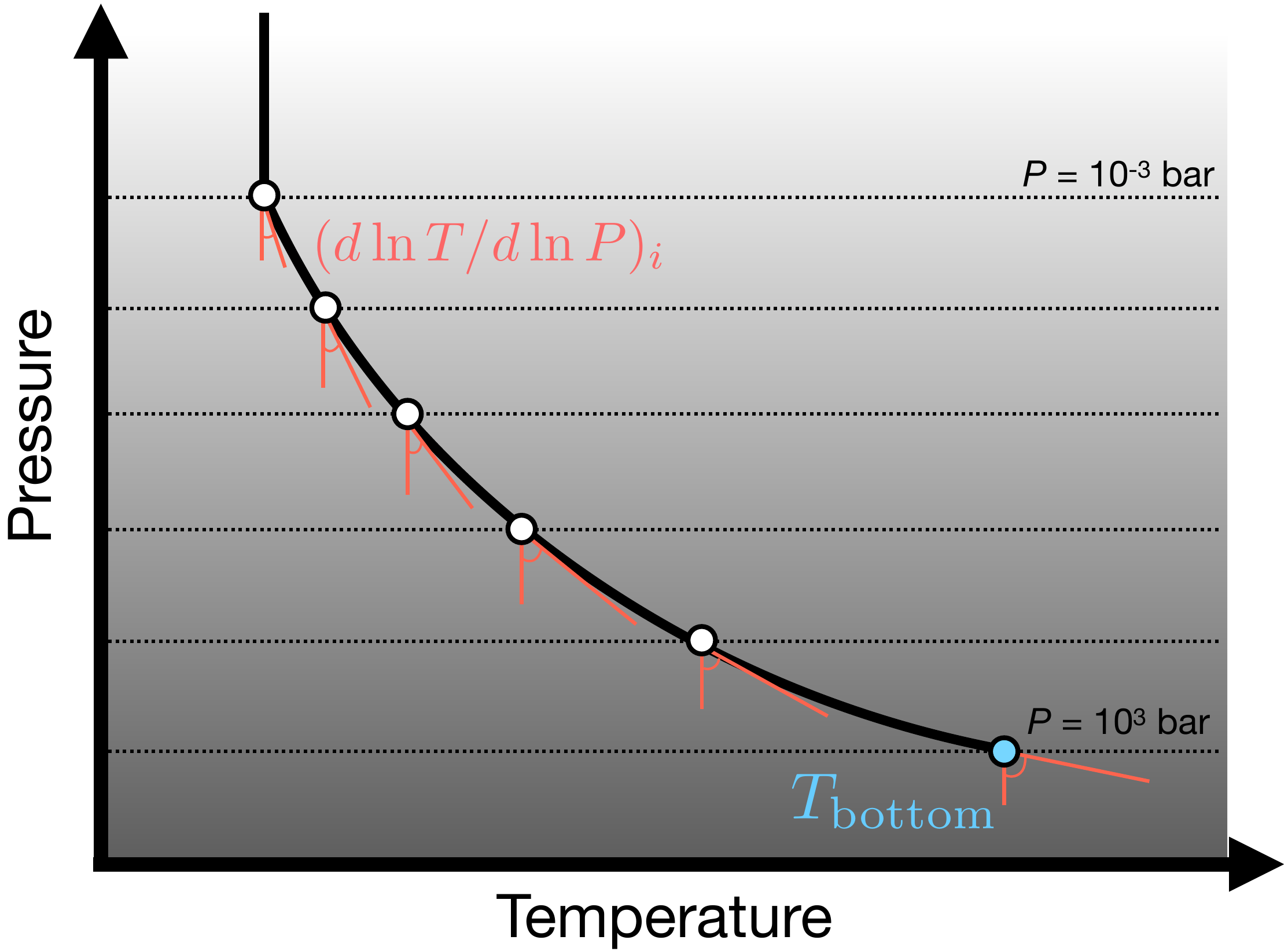}
\caption{Sketch of the temperature model described in Section~\ref{subsec:tp_profile}.}
\label{fig:thermal_structure}
\end{center}
\end{figure}

We model the thermal profile of AF~Lep~b by dividing its atmosphere into six layers that are evenly spaced in a logarithmic scale of pressure, ranging from $10^{3}$~bar to $10^{-3}$~bar. The temperature gradient, $d\ln{T}/d\ln{P}$, is fitted at each of these layers (Figure~\ref{fig:thermal_structure}). Another free parameter $T_{\rm bottom}$ is added for the temperature of the bottom layer at $10^{3}$~bar. With a given set of $d\ln{T}/d\ln{P}$ at six layers, a quadratic interpolation is performed to obtain the temperature gradient over a finer grid consisting of 1000 evenly spaced layers. The temperature, $T_{j}$, at each layer is then calculated as
\begin{equation} \label{eq:tp_profile}
\begin{aligned}
T_{1}  =&\ T_{\rm bottom} \\
T_{j+1}  =&\ \exp{ \left[ \ln{T_{j}} + (\ln{P_{j+1}} - \ln{P_{j}}) \times \left(\frac{d\ln{T}}{d\ln{P}}\right)_{j} \right]} \\
&\ {\rm with}\ j=1,2,3,\ldots,1000
\end{aligned}
\end{equation}
where a larger $j$ corresponds to a level with a higher altitude (or a lower pressure). The upper atmosphere with pressures below $10^{-3}$~bar is assumed to be isothermal. 

This new parameterization of the T-P profile differs from the typical thermal model employed in retrieval analyses of exoplanets and brown dwarfs, where the temperature is explicitly modeled as a function of pressure \citep[e.g.,][]{2015ApJ...807..183L, 2017ApJ...848...83L, 2017AJ....154...91L, 2017MNRAS.470.1177B, 2021MNRAS.506.1944B, 2019ApJ...877...24Z, 2022ApJ...936...44Z, 2020A&A...640A.131M, 2020AJ....160..150W, 2021ApJ...923...19G, 2021A&A...656A..76Z, 2021Natur.595..370Z, 2022ApJ...938...56G, 2022arXiv221114330B, 2022ApJ...937...54X, 2023MNRAS.521.5761G, 2023arXiv230304885H, 2023MNRAS.tmp.1071W}. \cite{2020MNRAS.497.5136P} developed a similar T-P parameterization as our approach, while their framework models the temperature difference among a pre-defined grid of pressure layers. As shown in the next subsection (Section~\ref{subsec:rce_prior}) and Section~\ref{subsec:advantages}, modeling T-P profiles via the temperature gradient enables the incorporation of the radiative-convective equilibrium as parameterized priors on the planet's thermal structure; this novel approach leads to a robust characterization of self-luminous atmospheres, especially those influenced by clouds.

\begin{deluxetable*}{lccccccccc} 
\setlength{\tabcolsep}{4pt} 
\tablecaption{Atmospheric Model Grids Investigated in Section~\ref{subsec:rce_prior}} \label{tab:atm_grids} 
\tablehead{ \multicolumn{1}{l}{Forward Model Grid} &  \multicolumn{6}{c}{Parameters} &  \multicolumn{1}{c}{} &  \multicolumn{2}{c}{Assumptions} \\ 
\cline{2-7} \cline{9-10} 
\multicolumn{1}{l}{} &  \multicolumn{1}{c}{$T_{\rm eff}$} &  \multicolumn{1}{c}{$\log{(g)}$} &  \multicolumn{1}{c}{[Fe/H]} &  \multicolumn{1}{c}{C/O} &  \multicolumn{1}{c}{$\log{(K_{\rm zz})}$} &  \multicolumn{1}{c}{cloud} &  \multicolumn{1}{c}{} &  \multicolumn{1}{c}{Clouds?} &  \multicolumn{1}{c}{Chem. Eq?} } 
\startdata 
ATMO2020  &  $[200, 3000]$  &  $[2.5, 5.5]$  &  $0$  &  $0.55$  &  multiple\tablenotemark{\scriptsize a}   &  --   &  &   Cloudless  &  CEQ + NEQ  \\  
Sonora Bobcat  &  $[200, 2400]$  &  $[3.0, 5.5]$  &  $[-0.5, +0.5]$  &  $[0.229, 0.687]$  &  --   &  --   &  &   Cloudless  &  CEQ  \\  
Sonora Cholla  &  $[500, 1300]$  &  $[3.5, 5.0]$  &  $0$  &  $0.458$  &  $[2, 9]$   &  --   &  &   Cloudless  &  NEQ  \\  
\cite{2022ApJ...938..107M}  &  $[400, 1000]$  &  $[4.5, 5.5]$  &  $0$  &  $0.458$  &  multiple\tablenotemark{\scriptsize b}   &  --   &  &   Cloudless  &  NEQ  \\  
\cite{2023arXiv230316295L}  &  $[200, 600]$  &  $[3.5, 5.0]$  &  $[-0.5, +0.5]$  &  $0.55$  &  6   &  multiple\tablenotemark{\scriptsize c}   &  &   Cloudy  &  CEQ + NEQ  \\  
\enddata 
\tablenotetext{a}{The $K_{\rm zz}$ of the ATMO2020 chemical dis-equilibrium models is a function of the surface gravity \citep[see Figure~1 of][]{2020AandA...637A..38P}.  }  
\tablenotetext{b}{The dis-equilibrium chemistry in the \cite{2022ApJ...938..107M} models are described in terms of (1) the varying $K_{\rm zz}$ in the radiative zones by factors of $100\times$, $1\times$, and $0.01\times$ from the \cite{2022ExA....53..279M} parameterization, and (2) the varying convective mixing lengths set by the $1\times$ and $0.1\times$ the scale height.  }  
\tablenotetext{c}{Water clouds of the \cite{2023arXiv230316295L} models are described by different shapes of the vertical opacity profiles. }  
\end{deluxetable*}

\begin{figure*}[t]
\begin{center}
\includegraphics[height=8.3in]{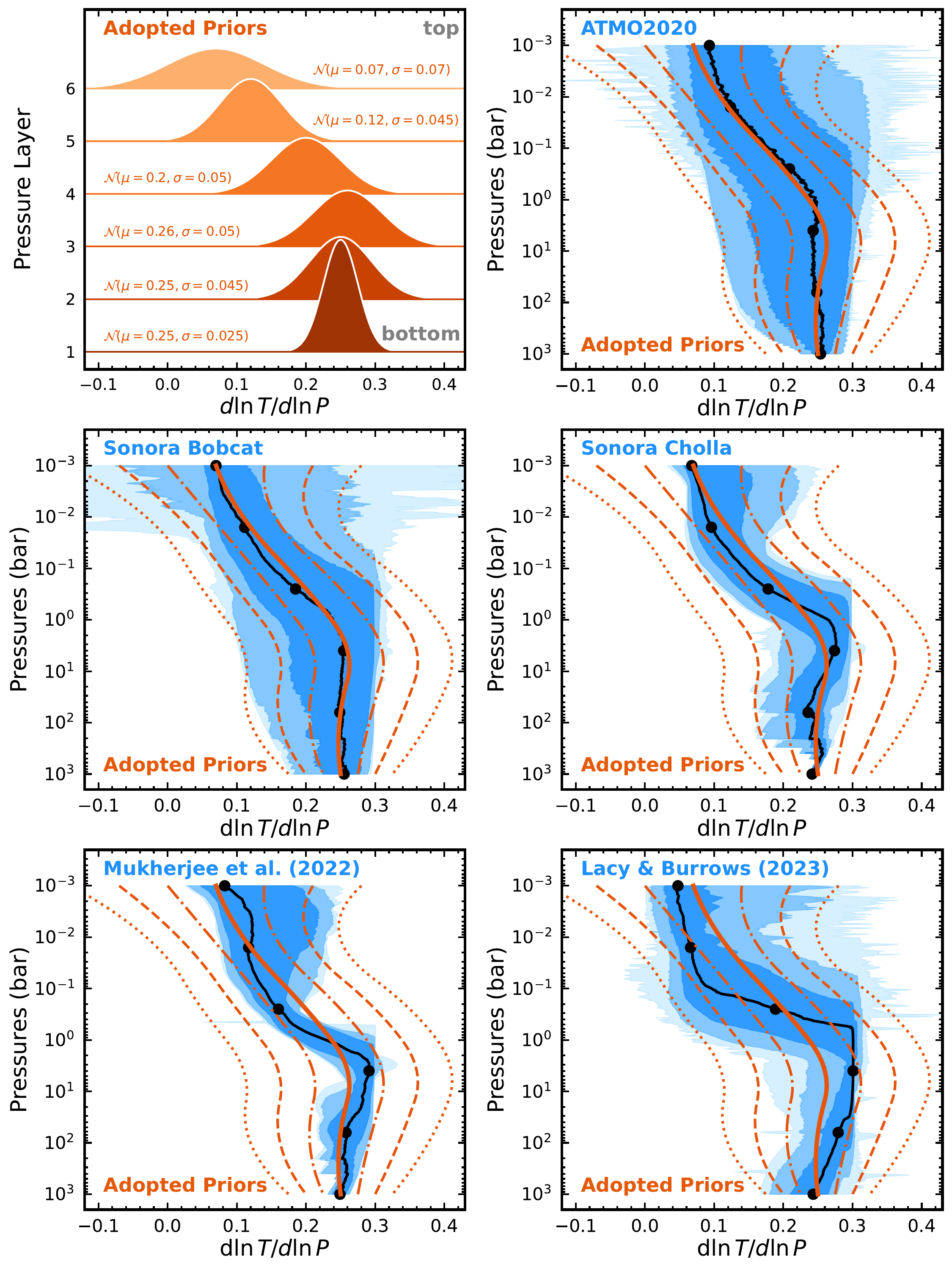}
\caption{ {\it Top Left}: Our adopted Gaussian priors for the temperature gradient $(d\ln{T}/d\ln{P})$ at six pressure layers, with layer number one and six corresponding to the bottom and top of the atmosphere with $P = 10^{3}$~bar and $10^{-3}$~bar, respectively. {\it Top Right}: The $(d\ln{T}/d\ln{P})_{\rm RCE}$ profile of the ATMO2020 model grid. The black line is the median profile among all grid points and the blue shaded regions represent (from dark to light) the $1\sigma$, $2\sigma$, and $3\sigma$ confidence intervals. Black circles mark the $(d\ln{T}/d\ln{P})_{\rm RCE}$ value of the six pressure layers. Overlaid orange lines (solid: median; dash-dotted: $1\sigma$ boundary; dashed: $2\sigma$ boundary; dotted: $3\sigma$ boundary) represent our adopted priors. For these prior profiles, the median and confidence intervals at six pressure layers are described in the top left panel, and the values between these layers are calculated via a quadratic interpolation (in logarithmic scale of pressure). {\it Middle and Bottom}: These panels share the format of the top right panel but with different forward model grids as summarized in Table~\ref{tab:atm_grids}. }
\label{fig:rce_dlnT_dlnP}
\end{center}
\end{figure*}

\subsection{Coupling T-P Profiles with Radiative-Convective Equilibrium}
\label{subsec:rce_prior}
Near the L/T transition, giant planets and brown dwarfs undergo significant changes in their spectrophotometric properties \citep[e.g.,][]{2004AJ....127.3516G, 2014ApJ...793...75R, 2019MNRAS.483..480V, 2021AJ....161...42B, 2021ApJS..253....7K}, which are thought to be related to the formation, condensation, and dissipation of clouds containing refractory species such as silicate and iron \citep[e.g.,][]{2006asup.book....1L, 2015ARA&A..53..279M}. However, retrievals of these cloudy objects sometimes result in a cloudless solution with a more isothermal T-P profile compared to the profile calculated under the assumption of radiative-convective equilibrium using the same $T_{\rm eff}$, $\log{(g)}$, and composition \citep[e.g.,][]{2017MNRAS.470.1177B, 2020A&A...640A.131M, 2022arXiv221114330B, 2023MNRAS.tmp.1071W}. This retrieved T-P profile converges with the RCE profile near the photosphere, but the former features relatively cooler temperatures in deeper atmospheres and warmer temperatures in upper atmospheres. The reduced temperature gradient of the retrieved T-P profiles mimics the effect of clouds by reddening the emergent spectra. Admittedly, the more isothermal T-P profile is consistent with a scenario proposed by \cite{2016ApJ...817L..19T, 2019ApJ...876..144T}, where thermo-compositional instabilities could explain the atmospheric properties of giant planets and brown dwarfs without invoking clouds. However, it is expected that cloud formation occurs in the ultracool, molecule-rich atmospheres of these objects. Observations from Spitzer and the recently launched JWST also probe the spectral features of silicate clouds near 10~$\mu$m across the L/T transition \citep[][]{2006ApJ...648..614C, 2022MNRAS.513.5701S, 2023ApJ...946L...6M}. 

To address the ``cloudless isothermal T-P problem'' encountered in retrievals, we propose a new approach, which involves coupling the T-P profiles with RCE during the retrieval process. Since the shape of T-P profiles and cloud properties are degenerate, our strategy is to add priors to the temperature gradient, $d\ln{T}/d\ln{P}$, that follow the shape expected by the RCE. These priors of the T-P profiles retrospectively constrain the cloud properties. 

To establish these priors, we examine the T-P profiles generated by various sets of atmospheric model grids that were all pre-computed under the RCE assumption. For each model set, we examine the T-P profile at each grid point within the corresponding parameter space and investigate the temperature gradient as a function of pressure. This investigation results in quantitative perspectives about the distribution of $\left(d\ln{T}/d\ln{P}\right)_{\rm RCE}$ at a given pressure layer. Our retrieval framework directly fits the $d\ln{T}/d\ln{P}$ values at six pressure layers throughout the atmosphere (Section~\ref{subsec:tp_profile}), thus, the $\left(d\ln{T}/d\ln{P}\right)_{\rm RCE}$ distributions in these six pressure levels provide priors to our $d\ln{T}/d\ln{P}$ parameters.

Five sets of atmospheric models are used to examine their $\left(d\ln{T}/d\ln{P}\right)_{\rm RCE}$ distributions:
\begin{enumerate}
\item[$\bullet$] The ATMO2020 models \citep[][]{2020AandA...637A..38P}.
\item[$\bullet$] The Sonora Bobcat models \citep[][]{2021ApJ...920...85M}.
\item[$\bullet$] The Sonora Cholla models \citep[][]{2021ApJ...923..269K}.
\item[$\bullet$] The models presented by \cite{2022ApJ...938..107M}.
\item[$\bullet$] The models presented by \cite{2023arXiv230316295L}.
\end{enumerate}
The parameter space and assumptions of these grids are listed in Table~\ref{tab:atm_grids}. Figure~\ref{fig:rce_dlnT_dlnP} presents the median and confidence intervals of the $\left(d\ln{T}/d\ln{P}\right)_{\rm RCE}$ profile by combining all grid points of each model set. While the temperature gradient profiles of these model grids are similar, they are not identical due to their different assumptions about clouds and chemical (dis-)equilibrium. 

We adopt the following Gaussian priors for $(d\ln{T}/d\ln{P})$ at each of six pressure layers,
\begin{equation} \label{eq:dlnT_dlnP_prior}
\begin{aligned}
&\frac{d\ln{T}}{d\ln{P}}\left(P = 10^{3}\ {\rm bar}\right) &\sim \mathcal{N}(\mu = 0.250, \sigma=0.025) \\
&\frac{d\ln{T}}{d\ln{P}}\left(P \approx 63\ {\rm bar}\right) &\sim \mathcal{N}(\mu = 0.250, \sigma=0.045) \\
&\frac{d\ln{T}}{d\ln{P}}\left(P \approx 4\ {\rm bar}\right) &\sim \mathcal{N}(\mu = 0.260, \sigma=0.050) \\
&\frac{d\ln{T}}{d\ln{P}}\left(P \approx 0.3\ {\rm bar}\right) &\sim \mathcal{N}(\mu = 0.200, \sigma=0.050) \\
&\frac{d\ln{T}}{d\ln{P}}\left(P \approx 0.02\ {\rm bar}\right) &\sim \mathcal{N}(\mu = 0.120, \sigma=0.045) \\
&\frac{d\ln{T}}{d\ln{P}}\left(P = 10^{-3}\ {\rm bar}\right) &\sim \mathcal{N}(\mu = 0.070, \sigma=0.070) 
\end{aligned}
\end{equation}
The mean and standard deviation of these Gaussian distributions are visually determined such that the confidence intervals of the temperature gradient profiles are qualitatively consistent between the grid models and the priors. The pressures of each layer are rounded in Equation~\ref{eq:dlnT_dlnP_prior}, but the exact pressure values are used in the retrievals.  

Our $(d\ln{T}/d\ln{P})$ priors have been established using grid models that encompass a much broader parameter space than the one centered on the properties of AF~Lep~b. Consequently, the priors presented in Equation~\ref{eq:dlnT_dlnP_prior} may be applied to a more diverse sample of directly imaged exoplanets and brown dwarfs. We also recommend users to customize the number of layers and the prior values by using different sets of forward models that align with their individual targets.

\begin{figure*}[t]
\begin{center}
\includegraphics[height=5.5in]{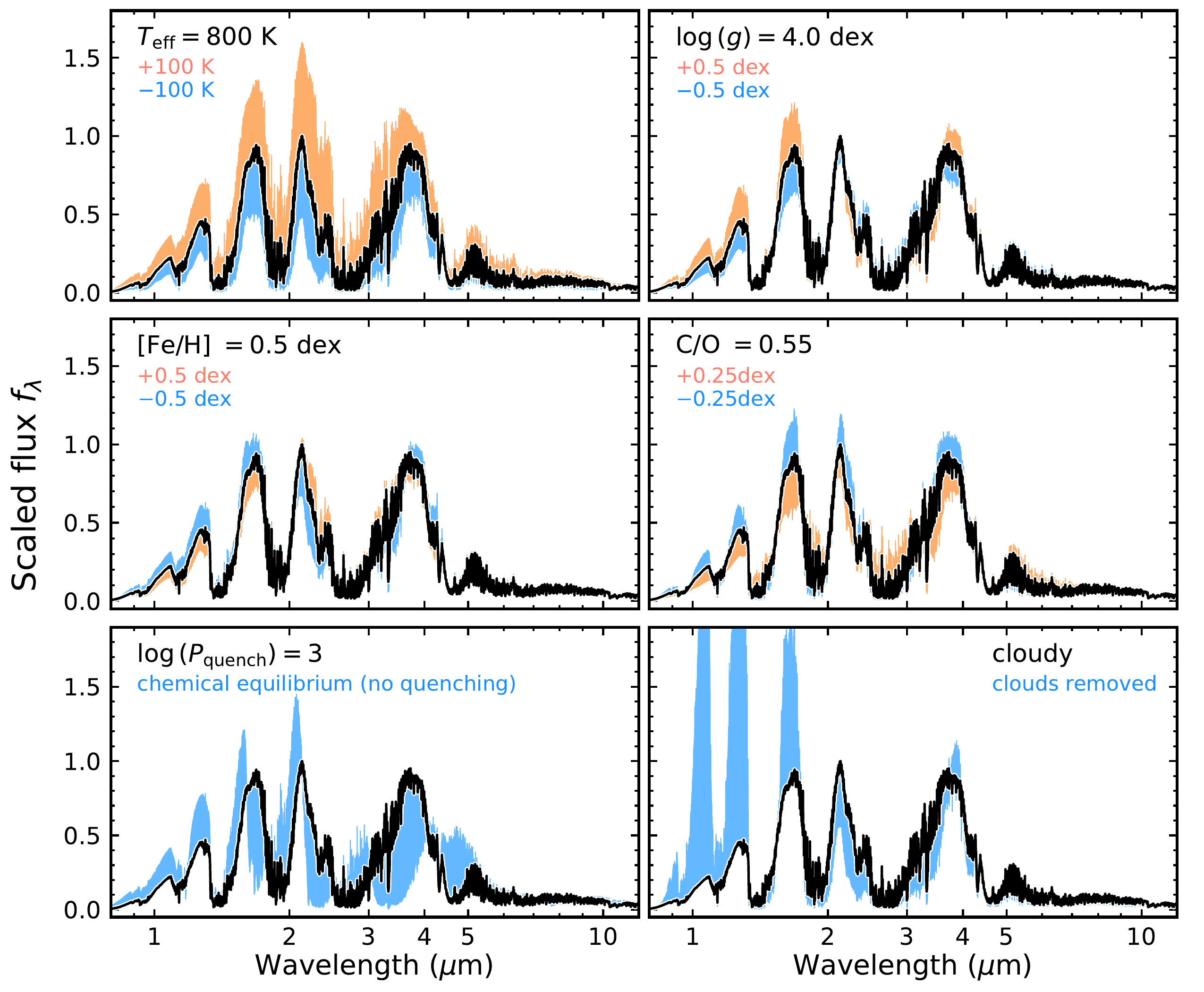}
\caption{The forward-modeled emission spectra ($R=1000$) over a mini-grid of parameter space in the vicinity of AF~Lep~b's properties (Section~\ref{subsec:emission_spec}). The black spectrum corresponds to a cloudy model with $T_{\rm eff}=800$~K, $\log{(g)}=4.0$~dex, [Fe/H]$=0.5$~dex, C/O$=0.55$~dex, $\log{(P_{\rm quench}/{\rm 1\ bar})}=3$, and $f_{\rm sed}=2$ and $\log{(K_{\rm zz})}=8$ for both MgSiO$_{3}$ and Fe clouds whose mass fractions at base pressures are $\log{(X_{0})} = -2$~dex and $-4$~dex, respectively. In each panel, we use orange/blue colors to show the spectral effect when a certain parameter is increased/decreased by the labeled amount (top and middle panels) or is removed (bottom panels).  }
\label{fig:mini_grid}
\end{center}
\end{figure*}

\subsection{Chemistry and Cloud Models}
\label{subsec:chem_cloud}

The chemistry and cloud models used in the retrievals are based on the approach described in \cite{2020A&A...640A.131M}, with a brief summary below. 

We first compute the abundances of all opacity sources at each atmospheric layer under the equilibrium chemistry. This calculation involves adding [Fe/H] and C/O as free parameters and combining them with T-P profiles in the retrievals \citep[also see][]{2017A&A...600A..10M}. To account for the effect of chemical dis-equilibrium, the logarithmic quench pressure $\log{(P_{\rm quench})}$ is added as a free parameter. The abundances (or mass fraction) of H$_{2}$O, CO, and CH$_{4}$ with $P < P_{\rm quench}$ are then re-set to the abundances with $P = P_{\rm quench}$ \citep[e.g.,][]{2014ApJ...797...41Z}. Thus, a higher $P_{\rm quench}$ value suggests that the dis-equilibrium chemistry impacts a wider vertical extent of the atmosphere.

The line species in our retrievals include: H$_{2}$O \citep{2018MNRAS.480.2597P}, CO \citep{1993sssp.book.....K, 2010JQSRT.111.2139R}, CO2 \citep{2020MNRAS.496.5282Y}, CH$_{4}$ \citep{2020ApJS..247...55H}, NH$_{3}$ \citep{2019MNRAS.490.4638C}, Na \citep{2019A&A...628A.120A}, K \citep{2016A&A...589A..21A}, PH$_{3}$ \citep{2015MNRAS.446.2337S}, VO and TiO \citep[B. Plez, private communication; see][]{2019A&A...627A..67M}, FeH \citep{2010A&A...523A..58W}, and H$_{2}$S \citep{2016MNRAS.460.4063A}. We also include H$_{2}$ and He as opacity sources of Rayleigh scattering \citep{1962ApJ...136..690D, 1965PPS....85..227C} and include H$_{2}$--H$_{2}$ and H$_{2}$--He as sources of the collision-induced absorption \citep[CIA;][]{1988ApJ...326..509B, 1989ApJ...336..495B, 1989ApJ...341..549B, 2001JQSRT..68..235B, 2002A&A...390..779B, 2012JQSRT.113.1276R}.

Cloud opacity is described by the mass fraction profile $X(P)$ of each cloud species, the mean size of the cloud particles $r_{g}$, and the width of the log-normal cloud particle size distribution $\sigma_{g}$ \citep{2019A&A...627A..67M}. For a given condensate, the cloud mass fraction profile is defined above the pressure of the cloud base $P_{\rm base}$, determined by the intersection between the T-P profile and the corresponding saturation vapor pressure curve. The $X(P)$ profile is computed as $X_{0} \left(P / P_{\rm base}\right)^{f_{\rm sed}}$, with $X_{0}$ being the cloud mass fraction at the base pressure and $f_{\rm sed}$ being the sedimentation efficiency. The mean size of cloud particles $r_{g}$ is determined by $f_{\rm sed}$ and the eddy diffusion coefficient $K_{\rm zz}$ (assumed to be independent of pressure) following the \cite{2001ApJ...556..872A} prescription. As explained by \cite{2020A&A...640A.131M}, the $K_{\rm zz}$ parameter in our framework only helps to determine the cloud particle size distribution; it might be inconsistent with $\log{(P_{\rm quench})}$, which is a dedicated parameter for the disequilibrium chemistry. 

The cloud species considered in our retrievals are MgSiO$_{3}$, Fe, and KCl, with irregular shapes and crystalline structures. The former two condensates are important for objects near the L/T transition \citep[e.g.,][]{1996A&A...308L..29T, 1986ApJ...310..238L, 2001ApJ...556..357A, 2002ApJ...568..335M, 2012ApJ...754..135M, 2006asup.book....1L}. Also, with the cool effective temperature of AF~Lep~b ($T_{\rm eff} \approx 650$~K; Table~\ref{tab:evoparams}), chloride and sulfide clouds become non-negligible \citep[e.g.,][]{2012ApJ...756..172M}. According to the microphysics models by \cite{2020NatAs...4..951G}, the KCl cloud formation is more efficient than the sulfide clouds given the fast nucleation rates of the former, so KCl clouds are added in our retrievals. All cloud condensates are assumed to share the $K_{\rm zz}$ and $\sigma_{g}$. Each species corresponds to an independent combination of ($X_{0}$, $f_{\rm sed}$), amounting to a total of eight free parameters in the cloud model.

{ 
\begin{deluxetable*}{lll}
\setlength{\tabcolsep}{15pt} 
\tablecaption{Free Parameters of Retrievals} \label{tab:retrieval_params} 
\tablehead{ \multicolumn{1}{l}{Parameter} &  \multicolumn{1}{l}{Prior} &  \multicolumn{1}{l}{Description} } 
\startdata 
\hline 
\multicolumn{3}{c}{Temperature Model (Sections~\ref{subsec:tp_profile}--\ref{subsec:rce_prior})}   \\ 
\hline 
$T_{\rm bottom}$  &  $\mathcal{U}(2\times10^{3}\ {\rm K}, 10^{4}\ {\rm K})$  &   Temperature at $P=10^{3}$ bar.   \\ 
$\left(d\ln{T}/d\ln{P}\right)_{1}$  &  $\mathcal{N}(\mu=0.250, \sigma=0.025)$  &   Temperature gradient at $P = 10^{3}$ bar.    \\ 
$\left(d\ln{T}/d\ln{P}\right)_{2}$  &  $\mathcal{N}(\mu=0.250, \sigma=0.045)$  &   Temperature gradient at $P \approx 63$ bar.    \\ 
$\left(d\ln{T}/d\ln{P}\right)_{3}$  &  $\mathcal{N}(\mu=0.260, \sigma=0.050)$  &   Temperature gradient at $P \approx 4$ bar.    \\ 
$\left(d\ln{T}/d\ln{P}\right)_{4}$  &  $\mathcal{N}(\mu=0.200, \sigma=0.050)$  &   Temperature gradient at $P \approx 0.3$ bar.    \\ 
$\left(d\ln{T}/d\ln{P}\right)_{5}$  &  $\mathcal{N}(\mu=0.120, \sigma=0.045)$  &   Temperature gradient at $P \approx 0.02$ bar.    \\ 
$\left(d\ln{T}/d\ln{P}\right)_{6}$  &  $\mathcal{N}(\mu=0.070, \sigma=0.070)$  &   Temperature gradient at $P = 10^{-3}$ bar.   \\  
\hline  
\multicolumn{3}{c}{Chemistry Model (Section~\ref{subsec:chem_cloud})}   \\ 
\hline  
[Fe/H]  &  $\mathcal{U}(-1\ {\rm dex}, 2\ {\rm dex})$  &   Iron abundance (relative to solar) of the exoplanet atmosphere.    \\ 
C/O  &  $\mathcal{U}(0.1, 1.6)$  &   Absolute carbon-to-oxygen ratio of the exoplanet atmosphere.    \\ 
$\log{(P_{\rm quench}/1\ {\rm bar})}$  &  $\mathcal{U}(-6\ {\rm dex}, 3\ {\rm dex})$  &   Quench pressure of H$_{2}$O, CH$_{4}$, and CO.   \\  
\hline  
\multicolumn{3}{c}{Cloud Model (Section~\ref{subsec:chem_cloud})}   \\ 
\hline   
$\log{(X_{0,{\rm MgSiO_{3}}})}$  &  $\mathcal{U}(-10\ {\rm dex}, 0\ {\rm dex})$  &   Mass fraction of the MgSiO$_{3}$ cloud at base pressure.   \\  
$f_{\rm sed,{\rm MgSiO_{3}}}$  &  $\mathcal{U}(0, 10)$  &   Sedimentation efficiency of the MgSiO$_{3}$ cloud.   \\  
$\log{(X_{0,{\rm Fe}})}$  &  $\mathcal{U}(-10\ {\rm dex}, 0\ {\rm dex})$  &   Mass fraction of the Fe cloud at base pressure.    \\  
$f_{\rm sed,{\rm Fe}}$  &  $\mathcal{U}(0, 10)$  &   Sedimentation efficiency of the Fe cloud.    \\ 
$\log{(X_{0,{\rm KCl}})}$  &  $\mathcal{U}(-10\ {\rm dex}, 0\ {\rm dex})$  &   Mass fraction of the KCl cloud at base pressure.   \\  
$f_{\rm sed,{\rm KCl}}$  &  $\mathcal{U}(0, 10)$  &   Sedimentation efficiency of the KCl cloud.    \\ 
$\log{(K_{\rm zz})}$  &  $\mathcal{U}(5\ {\rm dex}, 13\ {\rm dex})$  &   Vertical eddy diffusion coefficient of clouds.   \\   
$\sigma_{g}$  &  $\mathcal{U}(1.02, 3)$  &   Width of the log-normal cloud particle size distribution.     \\ 
\hline  
\multicolumn{3}{c}{Other Physical Parameters\tablenotemark{\scriptsize a} (Section~\ref{subsec:emission_spec})}   \\ 
\hline   
$\log{(g)}$  &  $\mathcal{U}(2.5\ {\rm dex}, 5.5\ {\rm dex})$  &   Surface gravity of the planet.    \\  
$R$  &  $\mathcal{U}(0.5\ {R_{\rm Jup}}, 2.5\ {R_{\rm Jup}})$  &   Radius of the planet.    \\  
\hline   
\multicolumn{3}{c}{Combined Spectral Dataset\tablenotemark{\scriptsize b} (Section~\ref{subsec:emission_spec})}   \\ 
\hline   
$\Delta{f}_{\rm M23}$  &  $\mathcal{U}(-3 \times f_{\rm max,M23}, 3 \times f_{\rm max,M23})$  &   Flux offset for the \cite{2023AandA...672A..93M} IFS spectrum.   \\
$\Delta{f}_{\rm D23}$  &  $\mathcal{U}(-3 \times f_{\rm max,D23}, 3 \times f_{\rm max,D23})$  &   Flux offset for the \cite{2023AandA...672A..94D} IFS spectrum.     
\enddata 
\tablenotetext{a}{Some of our retrieval runs adopt narrower and well-constrained priors on mass and radius (see Section~\ref{sec:retrieval}).}  
\tablenotetext{b}{The $f_{\rm max,D23}$ and $f_{\rm max,M23}$ represent the maximum flux of the \cite{2023AandA...672A..94D} and \cite{2023AandA...672A..93M} IFS spectra, respectively.}  
\end{deluxetable*} 
}

\subsection{Emission Spectroscopy}
\label{subsec:emission_spec}

In the retrievals, emission spectra are generated via \texttt{petitRADTRANS} by combining the T-P profile, line/continuum/cloud opacities, cloud scattering \citep[see][]{2020A&A...640A.131M}, and the planet's surface gravity $\log{(g)}$. The computed spectrum is then scaled by a factor of $(R/d)^{2}$ for comparison with the observed data, where $d = 26.8$~pc is the distance of AF~Lep~A and $R$ is the planet radius as a free parameter. When analyzing the two sets of IFS spectra of AF~Lep~b collected by \cite{2023AandA...672A..94D} and \cite{2023AandA...672A..93M}, an additive flux offset is implemented for each spectrum ($\Delta{f}_{\rm D23}$ and $\Delta{f}_{\rm M23}$) as free parameters (see discussions in Section~\ref{subsec:data_planet}). However, when analyzing only one spectrum, the originally observed flux is used. In addition, when photometric data are included in the retrievals, we compute the $K1/K2/L'$ photometry from the modeled emission spectrum by using the response curves obtained from the VLT/SPHERE\footnote{\url{https://www.eso.org/sci/facilities/paranal/instruments/sphere/inst/filters.html}} and Keck/NIRC2\footnote{\url{https://www2.keck.hawaii.edu/inst/nirc2/filters.html}} websites (also see Figure~\ref{fig:specphot}). 

Examples of emission spectra are shown in Figure~\ref{fig:mini_grid}. These forward-modeled spectra are calculated over a small parameter space around the AF~Lep~b's properties, with $T_{\rm eff} =$700--900~K. $\log{(g)}=$3.5--4.5~dex, [Fe/H]$=$0--1~dex, C/O$=$0.3--0.8, and $\log{(P_{\rm quench}/{\rm 1\ bar})}=3$. Clouds of MgSiO$_{3}$ and Fe are incorporated, with $\log{(X_{0})} = -2$~dex and $-4$~dex, respectively, and $f_{\rm sed} = 2$ and $\log{(K_{\rm zz})}=8$ for both condensates. Each spectrum is generated based on a self-consistent T-P profile computed under the RCE using the \texttt{petitCODE} \citep{2015ApJ...813...47M, 2017A&A...600A..10M}. These forward-modeled spectra are not used for the analysis of AF~Lep~b but are provided as examples to illuminate the effect of different physical parameters on spectral morphology.

\subsection{Free Parameters and the Nested Sampling}
\label{subsec:chem_clouds}

Table~\ref{tab:retrieval_params} summarizes all free parameters and their corresponding priors used in the retrievals. The \texttt{PyMultiNest} \citep{2014A&A...564A.125B}, building upon \texttt{MultiNest} \citep{2008MNRAS.384..449F, 2009MNRAS.398.1601F, 2019OJAp....2E..10F} is employed by \texttt{petitRADTRANS} for nested sampling. We adopt 4000 live points to sample the parameter posteriors and set a 0.05 sampling efficiency under the constant efficiency mode of \texttt{MultiNest}.

\begin{deluxetable*}{lccccccccccc}
\setlength{\tabcolsep}{1pt} 
\tablecaption{Retrieved Atmospheric Properties of AF Lep b} \label{tab:retrieval_params_key_runs} 
\tablehead{ \multicolumn{1}{l}{Parameters} &  \multicolumn{3}{c}{Default Priors (Table~\ref{tab:retrieval_params}) } &  \multicolumn{1}{c}{} &  \multicolumn{3}{c}{Plus Constrained $M$ Prior} &  \multicolumn{1}{c}{} &  \multicolumn{3}{c}{Plus Constrained $M$\&$R$ Priors} \\ 
\cline{2-4} \cline{6-8} \cline{10-12} 
\multicolumn{1}{l}{} &  \multicolumn{1}{c}{Phot. +} &  \multicolumn{1}{c}{Phot. +} &  \multicolumn{1}{c}{Phot. +} &  \multicolumn{1}{c}{} &  \multicolumn{1}{c}{Phot. +} &  \multicolumn{1}{c}{Phot. +} &  \multicolumn{1}{c}{Phot. +} &  \multicolumn{1}{c}{} &  \multicolumn{1}{c}{Phot. +} &  \multicolumn{1}{c}{Phot. +} &  \multicolumn{1}{c}{Phot. +} \\ 
\multicolumn{1}{l}{} &  \multicolumn{1}{c}{M23 Spec.} &  \multicolumn{1}{c}{D23 Spec.} &  \multicolumn{1}{c}{M23\&D23 Spec.} &  \multicolumn{1}{c}{} &  \multicolumn{1}{c}{M23 Spec.} &  \multicolumn{1}{c}{D23 Spec.} &  \multicolumn{1}{c}{M23\&D23 Spec.} &  \multicolumn{1}{c}{} &  \multicolumn{1}{c}{M23 Spec.} &  \multicolumn{1}{c}{D23 Spec.} &  \multicolumn{1}{c}{M23\&D23 Spec.}  }  
\startdata 
\multicolumn{12}{c}{Retrieved Physical and Chemical Properties}   \\ 
\hline   
[Fe/H] &  $1.2^{+0.4}_{-0.5}$ &  $-0.4^{+0.5}_{-0.3}$ &  $0.6^{+0.4}_{-0.4}$ & &  $1.2^{+0.3}_{-0.4}$ &  $1.0^{+0.2}_{-0.3}$ &  $1.3^{+0.4}_{-0.4}$ & &  $1.6^{+0.2}_{-0.2}$ &  $1.7^{+0.3}_{-0.3}$ &  $1.6^{+0.2}_{-0.2}$   \\ 
C/O &  $0.60^{+0.12}_{-0.26}$ &  $0.26^{+0.14}_{-0.09}$ &  $0.42^{+0.08}_{-0.10}$ & &  $0.57^{+0.15}_{-0.37}$ &  $0.75^{+0.06}_{-0.17}$ &  $0.34^{+0.19}_{-0.15}$ & &  $0.74^{+0.07}_{-0.23}$ &  $0.82^{+0.05}_{-0.58}$ &  $0.70^{+0.09}_{-0.12}$   \\ 
$\log{(g)}$ &  $3.40^{+1.37}_{-0.51}$ &  $2.87^{+0.37}_{-0.23}$ &  $2.95^{+0.36}_{-0.26}$ & &  $3.92^{+0.12}_{-0.12}$ &  $4.08^{+0.13}_{-0.12}$ &  $4.06^{+0.14}_{-0.11}$ & &  $3.65^{+0.04}_{-0.05}$ &  $3.66^{+0.03}_{-0.04}$ &  $3.65^{+0.04}_{-0.05}$   \\ 
$R$ &  $0.92^{+0.13}_{-0.10}$ &  $0.67^{+0.07}_{-0.06}$ &  $0.79^{+0.09}_{-0.08}$ & &  $0.93^{+0.13}_{-0.12}$ &  $0.78^{+0.10}_{-0.11}$ &  $0.79^{+0.11}_{-0.12}$ & &  $1.27^{+0.06}_{-0.04}$ &  $1.24^{+0.05}_{-0.03}$ &  $1.27^{+0.06}_{-0.04}$   \\ 
$\log{(P_{\rm quench}/1\ {\rm bar})}$ &  $2.0^{+0.7}_{-4.8}$ &  $-3.0^{+2.0}_{-1.6}$ &  $-3.0^{+1.5}_{-1.5}$ & &  $2.3^{+0.6}_{-6.1}$ &  $1.6^{+0.5}_{-0.7}$ &  $-3.4^{+2.3}_{-1.4}$ & &  $2.4^{+0.3}_{-1.9}$ &  $1.1^{+0.6}_{-5.6}$ &  $2.5^{+0.3}_{-1.1}$   \\ 
\hline   
\multicolumn{12}{c}{Derived Physical Properties}   \\ 
\hline   
$T_{\rm eff}$ &  $912^{+67}_{-52}$ &  $1120^{+41}_{-45}$ &  $1017^{+50}_{-52}$ & &  $910^{+60}_{-53}$ &  $1002^{+63}_{-50}$ &  $984^{+62}_{-46}$ & &  $806^{+26}_{-21}$ &  $848^{+13}_{-17}$ &  $789^{+22}_{-20}$   \\ 
$\log{(L_{\rm bol}/L_{\odot})}$ &  $-5.23^{+0.06}_{-0.11}$ &  $-5.18^{+0.03}_{-0.03}$ &  $-5.19^{+0.04}_{-0.04}$ & &  $-5.24^{+0.04}_{-0.04}$ &  $-5.24^{+0.04}_{-0.04}$ &  $-5.26^{+0.06}_{-0.04}$ & &  $-5.18^{+0.05}_{-0.04}$ &  $-5.11^{+0.03}_{-0.03}$ &  $-5.22^{+0.04}_{-0.04}$   \\ 
\hline   
\multicolumn{12}{c}{Cloud Properties}   \\ 
\hline   
$f_{\rm sed,{\rm MgSiO_{3}}}$ &  $2.9^{+1.3}_{-1.0}$ &  $5.9^{+2.4}_{-2.9}$ &  $5.3^{+2.0}_{-1.9}$ & &  $2.4^{+1.1}_{-0.9}$ &  $4.9^{+2.2}_{-2.1}$ &  $2.8^{+1.3}_{-1.1}$ & &  $2.2^{+1.4}_{-0.9}$ &  $5.4^{+2.0}_{-2.1}$ &  $2.9^{+1.2}_{-1.1}$   \\ 
$f_{\rm sed,{\rm Fe}}$ &  $5.3^{+2.4}_{-2.5}$ &  $6.4^{+2.1}_{-2.7}$ &  $5.3^{+2.5}_{-2.6}$ & &  $5.8^{+2.3}_{-2.8}$ &  $5.4^{+2.7}_{-2.8}$ &  $5.3^{+2.5}_{-2.6}$ & &  $5.6^{+2.4}_{-2.7}$ &  $5.5^{+2.7}_{-2.9}$ &  $5.8^{+2.3}_{-2.6}$   \\ 
$f_{\rm sed,{\rm KCl}}$ &  $4.6^{+2.7}_{-2.4}$ &  $4.7^{+2.9}_{-2.7}$ &  $4.5^{+2.7}_{-2.5}$ & &  $4.8^{+2.7}_{-2.4}$ &  $5.2^{+2.8}_{-2.9}$ &  $5.1^{+2.7}_{-2.8}$ & &  $4.6^{+2.8}_{-2.5}$ &  $5.2^{+2.9}_{-3.0}$ &  $5.1^{+2.6}_{-2.7}$   \\ 
$\log{(X_{0,{\rm MgSiO_{3}}})}$ &  $-1.3^{+0.7}_{-0.8}$ &  $-5.2^{+2.4}_{-2.7}$ &  $-1.4^{+0.8}_{-0.9}$ & &  $-1.3^{+0.7}_{-0.8}$ &  $-1.6^{+0.9}_{-1.1}$ &  $-1.3^{+0.7}_{-0.8}$ & &  $-1.8^{+0.9}_{-1.5}$ &  $-1.8^{+1.2}_{-4.7}$ &  $-1.4^{+0.8}_{-1.0}$   \\ 
$\log{(X_{0,{\rm Fe}})}$ &  $-5.6^{+2.7}_{-2.3}$ &  $-3.2^{+1.7}_{-3.4}$ &  $-6.1^{+2.7}_{-2.1}$ & &  $-5.1^{+2.9}_{-2.7}$ &  $-5.5^{+3.0}_{-2.6}$ &  $-5.3^{+2.7}_{-2.5}$ & &  $-5.6^{+2.8}_{-2.5}$ &  $-5.4^{+3.0}_{-2.8}$ &  $-4.6^{+2.6}_{-2.8}$   \\ 
$\log{(X_{0,{\rm KCl}})}$ &  $-6.1^{+2.2}_{-2.0}$ &  $-5.4^{+2.8}_{-2.6}$ &  $-4.6^{+2.5}_{-2.7}$ & &  $-5.6^{+2.2}_{-2.2}$ &  $-6.2^{+2.2}_{-2.1}$ &  $-5.8^{+2.3}_{-2.2}$ & &  $-6.3^{+2.2}_{-2.0}$ &  $-6.2^{+2.6}_{-2.3}$ &  $-6.4^{+2.1}_{-1.9}$   \\ 
$\log{(K_{\rm zz})}$ &  $8.1^{+1.6}_{-1.3}$ &  $8.8^{+1.8}_{-1.9}$ &  $8.6^{+1.9}_{-1.9}$ & &  $8.9^{+1.5}_{-1.8}$ &  $8.5^{+1.6}_{-1.8}$ &  $8.5^{+1.5}_{-1.7}$ & &  $8.7^{+1.5}_{-1.4}$ &  $8.0^{+2.7}_{-1.8}$ &  $8.5^{+1.6}_{-1.5}$   \\ 
$\sigma_{g}$ &  $1.2^{+0.4}_{-0.1}$ &  $1.2^{+0.5}_{-0.2}$ &  $1.2^{+0.5}_{-0.2}$ & &  $1.2^{+0.5}_{-0.2}$ &  $1.2^{+0.5}_{-0.2}$ &  $1.2^{+0.5}_{-0.2}$ & &  $1.2^{+0.5}_{-0.2}$ &  $1.3^{+0.6}_{-0.2}$ &  $1.3^{+0.5}_{-0.2}$   \\ 
\hline   
\multicolumn{12}{c}{Flux Offsets of Spectroscopic Dataset}   \\ 
\hline   
$\Delta{f}_{\rm M23} \times 10^{18}$ &  -- &  -- &  $1.3^{+0.9}_{-0.7}$ & &  -- &  -- &  $1.3^{+0.8}_{-0.6}$ & &  -- &  -- &  $0.4^{+0.5}_{-0.5}$   \\ 
$\Delta{f}_{\rm D23} \times 10^{18}$ &  -- &  -- &  $-3.6^{+0.9}_{-0.8}$ & &  -- &  -- &  $-3.6^{+0.8}_{-0.6}$ & &  -- &  -- &  $-4.6^{+0.6}_{-0.6}$   \\ 
\hline   
\multicolumn{12}{c}{T-P Profile Properties}   \\ 
\hline   
$T_{\rm bottom}$ &  $8257^{+984}_{-1996}$ &  $8719^{+721}_{-915}$ &  $8908^{+605}_{-728}$ & &  $8043^{+979}_{-990}$ &  $7113^{+1031}_{-845}$ &  $8578^{+728}_{-954}$ & &  $7205^{+1065}_{-1230}$ &  $8015^{+1285}_{-1100}$ &  $7392^{+908}_{-996}$   \\ 
$(d\ln{T}/d\ln{P})_{1}$ &  $0.25^{+0.02}_{-0.02}$ &  $0.25^{+0.02}_{-0.02}$ &  $0.25^{+0.02}_{-0.02}$ & &  $0.25^{+0.02}_{-0.02}$ &  $0.25^{+0.02}_{-0.02}$ &  $0.25^{+0.02}_{-0.02}$ & &  $0.25^{+0.02}_{-0.02}$ &  $0.25^{+0.02}_{-0.02}$ &  $0.25^{+0.02}_{-0.02}$   \\ 
$(d\ln{T}/d\ln{P})_{2}$ &  $0.24^{+0.02}_{-0.02}$ &  $0.25^{+0.03}_{-0.02}$ &  $0.24^{+0.02}_{-0.02}$ & &  $0.25^{+0.02}_{-0.02}$ &  $0.25^{+0.03}_{-0.03}$ &  $0.24^{+0.02}_{-0.02}$ & &  $0.25^{+0.03}_{-0.03}$ &  $0.24^{+0.03}_{-0.03}$ &  $0.25^{+0.02}_{-0.03}$   \\ 
$(d\ln{T}/d\ln{P})_{3}$ &  $0.27^{+0.03}_{-0.03}$ &  $0.24^{+0.03}_{-0.03}$ &  $0.26^{+0.02}_{-0.02}$ & &  $0.26^{+0.03}_{-0.03}$ &  $0.26^{+0.04}_{-0.04}$ &  $0.26^{+0.03}_{-0.03}$ & &  $0.25^{+0.03}_{-0.05}$ &  $0.27^{+0.04}_{-0.04}$ &  $0.26^{+0.03}_{-0.04}$   \\ 
$(d\ln{T}/d\ln{P})_{4}$ &  $0.18^{+0.05}_{-0.03}$ &  $0.15^{+0.02}_{-0.02}$ &  $0.16^{+0.02}_{-0.02}$ & &  $0.18^{+0.04}_{-0.03}$ &  $0.19^{+0.04}_{-0.04}$ &  $0.17^{+0.03}_{-0.03}$ & &  $0.19^{+0.04}_{-0.08}$ &  $0.20^{+0.02}_{-0.03}$ &  $0.20^{+0.03}_{-0.04}$   \\ 
$(d\ln{T}/d\ln{P})_{5}$ &  $0.10^{+0.03}_{-0.03}$ &  $0.10^{+0.03}_{-0.03}$ &  $0.08^{+0.02}_{-0.02}$ & &  $0.12^{+0.03}_{-0.04}$ &  $0.13^{+0.04}_{-0.04}$ &  $0.11^{+0.04}_{-0.05}$ & &  $0.13^{+0.03}_{-0.04}$ &  $0.13^{+0.03}_{-0.04}$ &  $0.13^{+0.03}_{-0.03}$   \\ 
$(d\ln{T}/d\ln{P})_{6}$ &  $0.07^{+0.05}_{-0.05}$ &  $0.07^{+0.05}_{-0.05}$ &  $0.06^{+0.05}_{-0.05}$ & &  $0.07^{+0.05}_{-0.05}$ &  $0.07^{+0.06}_{-0.06}$ &  $0.08^{+0.05}_{-0.05}$ & &  $0.07^{+0.05}_{-0.05}$ &  $0.07^{+0.06}_{-0.06}$ &  $0.06^{+0.05}_{-0.05}$   \\ 
\enddata 
\tablecomments{The properties listed in the last column is recommended as the nominal atmospheric properties of AF~Lep~b, given that these results are determined by combining the planet's all available spectrophotometry with its independently measured dynamical mass and age incorporated. However, we note that, if any atmospheric properties have vastly inconsistent inferred values among all these nine retrieval runs (e.g., C/O), then they should be interpreted with caution.}
\end{deluxetable*}

\begin{figure*}[t]
\begin{center}
\includegraphics[height=7.in]{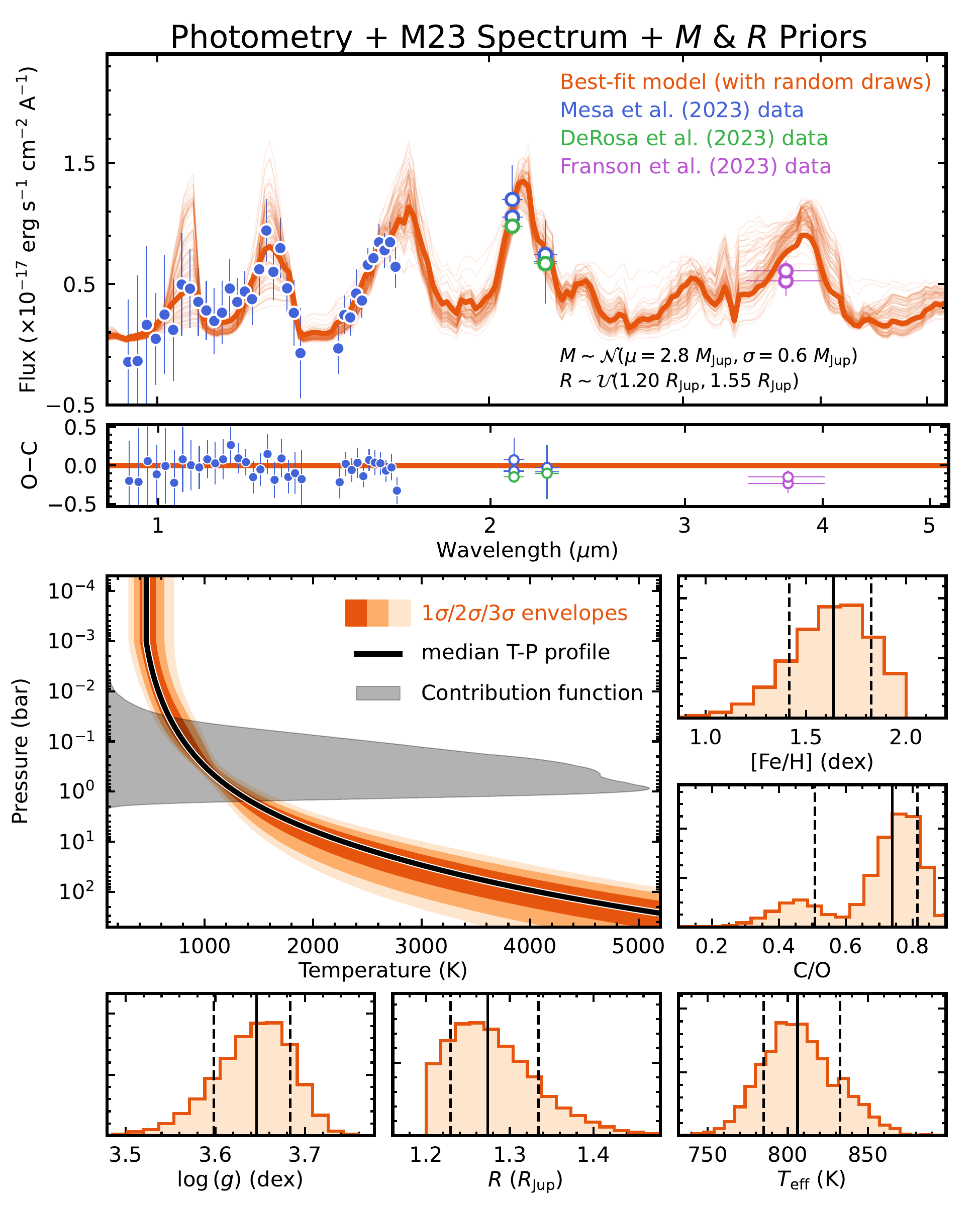}
\caption{Results of the retrieval analysis on $K1/K2/L'$ photometry and the \cite{2023AandA...672A..93M} spectrum of AF~Lep~b (Section~\ref{subsec:phot_m23spec}). In the top two panels, we compare the observed spectrophotometry (colored circles with the same format as Figure~\ref{fig:specphot}; the wavelength errorbars of $K1/K2/L'$ photometric data represent the half effective widths of the corresponding filters) with the emission spectrum corresponding to the best-fit model (thick orange line). Emission spectra generated at 100 random draws from the parameter posteriors (thin orange lines) are overlaid. In the middle panel on the left, we present the $1\sigma/2\sigma/3\sigma$ confidence intervals (orange shades) of our retrieved T-P profiles. A profile with median T-P parameters is shown in black and the corresponding weighted contribution function (computed over the same wavelength range as the input data) is shown as a grey shade. The remaining panels present the posterior distributions of key physical parameters, including [Fe/H], C/O, $\log{(g)}$, $R$, and $T_{\rm eff}$. The median and confidence intervals of all parameters are summarized in Table~\ref{tab:retrieval_params_key_runs}.  }
\label{fig:results_m23}
\end{center}
\end{figure*}

\begin{figure*}[t]
\begin{center}
\includegraphics[height=7.in]{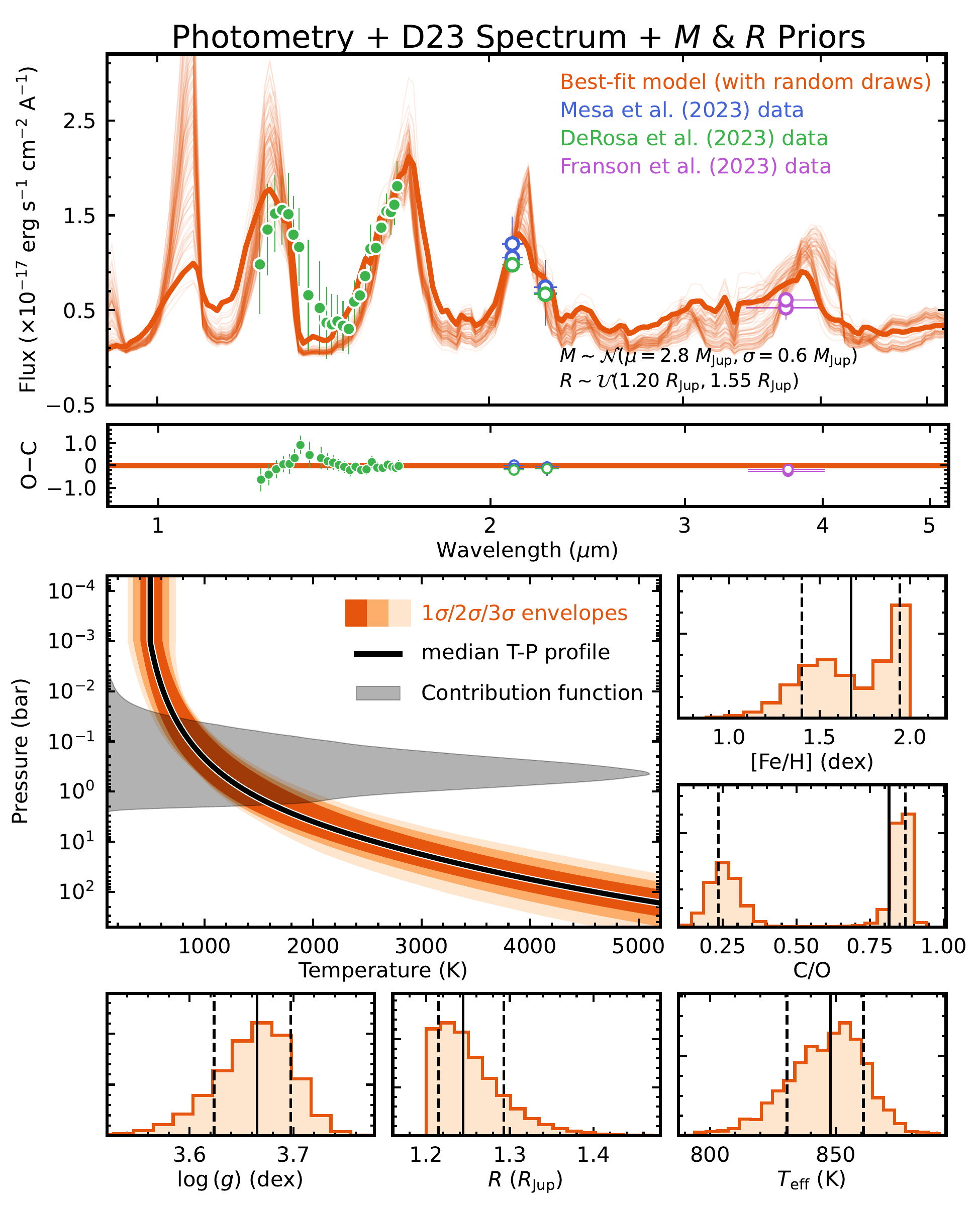}
\caption{Results of the retrieval analysis on $K1/K2/L'$ photometry and the \cite{2023AandA...672A..94D} spectrum of AF~Lep~b (Section~\ref{subsec:phot_d23spec}), with the same format as Figure~\ref{fig:results_m23}.}
\label{fig:results_d23}
\end{center}
\end{figure*}

\section{Retrieval Analysis of AF Lep \lowercase{b}} 
\label{sec:retrieval}

We perform retrievals on three sets of input data: 
\begin{enumerate}
\item[(1)] All published $K1/K2/L'$ photometry and the \cite{2023AandA...672A..93M} spectrum, 
\item[(2)] All $K1/K2/L'$ photometry and the \cite{2023AandA...672A..94D} spectrum, and 
\item[(3)] All $K1/K2/L'$ photometry and the spectra from both \cite{2023AandA...672A..93M} and \cite{2023AandA...672A..94D}. 
\end{enumerate}
For each dataset, three sets of parameter priors are adopted: 
\begin{enumerate}
\item[$\bullet$] We first use the default priors listed in Table~\ref{tab:retrieval_params}.
\item[$\bullet$] On the basis of Table~\ref{tab:retrieval_params}, a Gaussian prior is adopted for the planet mass ($M$) based on the directly measured dynamical mass (Section~\ref{sec:orbit}), as $\mathcal{N}(\mu = 2.8\ M_{\rm Jup}, \sigma=0.6\ M_{\rm Jup})$. This new prior is more constrained compared to the default mass prior of $\mathcal{U}(0.03\ M_{\rm Jup}, 797\ M_{\rm Jup})$ in Table~\ref{tab:retrieval_params}, as converted from the default $\log{(g)}$ and $R$ priors.
\item[$\bullet$] On the basis of Table~\ref{tab:retrieval_params}, we add constrained priors on both the planet mass as $\mathcal{N}(\mu = 2.8\ M_{\rm Jup}, \sigma=0.6\ M_{\rm Jup})$, and the planet radius as $\mathcal{U}(1.20\ R_{\rm Jup}, 1.55\ R_{\rm Jup})$. This new radius prior is contextualized by various evolution models (Section~\ref{sec:evo_params}).
\end{enumerate}
In total, we perform nine retrieval runs (given three datasets and three sets of priors). The bolometric luminosity and effective temperature are further derived after each retrieval run. Specifically, we use \texttt{petitRADTRANS} to generate emission spectra over a wavelength range of $0.6-250$~$\mu$m based on parameters sampled from the inferred posteriors. Then we extrapolate each spectrum to zero flux at $0$~$\mu$m and append a Rayleigh-Jeans tail toward longer wavelengths up to 1000~$\mu$m. The $L_{\rm bol}$ is computed by integrating the spectrum and $T_{\rm eff}$ is derived by combining $L_{\rm bol}$ with the retrieved $R$ following the Stefan-Boltzmann law. 

\begin{figure*}[t]
\begin{center}
\includegraphics[height=7.in]{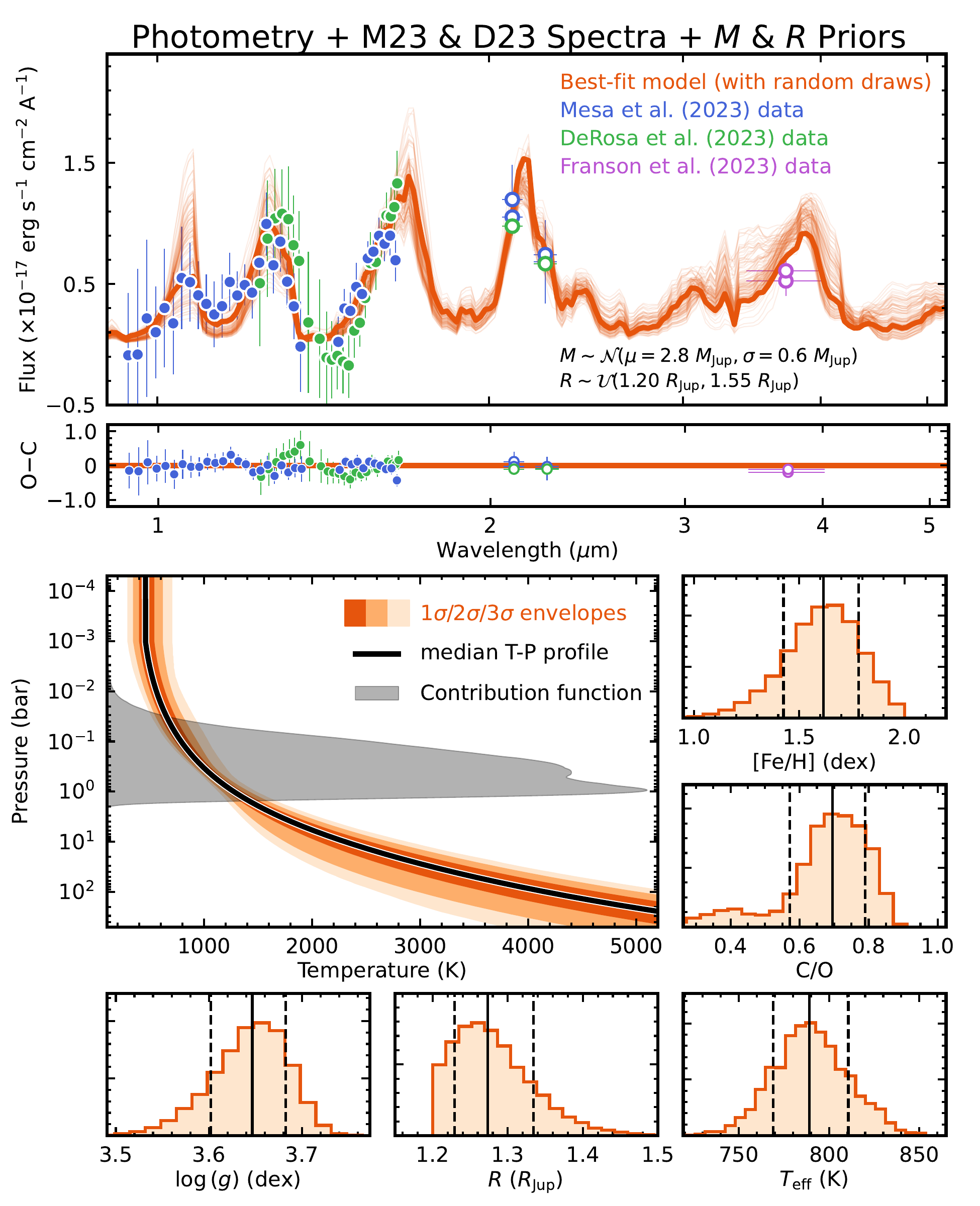}
\caption{Results of the retrieval analysis on $K1/K2/L'$ photometry and both \cite{2023AandA...672A..93M} and \cite{2023AandA...672A..94D} spectra of AF~Lep~b (Section~\ref{subsec:phot_d23spec}), with the same format as Figure~\ref{fig:results_m23}.}
\label{fig:results_m23d23}
\end{center}
\end{figure*}

In Figures~\ref{fig:results_m23}--\ref{fig:results_m23d23}, we present the results of our retrieval analysis for each input dataset, with the constrained $M$ and $R$ priors incorporated. Similar figures for other combinations of input datasets and priors can be found in Appendix~\ref{app:retrieval_with_phot}. Figure~\ref{fig:example_corner} presents the parameter posteriors for two selected retrieval runs. Figure~\ref{fig:violin_key} summarizes and compares the posteriors of key physical parameters from all these nine retrievals. The median and $1\sigma$ confidence intervals of all parameters are listed in Table~\ref{tab:retrieval_params_key_runs}. For the rest of this section, we discuss our retrieval results on each input dataset with each set of priors.

\begin{figure*}[t]
\begin{center}
\includegraphics[height=6.5in]{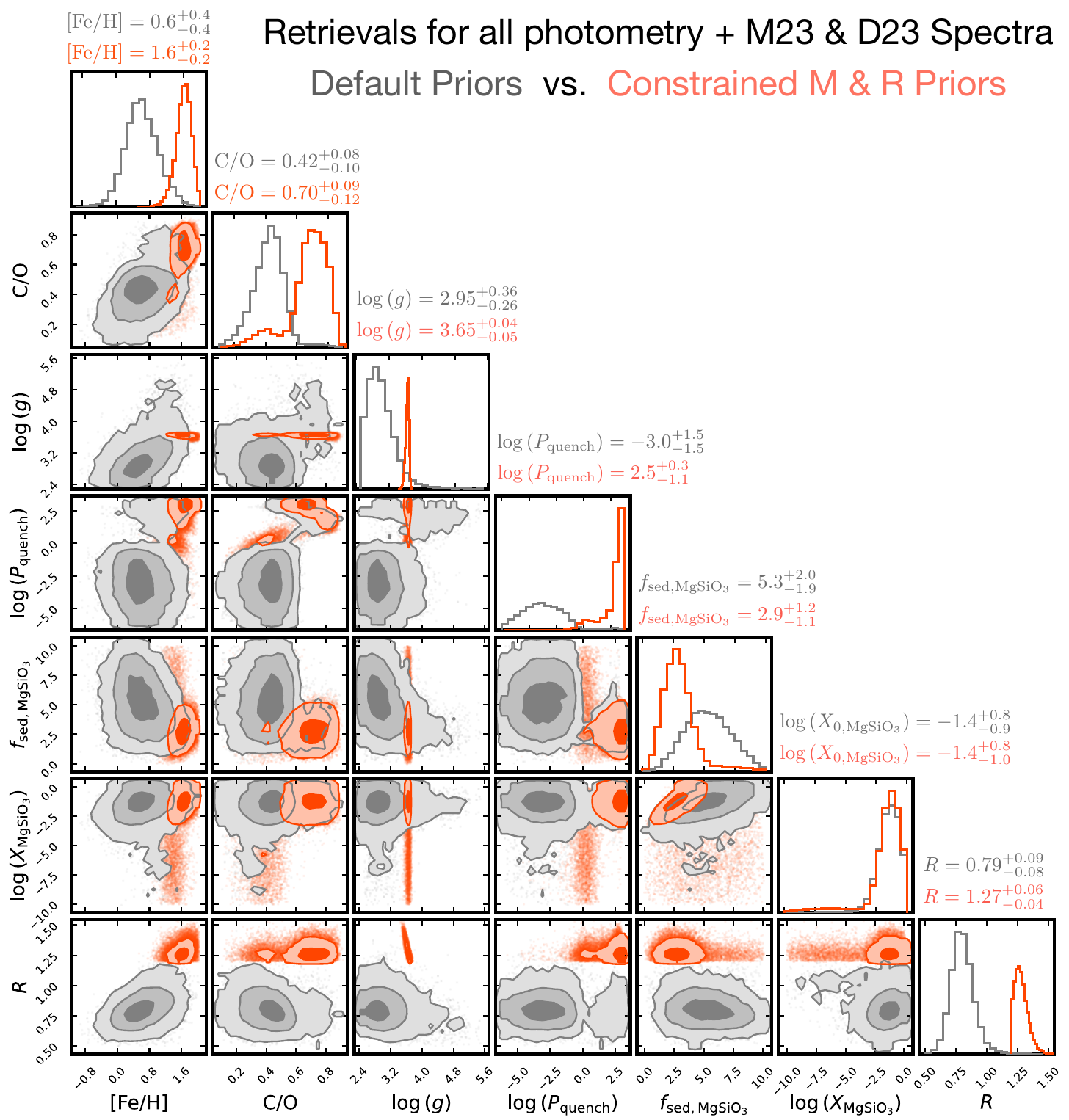}
\caption{Corner plot for the retrieval runs performed with default parameter priors (grey) and with constrained priors on the planet's mass as $\mathcal{N}(\mu = 2.8\ {M_{\rm Jup}}, \sigma = 0.6\ {M_{\rm Jup}})$ and radius as $\mathcal{U}(1.20\ {R_{\rm Jup}}, 1.55\ {R_{\rm Jup}})$ (orange). We show $1\sigma/2\sigma/3\sigma$ confidence intervals for parameters inferred by the former retrieval (with default priors), but only present $1\sigma/2\sigma$ confidence intervals for the latter in order to prevent visual overlap. Both retrieval runs are performed on $K1/K2/L'$ photometry and both \cite{2023AandA...672A..93M} and \cite{2023AandA...672A..94D} spectra (Section~\ref{subsec:phot_m23d23spec}). }
\label{fig:example_corner}
\end{center}
\end{figure*}

\begin{figure*}[t]
\begin{center}
\includegraphics[height=8.2in]{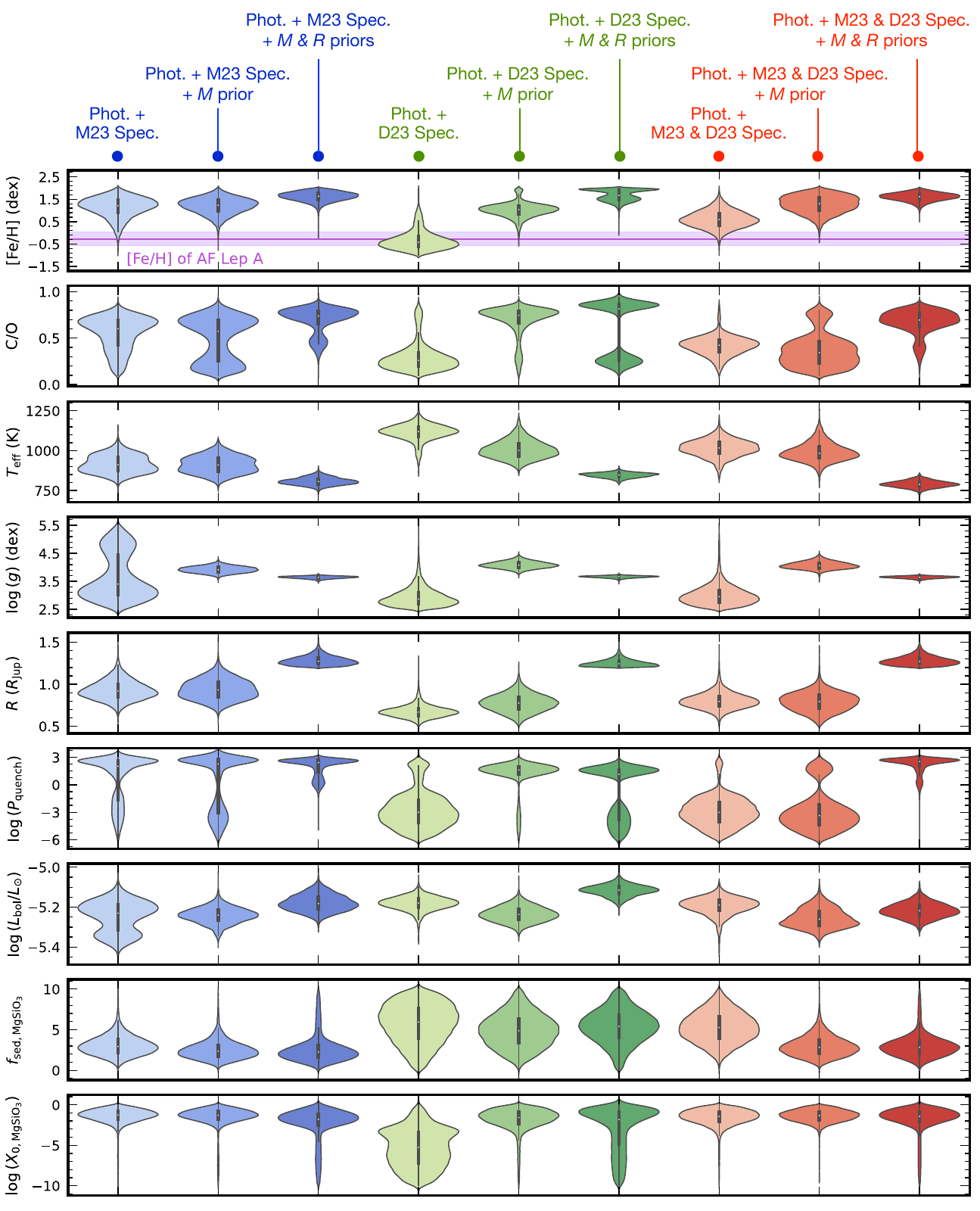}
\caption{Violin plot for the posteriors of several key physical and chemical parameters derived from all the nine retrieval runs (Section~\ref{sec:retrieval}). The left three columns (blue) correspond to the retrievals on the $K1/K2/L'$ photometry and the \cite{2023AandA...672A..93M} spectrum (Section~\ref{subsec:phot_m23spec}). The middle three columns (green) correspond to the retrievals on the $K1/K2/L'$ photometry and the \cite{2023AandA...672A..94D} spectrum (Section~\ref{subsec:phot_d23spec}). The right three columns (red) correspond to the retrievals on the $K1/K2/L'$ photometry and both \cite{2023AandA...672A..93M} and \cite{2023AandA...672A..94D} spectra. Plots with darker blue/green/red colors suggest the addition of narrower and constrained priors on $M$ (slightly darker) or both $M$ and $R$ (much darker). The measured [Fe/H] of the host star AF~Lep~A ($-0.27 \pm 0.31$~dex; Section~\ref{sec:host_param_abund}) is shown as the purple shade in the top panel. }
\label{fig:violin_key}
\end{center}
\end{figure*}

\begin{figure}[t]
\includegraphics[height=2.3in]{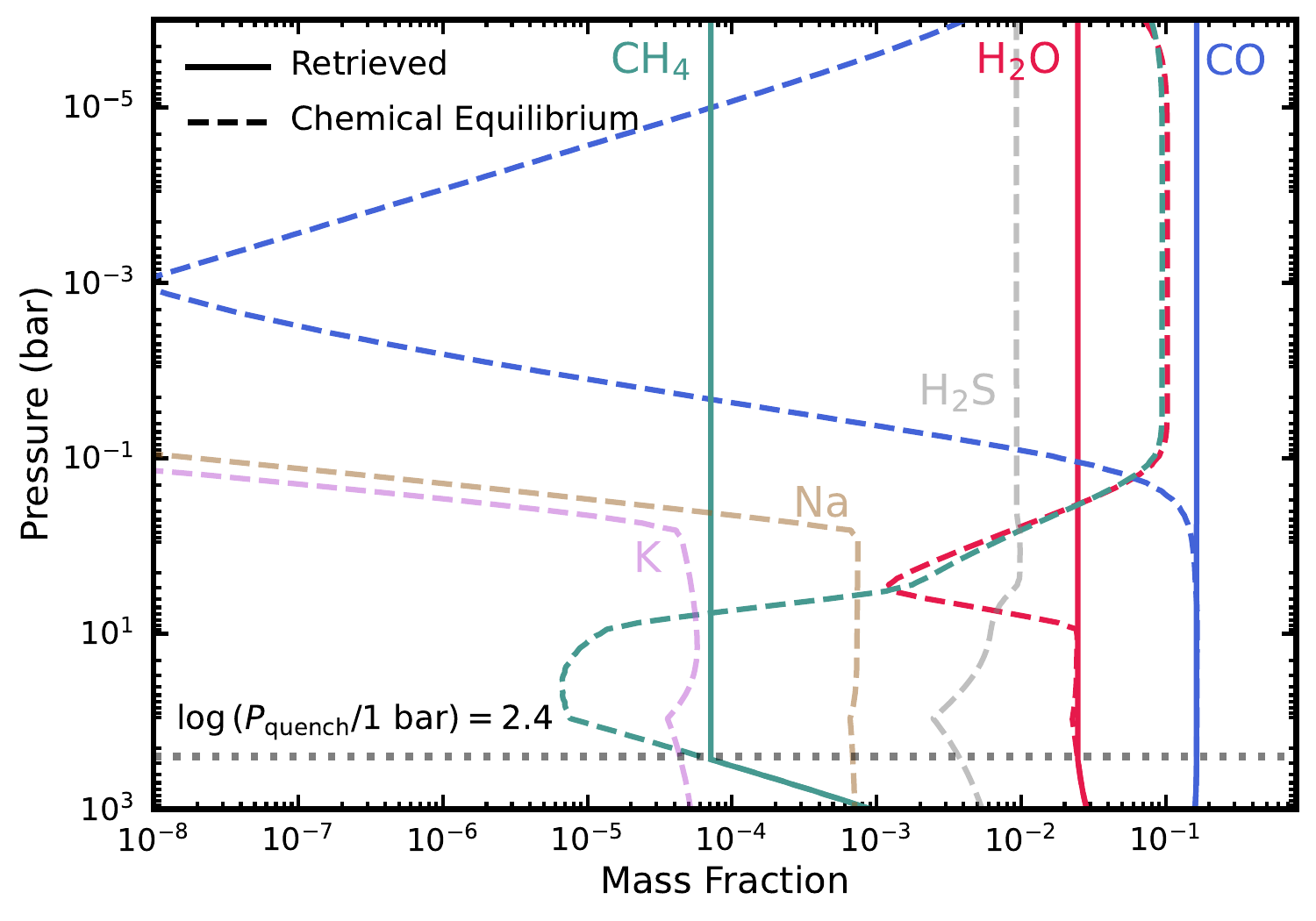}
\caption{The abundance of several key species (solid) inferred from the retrieval performed on the $K1/K2/L'$ photometry and the \cite{2023AandA...672A..93M} spectrum, along with constrained priors on the planet's mass and radius (Section~\ref{subsec:phot_m23spec}). These profiles correspond to the median values of parameters as summarized in Table~\ref{tab:retrieval_params_key_runs}, including a quenching pressure at $10^{2.4}$~bar. Dashed lines present the abundances under chemical equilibrium. For H$_{2}$O, CO, and CH$_{4}$, our retrieved profiles suggest the evidence of dis-equilibrium chemistry in the atmosphere of AF~Lep~b; for the other species (e.g., Na, K, H$_{2}$S), we assume their abundances follow the equilibrium chemistry in our retrievals (see Section~\ref{subsec:chem_cloud}).   } 
\label{fig:example_abundance_profile}
\end{figure}

\subsection{Retrievals on the $K1/K2/L'$ photometry and the \cite{2023AandA...672A..93M} spectrum}
\label{subsec:phot_m23spec}

When analyzing all available photometry and the \cite{2023AandA...672A..93M} spectrum of AF~Lep~b, regardless of whether the constrained priors on $M$ and $R$ are adopted, the retrievals consistently predict higher [Fe/H] values of the planet compared to its host star (Figure~\ref{fig:violin_key}). The enhanced metallicity of the planet is potentially linked to its formation history and will be discussed in Section~\ref{subsec:formation}. The retrieved C/O values all have large uncertainties and are consistent with solar abundance. The relative C/O between the planet and its host star is unknown, given that the stellar C/O cannot be reliably measured due to the fast stellar rotation (Section~\ref{sec:host_param_abund}). The precision of retrieved $\log{(g)}$ improves when a constrained $M$ prior is included. When a narrow $R$ prior is also adopted, the resulting $\log{(g)}$ posterior primarily reflects the $M$ and $R$ priors.

With a default radius prior of $\mathcal{U}{(0.5\ {\rm R_{\rm Jup}}, 2.5\ {\rm R_{\rm Jup}})}$, the retrieved $R$ of the planet falls in the range of $0.8$--$1.1$~R$_{\rm Jup}$. This radius is considered slightly small given the planet's dynamical mass and its young age based on evolution models (Section~\ref{sec:evo_params}). This ``small radius problem'' also occurred in other spectroscopic studies of brown dwarfs and giant planets \citep[e.g.,][]{2020ApJ...905...46G, 2021MNRAS.506.1944B, 2021ApJ...921...95Z, 2022ApJ...930..136L, 2022ApJ...937...54X, 2023arXiv230304885H}. The underestimated radius of the planet leads to an overestimated effective temperature, given that the derived bolometric luminosity is mostly consistent among different retrieval runs (Figure~\ref{fig:violin_key}). After incorporating a constrained and narrower $R$ prior, the resulting $T_{\rm eff}$ decreases from $\approx 900$~K to $\approx 800$~K. 

Retrievals of the input dataset also imply that the atmosphere of AF~Lep~b is likely impacted by the disequilibrium chemistry, as the retrieved $\log{(P_{\rm quench}/1\ {\rm bar})}$ consistently hovers around 2.3~dex in all three runs with different sets of parameter priors. Turning off the dis-equilibrium chemistry in our best-fit model spectrum results in an excessive CH$_{4}$ absorption that is incompatible with the observed spectrophotometry. The effect of dis-equilibrium chemistry on the resulting profiles of H$_{2}$O, CO, and CH$_{4}$ abundances is also demonstrated by Figure~\ref{fig:example_abundance_profile}. 

In addition, AF~Lep~b likely has silicate clouds (e.g., MgSiO$_{3}$) in the atmosphere with a sedimention effeciency of $f_{\rm sed} = 2$--$3$ and a high mass fraction of $\log{(X_{0})} \approx -1.5$. The properties of the other two cloud species, Fe and KCl, cannot be well-constrained, as their retrieved $f_{\rm sed}$ and $\log{(X_{0})}$ posteriors remain close to the adopted priors. The presence of clouds is also supported by the very red slope of the spectrophotometry, especially considering the relatively flat slope of the \cite{2023AandA...672A..93M} spectrum in $Y$ and $J$ bands, as well as the comparable or brighter $K1/K2$ photometry than fluxes at shorter wavelengths (Figure~\ref{fig:results_m23}).

\subsection{Retrievals on the $K1/K2/L'$ photometry and the \cite{2023AandA...672A..94D} spectrum}
\label{subsec:phot_d23spec}

When analyzing $K1/K2/L'$ photometry and the \cite{2023AandA...672A..94D} spectrum with default $M$ and $R$ priors, our retrieval results show that the atmospheric [Fe/H] of AF~Lep~b is consistent with the metallicity of its host star. However, in this retrieval run, the derived planet radius of $R = 0.67^{+0.07}_{-0.06}$~R$_{\rm Jup}$ is unphysically small. When combined with the derived $\log{(g)} = 2.87^{+0.37}_{-0.23}$~dex, this radius implies a very low planet mass ($0.13^{+0.17}_{-0.05}$~M$_{\rm Jup}$) that contradicts the observed astrometric properties of the system (Section~\ref{sec:orbit}). By incorporating more constrained and physics-driven priors for $M$, or both $M$ and $R$, the retrieved [Fe/H] of AF~Lep~b becomes higher than that of its host star, as seen from the retrievals obtained with the \cite{2023AandA...672A..93M} spectrum (Section~\ref{subsec:phot_m23spec}). Also, the impact of disequilibrium chemistry is still likely, given that $\log{(P_{\rm quench}/1\ {\rm bar})}$ spans $1.1-1.6$~dex with constrained $M$ and/or $R$ priors.

When default parameter priors are adopted, the presence of any type of cloud is not suggested by this dataset. Unlike the spectrum by \cite{2023AandA...672A..93M}, the \cite{2023AandA...672A..94D} spectrum has overall higher fluxes (Figure~\ref{fig:specphot}) and the flux near the peaks of $J$ and $H$ bands are slightly higher than the $K1/K2$ photometry, leading to a spectral slope indicative of a cloudless atmosphere. However, with constrained priors on $M$ and/or $R$, the mass fraction of the MgSiO$_{3}$ cloud significantly increases from $-5.2$~dex to $-1.8$~dex at the base pressure. This result suggests that the silicate cloud plays a more crucial role than Fe and KCl in shaping the planet's emission spectrum, as also suggested by the retrievals described in Section~\ref{subsec:phot_m23spec}.

\subsection{Retrievals on the $K1/K2/L'$ photometry and both \cite{2023AandA...672A..93M} and \cite{2023AandA...672A..94D} spectra}
\label{subsec:phot_m23d23spec}

By including an additive flux offset to each IFS spectrum, our retrieval analysis successfully explains all available photometry and spectra of AF~Lep~b with the fitted model spectra (Figure~\ref{fig:results_m23d23}). Similar to the retrievals described in Sections~\ref{subsec:phot_m23spec} and \ref{subsec:phot_d23spec}, the retrieved [Fe/H] of AF~Lep~b is enhanced compared to its host star's metallicity and this planet's atmosphere is likely influenced by silicate clouds (e.g., MgSiO$_{3}$), with a mass fraction around $-1.4$~dex at base pressure and a $f_{\rm sed}$ around $3$. The presence of disequilibrium chemistry is strongly suggested only when the constrained priors on both $M$ and $R$ are adopted. In addition, our retrievals suggest that a positive flux offset for the \cite{2023AandA...672A..93M} spectrum and a negative flux offset for the \cite{2023AandA...672A..94D} spectrum are required for the models to simultaneously explain all the spectrophotometry of AF~Lep~b.

For the interpretation of our analysis in the following section, we adopt the retrieval results inferred by combining all archival photometry and spectroscopy, with constrained priors for both $M$ and $R$ incorporated (unless otherwise noted). However, we note that, if any atmospheric properties have vastly inconsistent inferred values among all the nine retrieval runs (e.g., C/O), then these parameters should be interpreted with caution.

\section{Discussion} 
\label{sec:discussion}

\subsection{Formation Pathway of AF~Lep~b} 
\label{subsec:formation}

\subsubsection{Potential Metal Enrichment}
\label{subsubsec:metal_enrichment}

By investigating the population-level trends between the masses of gas-giant exoplanets and the metallicities of their host stars, \cite{2018ApJ...853...37S} suggested that core accretion and gravitational instabilities --- the two dominant planet formation mechanisms --- might operate at distinct mass regimes. They found gas giants with masses below $4-10$~$M_{\rm Jup}$ likely formed via core accretion, while those with higher masses likely formed via gravitational instabilities. Although such critical planet mass differs in other studies \citep[e.g.,][]{2011IAUS..276..117S, 2017A&A...603A..30S}, the range of $4-10$~M$_{\rm Jup}$ is among the lowest values. Therefore, based on the dynamical mass measurement of $2.8^{+0.6}_{-0.5}$~M$_{\rm Jup}$, AF~Lep~b is likely a product of core accretion according to \cite{2018ApJ...853...37S}. 

The enhanced metallicity of AF~Lep~b compared to its host star, as revealed by our study, also lines up with the core accretion formation scenario. As shown in Figure~\ref{fig:violin_key}, our retrieved [Fe/H] of AF~Lep~b is higher than the [Fe/H] of AF~Lep~A at a $1.9\sigma-5.3\sigma$ significance, regardless of (1) whether the individual spectrum from \cite{2023AandA...672A..93M} or \cite{2023AandA...672A..94D}, or both spectra are included in the retrievals, and (2) whether constrained and physically driven priors on the planet's mass, or both mass and radius are adopted.\footnote{As a reminder, the constrained mass prior is a Gaussian prior based on the directly measured dynamical mass of $\mathcal{N}(\mu=2.8\ M_{\rm Jup}, \sigma=0.6\ M_{\rm Jup})$; the constrained radius prior is a uniform prior contextualized by the evolution model analysis as $\mathcal{U}(1.20\ R_{\rm Jup}, 1.55\ R_{\rm Jup})$; these are introduced in Section~\ref{sec:retrieval}. Incorporating these priors allows the retrieval results to become consistent with observations beyond spectrophotometry and match the physical expectations. } The only exception occurs when we perform a retrieval run using all the photometry and only the \cite{2023AandA...672A..94D} spectrum, without constrained priors on the planet's mass or radius; in this case, the inferred [Fe/H] of the planet and the host star is consistent within $0.4\sigma$. However, this run also predicts a mass of $0.13^{+0.17}_{-0.05}$~M$_{\rm Jup}$ based on the retrieved $\log{(g)}$ and $R$, which is $4.3\sigma$ lower than the planet's measured dynamical mass, thus undermining the accuracy of its inferred planet's metallicity. The enhanced [Fe/H] of AF~Lep~b is also supported by the brighter $K$-band flux than the fluxes at shorter wavelengths, as seen in the retrievals for the \cite{2023AandA...672A..93M} spectrum or both spectra (with offsets applied). The photosphere with a metal-rich composition resides at lower pressures, where the collision-included absorption of H$_{2}-$H$_{2}$ and H$_{2}-$He becomes weaker, leading to higher $K$-band fluxes (e.g., \citealt{2008ApJ...683.1104F}; also see Figure~\ref{fig:mini_grid}).

Metal enrichment is consistent with the predictions of core-accretion models of planet formation and has been suggested for the solar system's giant planets \citep[e.g.,][]{2005ApJ...622L.145A, 2014ApJ...789...69H}, as well as extrasolar planets \citep[e.g.,][]{2011ApJ...736L..29M, 2014A&A...566A.141M, 2016ApJ...831...64T, 2019ApJ...874L..31T}. Building upon the work of \cite{2011ApJ...736L..29M}, \cite{2016ApJ...831...64T} studied the composition of a sample of transiting planets with directly measured masses and radii, finding an anti-correlation between the metal-enrichment of gas-giant planets ($Z_{\rm planet}/Z_{\rm star}$) and the planet's mass \citep[also see][]{2019AJ....158..239T}. The enhanced metallicity of gas giants can result from the accretion of pebbles and planetesimals \citep{1999Natur.402..269O, 2005ApJ...622L.145A, 2007ApJ...666..447Z, 2014ApJ...789...69H, 2016arXiv160604510A, 2021ApJ...918L..23M, 2021A&A...654A..72S}, which are composed of volatile and refractory materials with slightly different characteristics. Pebbles are coupled to gas and can be accreted into gas-giant's atmospheres along with gas until the planet is sufficiently massive to open a gap. Also, pebbles can drift inward across ice lines within the protoplanetary disk, followed by the evaporation that alters the chemical content of the disk gas and thereby the gas-giant planets' atmospheres \citep[e.g.,][]{2021A&A...654A..71S}. In contrast, planetesimals have larger sizes and are less affected by the aerodynamic gas drag, allowing planetesimal accretion to occur during the late stage of planet formation when pebble accretion is halted \citep[see reviews by][]{2014prpl.conf..643H, 2022arXiv220309759D}. 

Indeed, planetesimal accretion and the resulting metal-enrichment of planets are not an exclusive outcome of core accretion, but can also occur for planets formed via gravitational instabilities \citep[e.g.,][]{2000ASPC..219..475G, 2006Icar..185...64H, 2010Icar..207..503H, 2009ApJ...697.1256H, 2010ApJ...724..618B}. In addition, as discussed in \cite{2016ApJ...831...64T}, the metal enrichment might also result from the gap opening of planets and the subsequent starvation of gas accretion. 

Based on our analysis, AF~Lep~b has an enhanced metallicity than its host star by a factor of $Z_{\rm planet}/Z_{\rm star} = 75^{+94}_{-42}$.\footnote{To derive this metal-enrichment factor, we use the [Fe/H] chain retrieved from a run that incorporates all archival photometry and both spectra of AF~Lep~b, with constrained priors on the planet's mass and radius. We randomly draw the host star's [Fe/H] from a Gaussian distribution with an equal sample size to the planet's [Fe/H]. } At a mass of approximately $2.8$~M$_{\rm Jup}$, this inferred metal enrichment of AF~Lep~b is higher than the median level of the exoplanet sample studied in \cite{2016ApJ...831...64T}, although several planets in this work exhibited similarly high metal enrichment as seen in AF~Lep~b. It is possible that both pebble and planetesimal accretion impact the formation and early evolution of AF~Lep~b, leading to its enhanced metallicity. In particular, the late-stage planetesimal accretion also coincides with the presence of the debris disk in the planetary system at $40-60$~au \citep{2021MNRAS.502.5390P, 2022A&A...659A.135P}, which suggests a potentially large metal reservoir in the disk. As discussed by \cite{2023ApJ...950L..19F}, at its currently observed orbit, AF~Lep~b has a sufficient mass to dynamically stir the debris disk, triggering planetesimal collisions that potentially replenish the dust. We note that some of these planetesimals may be scattered inward and bombard the planet's atmosphere to enrich its atmospheric metallicity. 

An alternative explanation for AF~Lep~b's enhanced atmospheric metallicity is the possibility of giant impacts and planetary mergers \citep[e.g.,][]{2020MNRAS.498..680G, 2022MNRAS.509.1413A}. These events could be common occurrences during the advanced evolution stages of protoplanetary disks and might also lead to core erosion, the stripping of planets' hydrogen and helium envelopes, and altered orbital architecture of planetary systems \citep[e.g.,][]{1997ApJ...477..781L, 2010ApJ...720.1161L, 2015MNRAS.446.1685L, 2019MNRAS.485.4454B, 2019ApJ...884L..47F, 2019Natur.572..355L}.   

In addition, it is likely that AF~Lep~b has a diluted core, similar to gas and ice giants in solar system \citep[e.g.,][]{1995JGR...10023349M, 2011ApJ...726...15H, 2017GeoRL..44.4649W, 2019ApJ...872..100D} and likely exoplanets as well \citep[e.g.,][]{2019ApJ...874L..31T}. In this scenario, heavy elements from the planetary interior mix with the atmospheres, leading to enhanced atmospheric metallicity. \cite{2019ApJ...874L..31T} studied the metal enrichment of exoplanets as a function of planet mass with different levels of interior-atmosphere mixing. At a mass of 2.8~M$_{\rm Jup}$, our inferred $Z_{\rm planet}/Z_{\rm star}$ of AF~Lep~b lines up with the maximum metal-enrichment values, as shown in \cite{2019ApJ...874L..31T}, when assuming the interior and the atmosphere are fully mixed.

Beyond [Fe/H], comparing the C/O of AF~Lep~b and AF~Lep~A provides additional constraints to the planet formation pathways, including the initial formation location and the relative gas and dust accreted to assemble the planet's mass \citep[e.g.,][]{2011ApJ...743L..16O, 2014ApJ...794L..12M, 2017MNRAS.469.4102M, 2021A&A...654A..72S, 2022ApJ...934...74M}. However, studying the C/O ratio for this particular system is complicated by the difficulty in constraining the stellar C/O due to strong stellar rotation, which weakens and blends the characteristic lines of C and O (Section~\ref{subsec:stellar_abund}). In addition, the inferred C/O values of AF~Lep~b are less consistent among different retrieval runs compared to the case of [Fe/H] (Figure~\ref{fig:violin_key}). Also, when retrieving the \cite{2023AandA...672A..94D} spectrum with constrained priors on the planet's mass and radius, a bimodal distribution is seen in the C/O posterior. These two modes correspond to a super-solar and a sub-solar C/O, even though the host star does not necessarily have a solar C/O ratio. Extending the spectrophotometry of AF~Lep~b to a wider range of wavelengths with a higher S/N and/or spectral resolution may help to further constrain the planet's C/O and refine other atmospheric parameters.

\begin{figure*}[t]
\begin{center}
\includegraphics[height=3.8in]{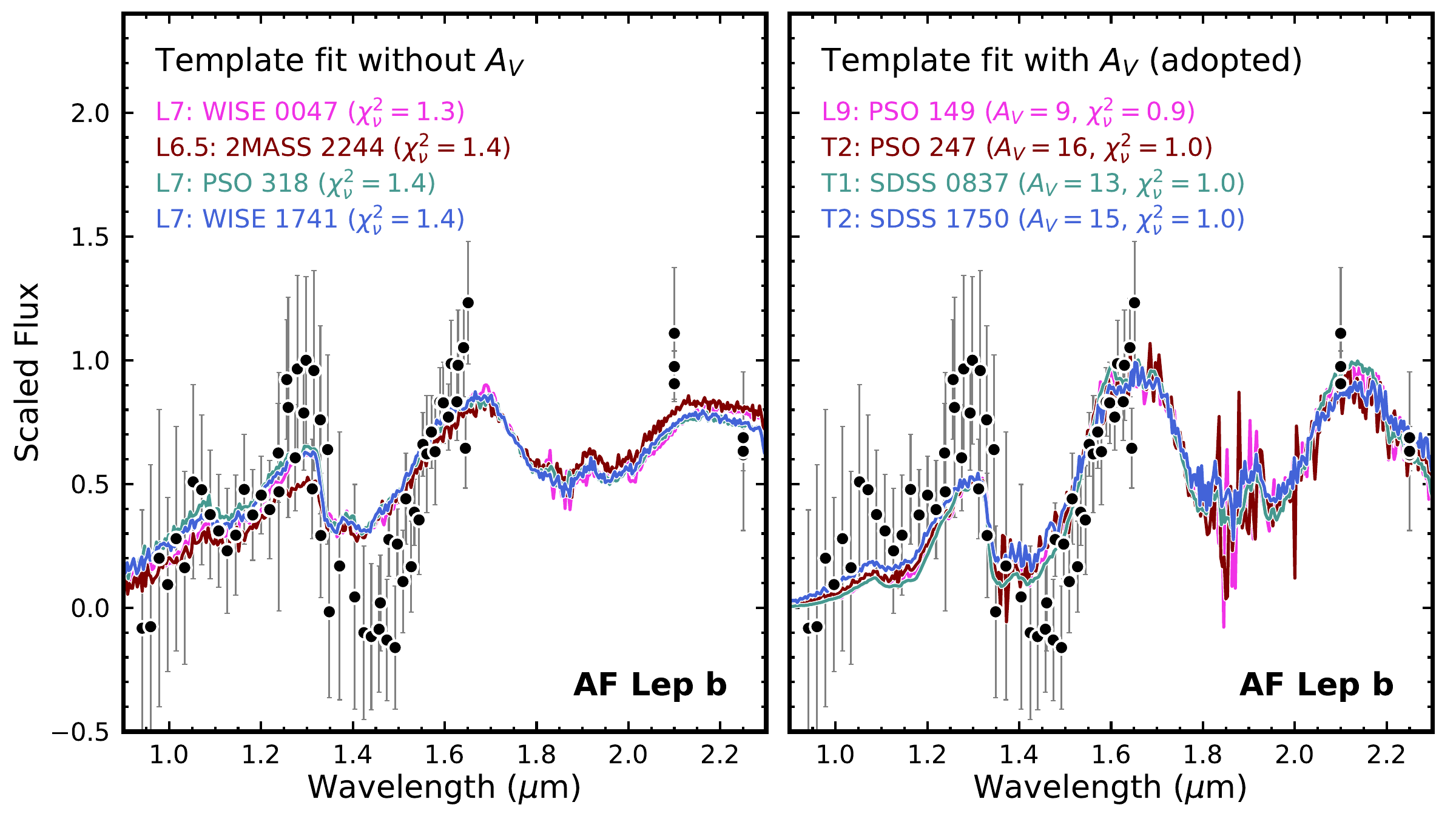}
\caption{Spectrophotometry of AF~Lep~b (black) compared to the top four best-matched ultracool dwarfs obtained from the template fitting that does not (left) and does (right) incorporate the reddening law of the interstellar medium, which we use to mimic the cloud effect for the purpose of spectral typing. }
\label{fig:spt}
\end{center}
\end{figure*}

\subsubsection{Planet Formation at a Later Epoch?}
\label{subsubsec:late_epoch}

Here we compare the physical properties of AF~Lep~b inferred by the evolution models (Section~\ref{sec:evo} and Table~\ref{tab:evoparams}) and atmospheric retrievals (Section~\ref{sec:retrieval} and Table~\ref{tab:retrieval_params_key_runs}). As discussed in Section~\ref{sec:evo_motivation}, discrepancies between evolution models and atmospheric models are well-recognized in the field. One specific example of the discrepancy is that the radii and surface gravities inferred by atmospheric models can be inconsistent with the object's independently known age and mass. Indeed, as shown in Table~\ref{tab:retrieval_params_key_runs}, if constrained priors on the planet's mass and radius are not incorporated, then our retrievals can derive unphysically low radii (down to $0.67$~R$_{\rm Jup}$) or too small $\log{(g)}$ that lead to a Saturn-like mass which is incompatible with the planet's orbit. 

It is these discrepancies that motivated us to adopt constrained priors on the planet's $M$ and $R$, in order to reliably characterize the atmospheric properties. However, even with these constrained priors, there are still differences between the retrieved effective temperature and bolometric luminosity and those estimated by various evolution models. Specifically, our retrievals infer $100-350$~K hotter effective temperature and $0.2-1.0$~dex brighter bolometric luminosity, when compared to the evolution models analyzed in Section~\ref{sec:evo} (i.e., comparing the last three columns of Table~\ref{tab:retrieval_params_key_runs} to Table~\ref{tab:evoparams}).

One possibility to reconcile these differences is if AF~Lep~b formed later than its host star, which would lead to a younger age of the planet and consequently increase the evolution-based $T_{\rm eff}$ and $L_{\rm bol}$. For hot-start evolution models, a 10~Myr younger age of AF~Lep~b significantly resolves the discrepancies (see Table~\ref{tab:younger_evoparams}), even though such a delayed planet formation timescale exceeds the typical lifetime of protoplanetary disks \citep[e.g.,][]{2009AIPC.1158....3M}. This scenario was also previously suggested by \cite{2023ApJ...950L..19F} albeit based on comparisons among a different set of parameters. Specifically, they found that if AF~Lep~b formed 5--15~Myr after its host star, then this planet's directly measured dynamical mass would be consistent with the mass predicted by several hot-start evolution models at the planet's age and an estimated bolometric luminosity.\footnote{\cite{2023ApJ...950L..19F} estimated the bolometric luminosity of AF~Lep~b as $\log{(L_{\rm bol}/L_{\odot})} = -4.81 \pm 0.13$~dex, using the $L'$-band absolute magnitude, with a bolometric correction assumed to match that of HR~8799~b. Their estimated bolometric luminosity is significantly brighter than the values inferred by our retrievals based on different portions of the planet's spectrophotometry (e.g., $-5.22 \pm 0.04$~dex; Table~\ref{tab:retrieval_params_key_runs}); it is possible that the bolometric corrections required by HR~8799~b and AF~Lep~b are different. New spectroscopy or photometry in longer wavelengths will be useful to refine the planet's bolometric luminosity.} For cold-start evolution models (\citealt{2012ApJ...745..174S, 2021ApJ...920...85M}; see Table~\ref{tab:younger_evoparams}), however, a 10~Myr younger age for the planet only slightly reduces the discrepancies between our retrieved and evolution-based $T_{\rm eff}$ and $L_{\rm bol}$ by about $30\%$. 

Another possibility is the occurrence of giant impacts and planetary mergers during the evolution history of AF~Lep~b, which is also a candidate explanation of the planet's metal enrichment (see Section~\ref{subsubsec:metal_enrichment}). In this scenario, the dissipation of kinematic energy into the planet's atmosphere and interior might act as a mechanism of rejuvenation, by altering its entropy state \citep[also see][]{2020MNRAS.498..680G}. This process can potentially result in elevated values for both the bolometric luminosity and effective temperature that deviate from the predictions of evolution models at the planet's current age. A thorough quantitative analysis would be useful to assess the viability of this hypothesis.

Acquiring new spectrophotometry of AF~Lep~b with a wider wavelength coverage than the existing data will provide more detailed information about the spectrum and bolometric flux of this planet. Such observation might in turn help to refine the $T_{\rm eff}$ and $\log{(L_{\rm bol}/L_{\odot})}$ inferred by retrievals. Also, ongoing theoretical advancements in the evolution models for planets formed via core accretion, incorporating different assumptions about the planetary atmospheres, will lead to a better understanding of the physical properties of planets formed via core accretion \citep[e.g.,][]{2017A&A...608A..72M, 2021A&A...656A..69E}. The combination of observational and theoretical progress of exoplanets will be a topic of continuing research, and bring insights into the discrepancies between the predictions of retrievals and evolution models as seen in AF~Lep~b.

\begin{figure*}[t]
\begin{center}
\includegraphics[height=3.4in]{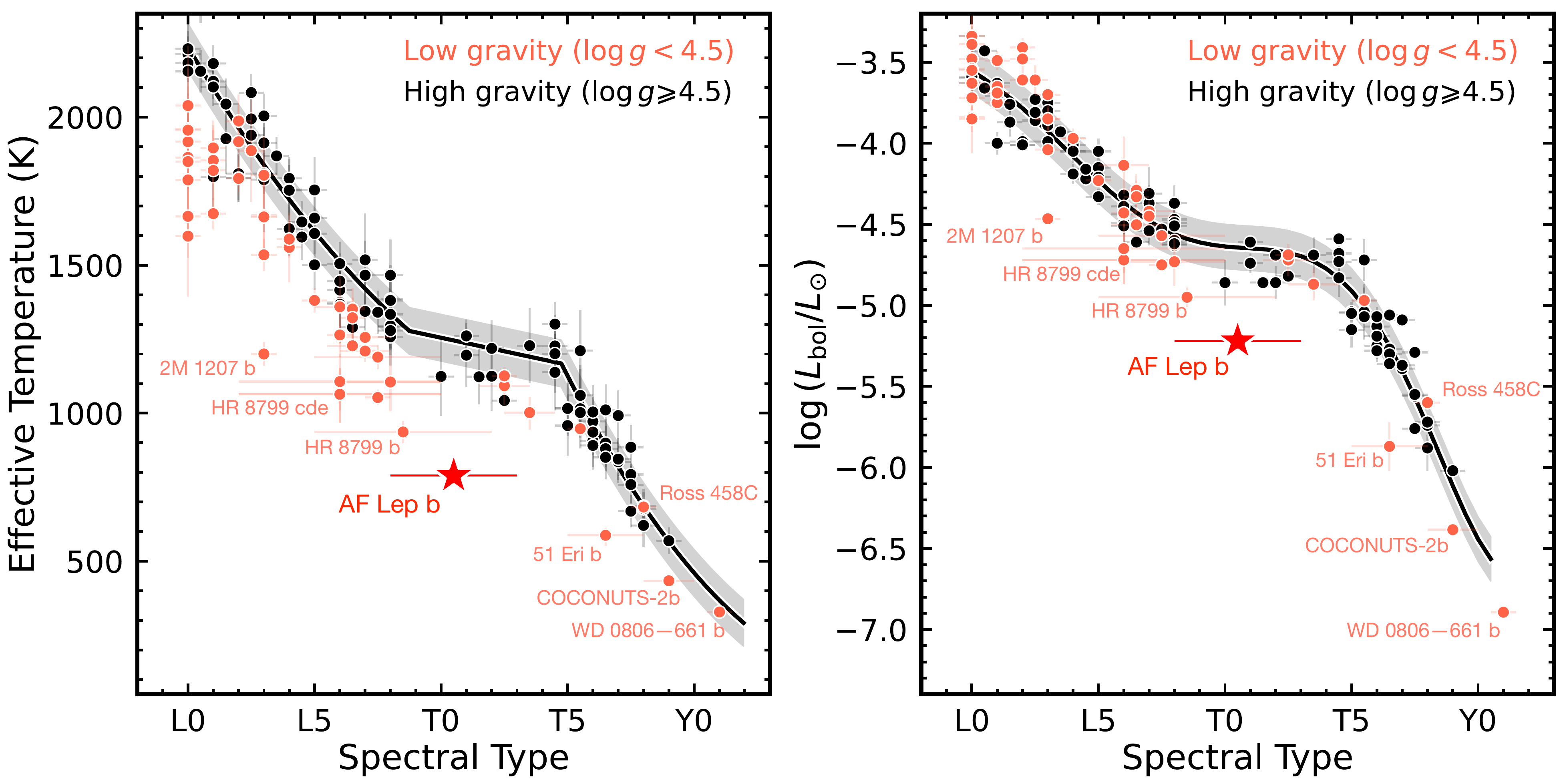}
\caption{Effective temperature (left) and bolometric luminosity (right) of AF~Lep~b (red star) compared to other directly imaged planets, free-floating planets, and brown dwarfs that are color-coded by their surface gravities (orange and black for $\log{(g)}$ below and above 4.5~dex, respectively); these $\log{(g)}$ of the comparison sample are obtained from \cite{2015ApJ...810..158F}, \cite{2020ApJ...891..171Z}, \cite{2021ApJ...911....7Z}, and \cite{2021ApJ...916L..11Z}. The plotted properties of AF~Lep~b are inferred by our retrieval analysis performed for the full spectrophotometry with constrained priors for the planet's mass and radius (i.e., the last column of Table~\ref{tab:retrieval_params_key_runs}). In both panels, we use black lines and grey shades to show the polynomials that convert spectral types to $T_{\rm eff}$ \citep[Table~13 of][]{2021ApJS..253....7K} or to $L_{\rm bol}$ \citep[footnote~16 of][]{2020ApJ...891..171Z} in the high-gravity regime. AF~Lep~b is thus far the coldest exoplanet with suggested evidence of silicate clouds. }
\label{fig:teff_lbol_vs_spt}
\end{center}
\end{figure*}

\subsection{AF~Lep~b as an Exceptional Planet Straddling the L/T Transition}
\label{subsec:context}

\subsubsection{Refined Spectral Type of AF~Lep~b}
\label{subsubsec:spt}
To derive the spectral type of AF~Lep~b, \cite{2023AandA...672A..93M} compared their observed SPHERE/IFS spectrum and the $K1/K2$ photometry to a spectral library of ultracool dwarfs, finding that spectral templates with L6--L6.5 types yield the best match. \cite{2023AandA...672A..94D} performed a similar analysis using their own observations and found a spectral type of L9--T0.5. The different spectral type inferred by these two studies is mainly caused by the distinct fluxes of their observed spectra (Figure~\ref{fig:specphot}). By accounting for the directly measured dynamical mass of the planet and the age of the planetary system, our retrieval analysis has now determined flux offsets that effectively combine the two sets of spectra. Therefore, in this section, we refine the spectral type of AF~Lep~b by combining all its observed spectrophotometry. The flux offset of each IFS spectrum is set to the best-fit value determined by the retrieval run that uses all observations and adopts constrained priors on the planet's mass and age (see Section~\ref{subsec:phot_m23d23spec}). 

We compare the SED of AF~Lep~b to IRTF/SpeX spectra of M5--T9 ultracool dwarfs over a wavelength range of 0.9--2.3~$\mu$m following \cite{2021ApJ...911....7Z}. A total of 930 ultracool dwarfs, or spectral templates, are selected from the UltracoolSheet \citep[][]{ultracoolsheet} as long as they are not resolved/candidate binaries and have good-quality spectra with S/N$>20$ per pixel in $J$ band. A scale factor for each template is computed to minimize the $\chi^{2}$. As shown in Figure~\ref{fig:spt}, the fitted templates with the lowest $\chi^{2}$ have L6.5--L7 types and are all unusually red and young planetary-mass objects, including WISEP~J004701.06+680352.1 \citep[][]{2012AJ....144...94G}, 2MASSW~J2244316+204343 \citep[][]{2002AJ....124.1170D}, PSO~J318.5338$-$22.8603 \citep{2013ApJ...777L..20L}, and WISE~J174102.78$-$464225.5 \citep{2014AJ....147...34S}. 

However, mid-L templates cannot reproduce all the spectral features of AF~Lep~b. The H$_{2}$O absorption between $J$ and $H$ bands of AF~Lep~b appears to be much deeper, suggestive of a colder effective temperature or a later spectral type. Also, the significant drop in flux from $K1$ to $K2$ is not well-explained by the relatively flat spectral shape of the L6.5--L7 templates. This blue $K1-K2$ color lines up with a strong CH$_{4}$ absorption band head at 2.2~$\mu$m, indicative of a late-L and early-T type. These properties, along with the overall red spectral morphology of AF~Lep~b, suggest a cloudy object at approximately T0 spectral type. However, such empirical spectral templates with high-quality S/N are lacking, particularly near the L/T transition where ultracool dwarfs are inherently rare \citep[e.g.,][]{2021AJ....161...42B}. 

To account for the cloud effect in the spectral typing process, the above template fitting is repeated with the \cite{2011ApJ...737..103S} reddening law incorporated. This extinction law is developed for the interstellar medium and might qualitatively (though might not quantitatively) demonstrates the effect of clouds on the emission spectroscopy (e.g., Figure~\ref{fig:mini_grid}). For each template, we explore a grid of $V$-band extinction spanning $0-30$~mag with steps of 0.1~mag and identify the $A_{V}$ that leads to the minimum $\chi^{2}$. The inclusion of reddening leads to best-fit templates with L9--T2 types (Figure~\ref{fig:spt}), which match better with AF~Lep~b in terms of the blue wing of the $H$ band (shaped by H$_{2}$O absorption) and the blue $K1-K2$ color (potentially shaped by the CH$_{4}$ absorption). In $Y$ and $J$ bands, the reddened templates appear to be systematically fainter, likely due to the extinction law of the interstellar medium not fully accounting for the cloud effect on self-luminous gas-giant planets. Nevertheless, the qualitatively good match in $H$ and $K$ bands (which are less affected by clouds than shorter wavelengths) suggests that a spectral type of AF~Lep~b near T0. 

Based on the observed spectrophotometry, we adopt a spectral type of L8--T3 for AF~Lep~b. This range covers the top $5\%$ of best-matched (reddened) spectral templates. This spectral type range is also later than the result inferred by the template fit without incorporating $A_{V}$, which cannot fully explain the deep H$_{2}$O absorption and blue $K1-K2$ color.

\subsubsection{The Unusual Atmospheric Properties of AF~Lep~b}
\label{subsubsec:unusual_teff_lbol}

AF~Lep~b is an exceptional giant planet straddling the L/T transition. As shown in Figure~\ref{fig:teff_lbol_vs_spt}, compared to the older ultracool dwarfs with similar spectral types but higher surface gravities and larger masses, this planet has a 450--600~K colder effective temperature and a 0.6--0.9~dex fainter bolometric luminosity. Notably, AF~Lep~b is thus far the coldest object with suggested evidence of silicate clouds.

The peculiar properties of AF~Lep~b line up with its very low surface gravity (Table~\ref{tab:evoparams}). As discussed in \cite{2020ApJ...891..171Z}, the properties of objects near the L/T transition depend on their surface gravities. Objects with lower surface gravities tend to have colder $T_{\rm eff}$, fainter absolute magnitudes in $J$ and $H$ bands, and slightly fainter bolometric luminosities than their older counterparts. This gravity dependence is particularly significant in late-L types but becomes weaker towards early-T types. With an L8--T3 spectral type (Section~\ref{subsubsec:spt}), AF~Lep~b maintains significant peculiarities, likely due to its remarkably low surface gravity of $\log{(g) \approx 3.6}$~dex. This $\log{(g)}$ is even 0.7~dex lower than those of previously known young and low-gravity T0--T5 ultracool dwarfs, including SIMP~J013656.5+093347.3, 2MASS~J13243553+6358281, GU~Psc~b, and SDSSp~J111010.01+011613.1. Across the L and T spectral types, gas-giant planets that have the coldest $T_{\rm eff}$ and the faintest $L_{\rm bol}$ are consistently associated with the lowest surface gravities. Examples include 2MASS~1207~b \citep[$\log{g} = 3.8 \pm 0.1$~dex;][]{2015ApJ...810..158F}, HR 8799~b \citep[$\log{(g)} = 4.1 \pm 0.2$~dex;][]{2020ApJ...891..171Z} and cde \citep[$\log{(g)} = 4.2 \pm 0.2$~dex;][]{2020ApJ...891..171Z}, AF~Lep~b ($\log{(g)} \approx 3.6$~dex; Table~\ref{tab:evoparams}), 51~Eri~b \citep[$\log{(g)} = 3.55^{+0.55}_{-0.03}$~dex;][]{2020ApJ...891..171Z}, and COCONUTS-2b \citep[$\log{(g)} = 4.11^{+0.11}_{-0.18}$~dex;][]{2021ApJ...916L..11Z}. The unusual physical properties of AF~Lep~b might be also linked to its metal-enriched atmosphere, suggesting the L/T transition is potentially metallicity dependent.

\begin{figure*}[t]
\begin{center}
\includegraphics[height=6.in]{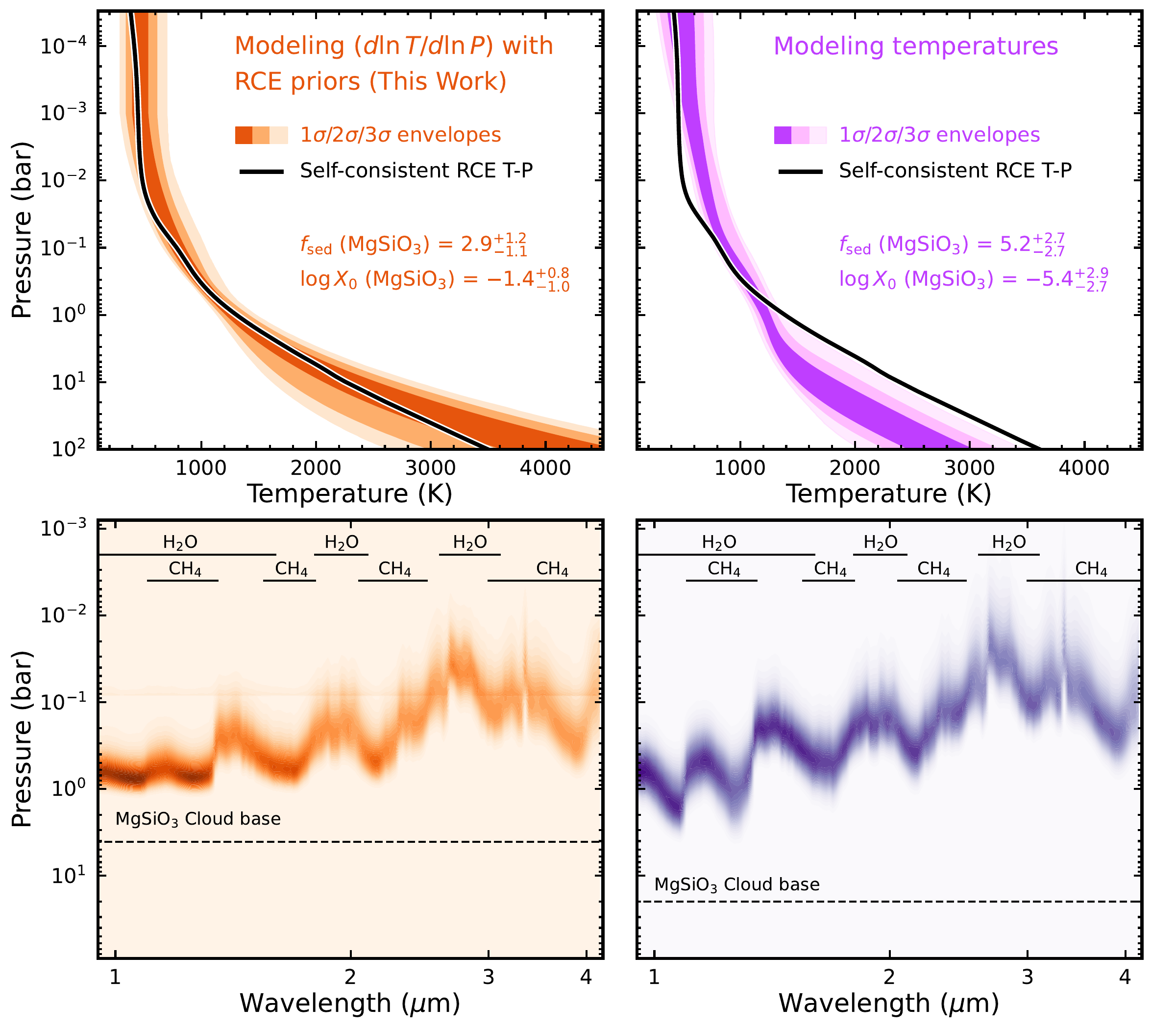}
\caption{{\it Top left}: The retrieved T-P profiles (orange) with the atmospheric thermal structure modeled via the temperature gradient. The input data include the $K1/K2/L'$ photometry and both spectra of AF~Lep~b, with constrained $M$ and $R$ priors; thus, the retrieved T-P profiles shown here are exactly the same as the middle left panel of Figure~\ref{fig:results_m23d23}. The $1\sigma/2\sigma/3\sigma$ confidence intervals are shown as orange shades. Overlaid is an RCE T-P profile (black) created at the median values of all parameters, which is consistent with the retrieved T-P profiles. {\it Bottom left}: Contribution function of the best-fit model for the retrieval shown in the top left panel. The 1--4~$\mu$m photosphere is located above $1$~bar due to the cloud effect; the retrieved silicate cloud has a base pressure near $3$~bar with a relatively small $f_{\rm sed}$ and large mass fraction. {\it Top and bottom right}: Retrieved T-P profiles and the contribution function of the best-fit model (purple), with the atmospheric thermal structure modeled in terms of the temperature. In comparison, modeling the thermal structure via temperature gradient offers the flexibility to incorporate an RCE prior and thereby better constraints on cloud properties. }
\label{fig:model_temperature_vs_temperature_gradient}
\end{center}
\end{figure*}

\subsection{Retrieving Thermal Structures of Cloudy Atmospheres: Modeling Temperature or Temperature Gradient?}
\label{subsec:advantages}

In the context of studying the thermal structure of imaged planets and brown dwarfs, retrieval studies often employ models that describe the temperature as a function of pressure to analyze observed data. Some studies parameterize the T-P profiles using explicit equations \citep[e.g.,][]{2017AJ....154...91L, 2017MNRAS.470.1177B, 2021MNRAS.506.1944B, 2020A&A...640A.131M, 2020AJ....160..150W, 2021ApJ...923...19G, 2022ApJ...938...56G, 2022arXiv221114330B, 2022ApJ...937...54X, 2023MNRAS.521.5761G, 2023MNRAS.tmp.1071W}, while others directly fit the temperature at different atmospheric layers \citep[e.g.,][]{2015ApJ...807..183L, 2017ApJ...848...83L, 2019ApJ...877...24Z, 2022ApJ...936...44Z, 2021A&A...656A..76Z, 2021Natur.595..370Z, 2023arXiv230304885H}. In this work, we adopt an approach that models the temperature gradient, $(d\ln{T}/d\ln{P})$, which allows for the novel incorporation of RCE during the retrievals (Sections~\ref{subsec:tp_profile}--\ref{subsec:rce_prior}). 

In this section, we conduct a comparative retrieval analysis for spectrophotometry of AF~Lep~b. There are two sets of retrievals: (1) we model the thermal structure in terms of the temperature gradient, as already established in Section~\ref{sec:retrieval}, and (2) we explicitly model the temperature. We then compare the resulting thermal structures and atmospheric properties from these two approaches. For this experiment, we use all available photometry and both spectra from \cite{2023AandA...672A..93M} and \cite{2023AandA...672A..94D}, with an additive flux offset applied for each spectrum. 

In the ``modeling temperatures'' approach, we use the thermal model of \cite{2020A&A...640A.131M}. The atmosphere is divided into the following three regions based on pressure $P$ or optical depth $\tau$, with a conversion of $\tau = \delta P^{\alpha}$; both $\delta$ and $\alpha$ are free parameters in this retrieval framework and the $P$ is in units of dyn~cm$^{-2}$.
\begin{enumerate}
\item[$\bullet$] The photosphere region spans from $\tau = 0.1$ to the radiative-convective boundary. The Eddington approximation is adopted, with the internal temperature $T_{\rm internal}$ added as a free parameter.
\item[$\bullet$] The low-altitude, troposphere region spans from the bottom of the photosphere region to the bottom of the atmosphere (assumed as $P = 10^{3}$~bar). The T-P profile is forced on the moist adiabat.
\item[$\bullet$] The high-altitude region spans from $P = 10^{-6}$~bar down to the top of the photosphere region and is divided into 3 layers evenly spaced in $\log{(P)}$. The temperatures of these layers, $T_{1}, T_{2}, T_{3}$ (from top to bottom), are all free parameters. A cubic spline interpolation is used to describe the thermal structure of this region. 
\end{enumerate}
There are six T-P parameters in total, $(\delta, \alpha, T_{\rm internal}, T_{1}, T_{2}, T_{3})$, and we set the same priors as \cite{2020A&A...640A.131M} in their retrieval analysis of the directly imaged planet HR~8799~e. The same chemistry and cloud models as described in Section~\ref{sec:retrieval_framework} are used. In addition, we adopt constrained priors for the planet's mass as $\mathcal{N}(\mu = 2.8\ {\rm M_{\rm Jup}}, \sigma = 0.6\ {\rm M_{\rm Jup}})$ and radius as $\mathcal{U}(1.20\ {\rm R_{\rm Jup}}, 1.55\ {\rm R_{\rm Jup}})$. 

Figure~\ref{fig:model_temperature_vs_temperature_gradient} presents the retrieved T-P profiles from two types of retrievals. While both retrieval analyses predict emission spectra that match the observed spectrophotometry of AF~Lep~b, the ``modeling temperatures'' approach returns a more isothermal T-P profile compared to a self-consistent RCE profile generated at median parameter values. In addition, properties of cloud condensates cannot be constrained under this approach, with $f_{\rm sed}$ and $\log{(X_{0})}$ primarily set by the priors. This result is consistent with the scenario proposed by \cite{2016ApJ...817L..19T}. Under this scenario, the atmospheres of L/T transition objects are impacted by the thermo-compositional instabilities arising from CO$\rightleftharpoons$CH$_{4}$ dis-equilibrium chemistry, leading to a more isothermal T-P profile without invoking clouds \citep[although this scenario has been challenged by][]{2018ApJ...853L..30L}. However, the retrieved $\log{(P_{\rm quench})} = -3.5^{+3.5}_{-1.5}$~dex from this approach does not strongly suggest the evidence of chemical dis-equilibrium in the photosphere of AF~Lep~b, which does not line up with the \cite{2016ApJ...817L..19T} scenario.\footnote{As a side note, this retrieval approach infers a [Fe/H]$=1.6 \pm 0.2$~dex for AF~Lep~b, consistent with the findings of our established analysis about this planet's potential metal-enrichment (Table~\ref{tab:retrieval_params_key_runs}).}

In contrast, the ``modeling temperature gradient'' approach (corresponding to the last column in Table~\ref{tab:retrieval_params_key_runs}) returns T-P profiles that are consistent with the RCE profile within 1--2$\sigma$ thanks to the RCE prior based on self-consistent forward models (Section~\ref{subsec:rce_prior}). This approach suggests the presence of silicate clouds and the top layers of these clouds likely truncate the retrieved contribution function at around $1$~bar (Figure~\ref{fig:model_temperature_vs_temperature_gradient}). 

This experiment does not aim to assess whether the \cite{2016ApJ...817L..19T} scenario is applicable to the atmosphere of AF~Lep~b, given that the S/N and wavelength span of this object's spectrophotometry is yet to be improved in the near future. Instead, this test indicates that when retrieving properties for a self-luminous planet or a brown dwarf that are expected to possess clouds, modeling the T-P profiles in terms of the temperature gradient $(d\ln{T}/d\ln{P})$ enables the incorporation of the RCE as parameterized priors for the object's thermal structure. These RCE priors can break the degeneracy between clouds and the shape of T-P profiles and lead to a robust characterization of the objects' cloud properties.

Another advantage of modeling the atmospheric thermal structure using temperature gradients is that the resulting T-P profiles are inherently smooth. Such smoothness is achieved because the temperature gradient throughout the atmosphere is obtained from the quadratic interpolation of the $(d\ln{T}/d\ln{P})$ at six input layers (Section~\ref{subsec:tp_profile}). In retrieval studies that directly fit temperatures at multiple layers \citep[e.g.,][]{2015ApJ...807..183L, 2017ApJ...848...83L, 2019ApJ...877...24Z, 2022ApJ...936...44Z, 2023arXiv230304885H}, a ``smoothing'' hyper-parameter is often used to prevent temperature oscillations as a function of pressures \citep[e.g., Equation~5 of][]{2015ApJ...807..183L}. Such a hyperparameter is not required when modeling T-P profiles using temperature gradients.

\subsection{Implications for Studies of Directly Imaged Exoplanets} 
\label{subsec:implications}

Here we list several valuable lessons learned from our work of AF~Lep~b that may have broader implications for the study of directly imaged exoplanets:

\noindent (1) Performing the end-to-end reduction for the same set of direct imaging data by using different pipelines, along with self-consistently computed covariance matrices, can provide a less-biased perspective about the planets' properties \citep[also see][]{2018AJ....155..226G, 2022ApJ...937...54X, 2023A&A...673A..98B, 2023arXiv230801343N}. For AF~Lep~b, the emission spectra observed and reduced by \cite{2023AandA...672A..93M} and \cite{2023AandA...672A..94D} are distinct. As seen in Figure~\ref{fig:violin_key}, the retrieval analysis performed for each spectrum can lead to different atmospheric properties including the $T_{\rm eff}$, cloud properties, presence of dis-equilibrium chemistry, and the tracers of the planet formation, such as C/O (also see Table~\ref{tab:retrieval_params_key_runs}). Some of these differences are pipeline-dependent and would be otherwise unknown if only one pipeline was adopted for data reduction.

\noindent (2) Dynamical mass provides key constraints to the planets' radii and surface gravities, and thereby other atmospheric properties. When retrieving each spectrum of \cite{2023AandA...672A..93M} and \cite{2023AandA...672A..94D} using default priors, several inferred parameters are very different between the two datasets, including [Fe/H] and $\log{(P_{\rm quench})}$. However, after incorporating a constrained mass prior based on the dynamical mass, the inferred parameters for these two spectra become consistent. In addition, the surface gravity and metallicity of gas-giant planets and brown dwarfs are often degenerate based on low-resolution spectral analysis \citep[e.g.,][]{2006ApJ...639.1095B, 2007ApJ...667..537L, 2007ApJ...660.1507L, 2009MNRAS.395.1237B, 2021ApJ...921...95Z}. Such degeneracy is also revealed by our retrieval analysis when adopting the default parameter priors (Figure~\ref{fig:example_corner}). The directly measured dynamical mass can provide informative priors on $\log{(g)}$, which can retrospectively constrain the objects' atmospheric metallicity.

\noindent (3) It is useful to contextualize the atmospheric characterization of directly imaged planets using evolution models. These evolution models provide expected ranges of planets' properties based on the objects' dynamical masses, ages, and/or bolometric luminosities. Incorporating these predictions can help suppress the ``small radius problem'' that often occurs in retrievals and forward-modeling analyses, thus improving the accuracy of other atmospheric properties. 

\noindent (4) Spectrophotometry with a wide wavelength coverage is essential to constrain the properties of directly imaged planets. In Appendix~\ref{app:retrieval_without_phot}, we present nine more retrieval runs of AF~Lep~b similar to those described in Section~\ref{sec:retrieval}, but with all $K1/K2/L'$ photometry excluded. Without these photometric data, a large scatter is seen in fitted model spectra with wavelengths beyond $2$~$\mu$m, which also lead to large uncertainties of the planet's [Fe/H] and C/O.
 
\noindent (5) Spectrophotometric monitoring of imaged planets is essential to reveal their atmospheric variability. Dedicated observations and analyses of time-series spectrophotometry will bring new insights into the formation and evolution of gas-giant planets.

\section{Summary}
\label{sec:summary}

AF~Lep A+b is a remarkable planetary system hosting a gas-giant planet, AF~Lep~b, with the lowest dynamical mass among directly imaged exoplanets. In order to investigate the formation pathway of AF~Lep~b, we have performed an in-depth analysis of the orbital and atmospheric properties of both the star and planet. Our main findings are summarized below.

\begin{enumerate}
\item[$\bullet$] Using our newly observed high-resolution spectroscopy of AF~Lep~A, we constrain a uniform set of stellar parameters and elemental abundances including a mass of $1.09 \pm 0.06$~M$_{\odot}$ and an iron abundance of $-0.27 \pm 0.31$~dex. Measurements of the stellar C/O are challenged by the fast stellar rotation that causes line broadening and blending of the characteristic C and O features (Section~\ref{sec:host_param_abund}).

\item[$\bullet$] We have updated the orbit and dynamical mass of AF~Lep~b, by combining published radial velocities, relative astrometry, and absolute astrometry, as well as the newly determined stellar mass (Section~\ref{sec:orbit}). The refined planet's dynamical mass is $2.8^{+0.6}_{-0.5}$~M$_{\rm Jup}$, with a semi-major axis of $8.2$~au. The architecture of the AF~Lep system is consistent with a spin-orbit alignment (or a misalignment) as previously suggested \citep{2023ApJ...950L..19F}.

\item[$\bullet$] Several evolution models are adopted to contextualize the atmospheric properties of AF~Lep~b using the planet's directly measured dynamical mass and the system's age. All these models predict consistent planet properties, although the hot-start models ($T_{\rm eff} = 610-680$~K, with $\log{(L_{\rm bol}/L_{\odot})}$ from $-5.7$~dex to $-5.4$~dex) suggest slightly hotter effective temperatures and brighter bolometric luminosities than those inferred by cold-start models ($T_{\rm eff} = 480-600$~K, with $\log{(L_{\rm bol}/L_{\odot})}$ from $-6.1$~dex to $-5.7$~dex). The radius of AF~Lep~b is estimated to span a range of $1.2-1.55$~R$_{\rm Jup}$ (Section~\ref{sec:evo}). 

\item[$\bullet$] We have performed chemically-consistent retrievals for AF~Lep~b using \texttt{petitRADTRANS} (Sections~\ref{sec:retrieval_framework} and \ref{sec:retrieval}) and developed a new retrieval approach that can lead to a robust characterization for cloudy self-luminous atmospheres (Section~\ref{subsec:advantages}). Specifically, we incorporate the radiative-convective equilibrium temperature profiles as parameterized priors on the planet's thermal structure during the retrievals. This novel approach is enabled by constraining the temperature-pressure profiles via the temperature gradient ($d\ln{T}/d\ln{P}$), a departure from previous studies that solely modeled the temperature. 

\item[$\bullet$] We have analyzed all published emission spectra and photometry of AF~Lep~b, spanning $0.9-4.2$~$\mu$m. Multiple runs are conducted using different portions of the planet's spectrophotometry, along with different priors on the planet's mass and radius. These retrievals consistently suggest that AF~Lep~b likely has a metal-enriched atmosphere ([Fe/H]$>1.0$~dex or $Z_{\rm planet}/Z_{\rm star} = 75^{+94}_{-42}$) compared to the metallicity of its host star. Our analysis also determines $T_{\rm eff} \approx 800$~K, $\log{(g)} \approx 3.7$~dex, and the presence of silicate clouds and the dis-equilibrium chemistry in the atmosphere of AF~Lep~b.

\item[$\bullet$] The potential metal enrichment of AF~Lep~b might be explained by planetesimal accretion, giant impacts, and/or core erosion. The first process also coincides with the presence of a debris disk in the system located at $40-60$~au. At its observed orbit, AF~Lep~b has a sufficient mass to dynamically excite the debris disk, causing planetesimal scattering and bombardment onto the planet's atmosphere (Section~\ref{subsec:formation}). 

\item[$\bullet$] If AF~Lep~b formed a few Myr later than its host star, then the planet's retrieved effective temperature and bolometric luminosity would be consistent with predictions of hot-start evolution models. When compared to predictions of cold-start evolution models, our retrieved $T_{\rm eff}$ and $L_{\rm bol}$ appear to be much higher regardless of whether the planet is coeval with the host star or formed 10~Myr later (Section~\ref{subsec:formation}). 

\item[$\bullet$] If giant impacts and planetary mergers occurred during the evolution history of AF~Lep~b, these processes might act as a mechanism of rejuvenation, by altering the planet's entropy state. In this scenario, the noted discrepancies bewteen retrieved and evolution-based $T_{\rm eff}$ and $L_{\rm bol}$ of AF~Lep~b can be potentially resolved.

\item[$\bullet$] We have refined the spectral type of AF~Lep~b to be L8--T3. Compared to higher-mass brown dwarfs with similar spectral types, AF~Lep~b has $450-600$~K colder $T_{\rm eff}$ and $0.6-0.9$~dex fainter $L_{\rm bol}$. These peculiar properties of AF~Lep~b are likely linked to its very low surface gravity and high atmospheric metallicity. Notably, AF~Lep~b is the coldest object with suggested evidence of silicate clouds to date (Section~\ref{subsec:context}). 

\item[$\bullet$] Our analysis of AF~Lep~b also leads to several valuable lessons that may have broader implications for the atmospheric study of directly imaged exoplanets. Specifically, we highlight the values of (1) performing the end-to-end reduction for the same set of direct imaging data by using different pipelines, (2) measuring the planet's dynamical mass based on orbital monitoring, (3) contextualizing the planets' atmospheric properties by using evolution models, (4) acquiring the planet's spectrophotometry with a broad wavelength coverage, and (5) studying the planet's top-of-atmosphere inhomogeneities based on the variability monitoring.

\end{enumerate}

For future work, it is essential to acquire spectrophotometry of AF~Lep~b with a higher S/N and/or spectral resolution over a broader wavelength range (e.g., VLTI/GRAVITY). Also, consistent reduction procedures should be applied to the existing SPHERE data (from \citealt{2023AandA...672A..93M} and \citealt{2023AandA...672A..94D}) to improve the quality of the emission spectra. These efforts will help improve the accuracy and precision of the inferred atmospheric properties and of AF~Lep~b, allowing for a reassessment of the formation pathway of this remarkable planetary system.

\begin{acknowledgments}
Z. Z. and P. M. thank Michiel Min for the idea of parameterizing the thermal structure using the temperature gradient. Z. Z. thanks Artem Aguichine, William Balmer, Bertram Bitsch, Douglas Lin, Evert Nasedkin, Kazumasa Ohno, and Daniel Thorngren for very helpful discussions. We thank the referee for suggestions that improved the manuscript. Support for this work was provided by NASA through the NASA Hubble Fellowship grant HST-HF2-51522.001-A awarded by the Space Telescope Science Institute, which is operated by the Association of Universities for Research in Astronomy, Inc., for NASA, under contract NAS5-26555. B.P.B. acknowledges support from the National Science Foundation grant AST-1909209, NASA Exoplanet Research Program grant 20-XRP20$\_$2-0119, and the Alfred P. Sloan Foundation. This work has benefited from The UltracoolSheet at http://bit.ly/UltracoolSheet, maintained by Will Best, Trent Dupuy, Michael Liu, Rob Siverd, and Zhoujian Zhang, and developed from compilations by \cite{2012ApJS..201...19D}, \cite{2013Sci...341.1492D}, \cite{2016ApJ...833...96L}, \cite{2018ApJS..234....1B}, and \cite{2021AJ....161...42B}. This work has made use of data from the European Space Agency (ESA) mission Gaia (\url{https://www.cosmos.esa.int/gaia}), processed by the Gaia Data Processing and Analysis Consortium (DPAC, \url{https://www.cosmos.esa.int/web/gaia/dpac/consortium}). Funding for the DPAC has been provided by national institutions, in particular, the institutions participating in the Gaia Multilateral Agreement.
\end{acknowledgments}

\facilities{Smith (Tull), Keck:I (HIRES), Keck:II (NIRC2), VLT:Melipal (SPHERE)}
\software{\texttt{DCR} \citep{2004PASP..116..148P}, \texttt{iSpec} \citep[][]{2014A&A...569A.111B, 2019MNRAS.486.2075B}, \texttt{BACCHUS} \citep{2016ascl.soft05004M},  \texttt{TURBOSPECTRUM} \citep{2012ascl.soft05004P}, \texttt{isochrones} \citep[][]{2015ascl.soft03010M}, \texttt{gaiadr3-zeropoint} \citep{2021A&A...649A...4L}, \texttt{orvara} \citep[][]{2021AJ....162..186B}, \texttt{petitRADTRANS} \citep[][]{2019A&A...627A..67M}, \texttt{petitCODE} \citep[][]{2015ApJ...813...47M, 2017A&A...600A..10M}, \texttt{PyMultiNest} \citep[][]{2014A&A...564A.125B}, \texttt{MultiNest} \citep[][]{2008MNRAS.384..449F, 2009MNRAS.398.1601F, 2019OJAp....2E..10F}, \texttt{corner.py} \citep[][]{corner}, \texttt{Astropy} \citep{2013A&A...558A..33A, 2018AJ....156..123A}, \texttt{IPython} \citep{PER-GRA:2007}, \texttt{Numpy} \citep{numpy},  \texttt{Scipy} \citep{scipy}, \texttt{Matplotlib} \citep{Hunter:2007}.}

\appendix

\section{Literature Comparison for our Stellar parameters of AF~Lep~A}
\label{app:stellar_params_compare_lit}
\begin{figure*}[h]
\begin{center}
\includegraphics[height=5.5in]{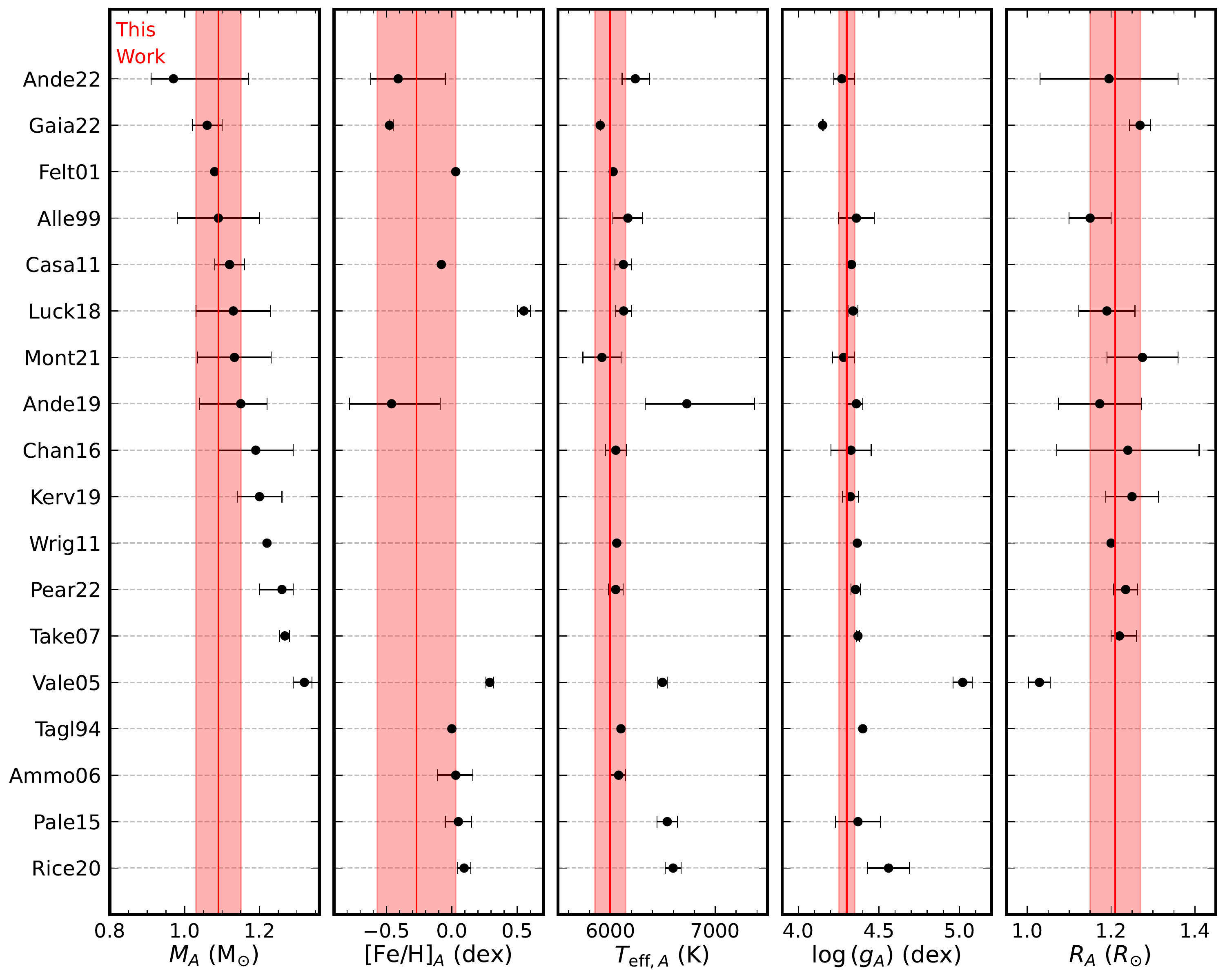}
\caption{Our measured stellar parameters of AF~Lep~A (verticle red line with shadow) compared to those in the literature (black; sorted by $M_{A}$, and then [Fe/H]$_{A}$ if $M_{A}$ is not reported). References for labels shown on the y axis are listed as follows. Tagl94: \cite{1994A&A...285..272T}; Alle99: \cite{1999A&A...352..555A}; Felt01: \cite{2001A&A...377..911F}; Vale05: \cite{2005ApJS..159..141V}; Ammo06: \cite{2006ApJ...638.1004A}; Take07: \cite{2007ApJS..168..297T}; Casa11: \cite{2011A&A...530A.138C}; Wrig11: \cite{2011ApJ...743...48W}; Pale15: \cite{2015A&A...573A..67P}; Chan16: \cite{2016AJ....151...59C}; Luck18: \cite{2018AJ....155..111L}; Ande19: \cite{2019A&A...628A..94A}; Kerv19: \cite{2019A&A...623A..72K}; Rice20: \cite{2020ApJ...898..119R}; Mont21: \cite{2021A&A...653A..98M}; 
Ande22: \cite{2022A&A...658A..91A}; Gaia22: \cite{2022arXiv220800211G}; Pear22: \cite{2022A&A...659A.135P}. }
\label{fig:stellar_params_compare_lit}
\end{center}
\end{figure*}

As shown in Figure~\ref{fig:stellar_params_compare_lit}, our inferred mass, iron abundance, effective temperature, surface gravity, and radius of AF~Lep~A (Table~\ref{tab:stellar_params}) are generally consistent with those determined in the literature. When collecting the literature values, we exclude the studies that cataloged or averaged the stellar parameters measured by other work. 

In this work, $M_{A}$ is a key parameter for constraining the orbit and dynamical mass of AF~Lep~b. Several studies measured slightly higher $M_{A}$ than our adopted value, including \cite{2005ApJS..159..141V}, \cite{2007ApJS..168..297T}, \cite{2011ApJ...743...48W}, \cite{2019A&A...623A..72K}, and \cite{2022A&A...659A.135P}; the latter four studies did not measure the stellar [Fe/H]. As detailed below, the different $M_{A}$ estimates between their and our works are likely due to the systematic differences of the adopted stellar model spectra and isochrones (with different atomic/molecular line lists), input photometry (single band versus multiple bands), the assumption about the stellar age, and whether all input parameters are simultaneously or separately constrained by the stellar models. 

\noindent $\bullet$ \cite{2005ApJS..159..141V} estimated the bolometric luminosity of AF~Lep~A via $V$-band magnitude with a bolometric correction and derived other stellar parameters by fitting the observed spectrum using the \cite{1992IAUS..149..225K} model atmospheres; they then updated the stellar mass by modeling the $L_{\rm bol}$ and $T_{\rm eff}$ via the Yonsei-Yale isochrones \citep[][]{2004ApJS..155..667D}. Their resulting $M_{A}$, [Fe/H]$_{A}$, $T_{\rm eff,A}$, and $\log{(g_{A})}$ are all among the highest (and their derived $R_{A}$ is among the lowest) in the literature. 

\noindent $\bullet$ The analysis of \cite{2007ApJS..168..297T} is tied to that of \cite{2005ApJS..159..141V}. Briefly, \cite{2007ApJS..168..297T} derived the mass and radius of AF~Lep~A (along with about $1000$ cool stars) by modeling this object's parallax, $V$-band magnitude, and the \cite{2005ApJS..159..141V} $T_{\rm eff}$, $\log{(g)}$, [Fe/H] measurements using the Yale Rotational Evolution Code. 

\noindent $\bullet$ \cite{2011ApJ...743...48W} first estimated a bolometric luminosity based on the $V$-band magnitude and a bolometric correction guided by the object's $V-K_{s}$ color. Assuming AF~Lep~A has a much older age of $1$~Gyr (compared to its age of $24 \pm 3$~Myr), they then used the bolometric luminosity and the \cite{2000A&A...358..593S} isochrones to derive stellar $M_{A}$, $R_{A}$, and $T_{\rm eff,A}$. 

\noindent $\bullet$ \cite{2019A&A...623A..72K} estimated the radius of AF~Lep~A using the $V$-band magnitude, $V-K_{s}$ color, and the parallax. Then they derived $M_{A}$ by using the \cite{2000A&AS..141..371G} isochrones.

\noindent $\bullet$ \cite{2022A&A...659A.135P} determined the effective temperature and bolometric luminosity of AF~Lep~A by fitting the PHOENIX \citep[][]{2013A&A...553A...6H} model atmospheres (with a blackbody component to account for the debris disk) to the observed multi-band photometry. They then modeled these inferred $T_{\rm eff, A}$ and $L_{\rm bol, A}$ using the MIST evolution models to derive $M_{A}$.

\section{Parameter Posteriors and Data--Model Comparison of Our Orbit Analysis}
\label{app:orbit_posterior}
\begin{figure*}[h]
\begin{center}
\includegraphics[height=7.5in]{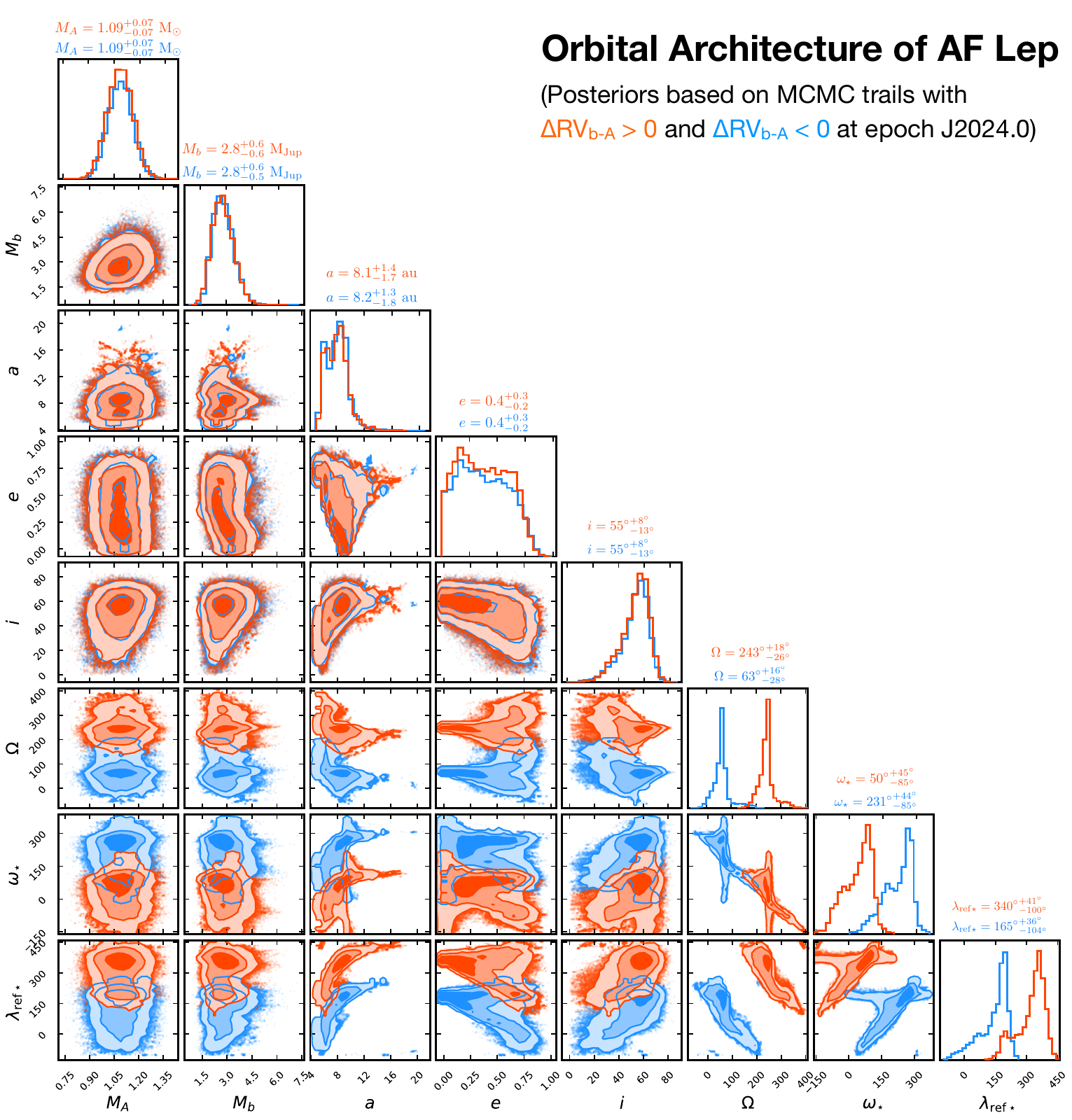}
\caption{Parameter posteriors for two modes of our derived orbital solution, corresponding to positive (red) and negative (blue) $\Delta$RV$_{\rm b-A}$ values at epoch J2024.0. The axes of $\Omega$, $\omega_{\star}$, and $\lambda_{\rm ref,\star}$ extend beyond the nominal range of [$0^{\circ}, 360^{\circ}$) so that each mode of the orbital solution does not split into multiple peaks due to the modulus of $360^{\circ}$.  }
\label{fig:orbit_posterior}
\end{center}
\end{figure*}

\begin{figure*}[t]
\begin{center}
\includegraphics[height=7.5in]{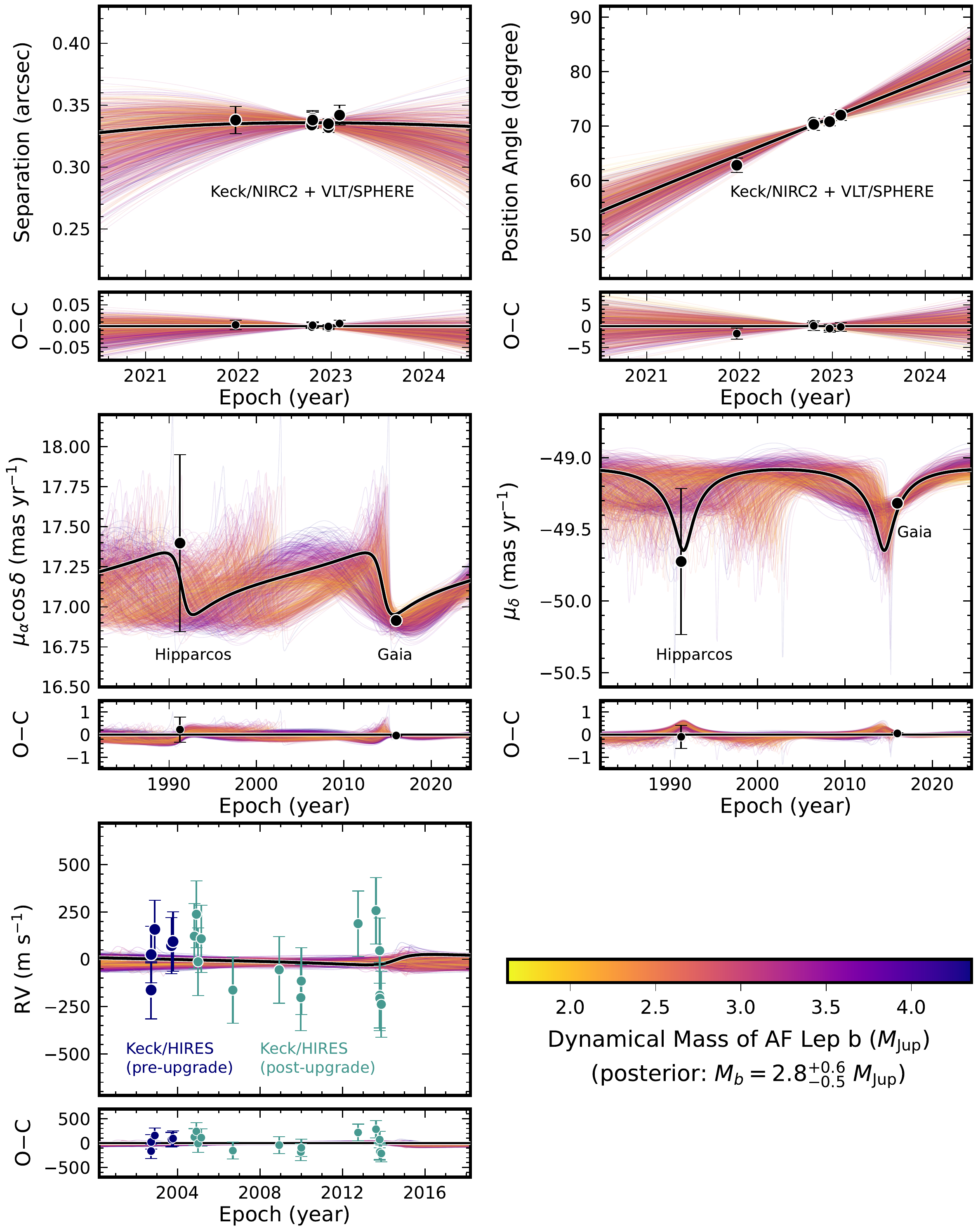}
\caption{Fitted orbits compared to the observed relative astrometry (top), absolute astrometry (middle), and the host star's relative RVs (bottom left). In each panel, we present the observed data (top) and residuals (bottom) as filled circles, and the best-fit orbital solution as black solid lines. Predictions of 1000 random orbits are color-coded by the dynamical mass of AF~Lep~b, estimated to be $2.7^{+0.6}_{-0.5}$~M$_{\rm Jup}$ by our analysis.}
\label{fig:data_vs_model}
\end{center}
\end{figure*}

In Figure~\ref{fig:orbit_posterior}, we present the posteriors for masses of AF~Lep~A and b, as well as the planet's orbital parameters. Figure~\ref{fig:data_vs_model} compares our input observational data with the fitted orbits.

\section{Evolution-based Properties of AF~Lep~\lowercase{b} Assuming a Younger Age than the Host Star}
\label{app:evo_params_young}
Table~\ref{tab:younger_evoparams} presents the physical properties of AF~Lep~b inferred by various evolution models. Input parameters of this analysis include the directly measured mass ($M = 2.8^{+0.6}_{-0.5}$~M$_{\rm Jup}$; Section~\ref{sec:orbit}) and an age of $14 \pm 3$~My that is 10~Myr younger than that of its host star, assuming a slightly later epoch of planet formation.

\section{Retrievals on All $K1/K2/L'$ Photometry with Different Spectroscopic Datasets and Parameter Priors}
\label{app:retrieval_with_phot}
Figure~\ref{fig:retrieval_with_phot} presents the results of all retrieval runs that incorporate all the $K1/K2/L'$ photometry, with different sets of spectra and parameter priors. As described in Section~\ref{sec:retrieval}, the three spectral sets include: (1) the \cite{2023AandA...672A..93M} spectrum, (2) the \cite{2023AandA...672A..94D} spectrum, and (3) both spectra. The three sets of parameter priors include: (1) the default priors summarized in Table~\ref{tab:retrieval_params}, (2) a constrained prior on the planet's mass and default priors for the remaining parameters, and (3) constrained priors on both the planet's mass and radius with default priors for the remaining parameters. Retrievals with constrained priors on both $M$ and $R$ are already shown in Figures~\ref{fig:results_m23}--\ref{fig:results_m23d23}, so Figure~\ref{fig:retrieval_with_phot} presents the results for the remaining six retrieval runs. Median values and confidence intervals of all parameters inferred by these retrievals are summarized in Table~\ref{tab:retrieval_params_key_runs}.

\section{Retrievals on Different Spectroscopic Datasets and Parameter Priors of AF~Lep~\lowercase{b}, with Photometry Excluded}
\label{app:retrieval_without_phot}
Figure~\ref{fig:retrieval_without_phot} presents the results of the additional nine retrieval runs that incorporate different sets of spectra and parameter priors (same as described in Appendix~\ref{app:retrieval_with_phot}), but with all the $K1/K2/L'$ photometry of AF~Lep~b excluded.

\end{CJK*}

%%%%%%%%%%%%%%%
%%% BIBLIOGRAPHY %%%
%%%%%%%%%%%%%%%
\clearpage
\bibliographystyle{aasjournal}
\bibliography{ms}{}

%%%%%%%%%%%%
%%% FIGURES %%%
%%%%%%%%%%%%
\clearpage

%%% spectra + photometry vs. models
\begin{figure*}[t]
\begin{center}
\includegraphics[height=7.5in]{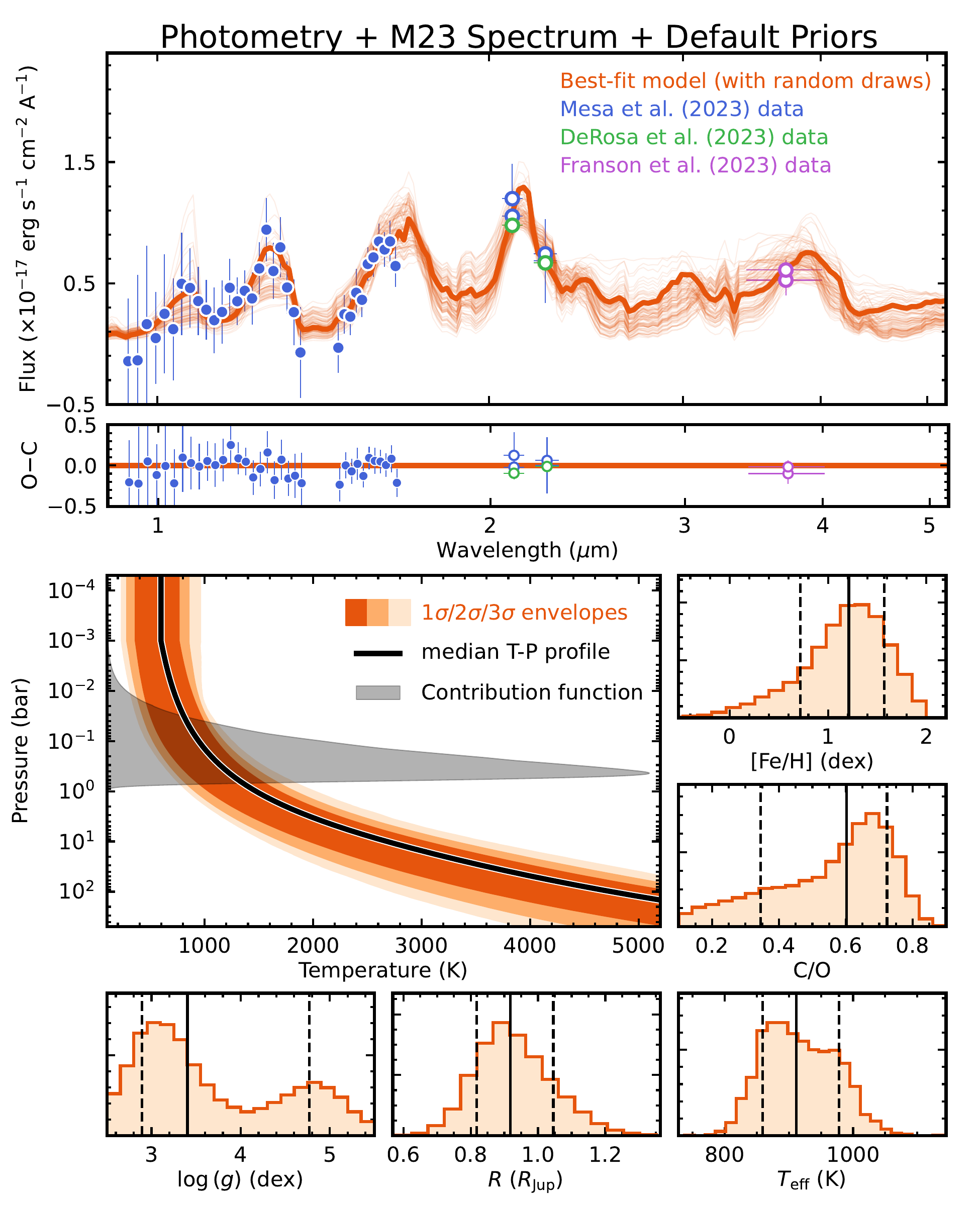}
\caption{Results of the retrieval analysis on $K1/K2/L'$ photometry and the \cite{2023AandA...672A..93M} spectrum of AF~Lep~b, with default parameter priors listed in Table~\ref{tab:retrieval_params}. The format is the same as Figure~\ref{fig:results_m23}.}
\label{fig:retrieval_with_phot}
\end{center}
\end{figure*}

\renewcommand{\thefigure}{\arabic{figure}}
\addtocounter{figure}{-1}
\begin{figure*}[t]
\begin{center}
\includegraphics[height=7.5in]{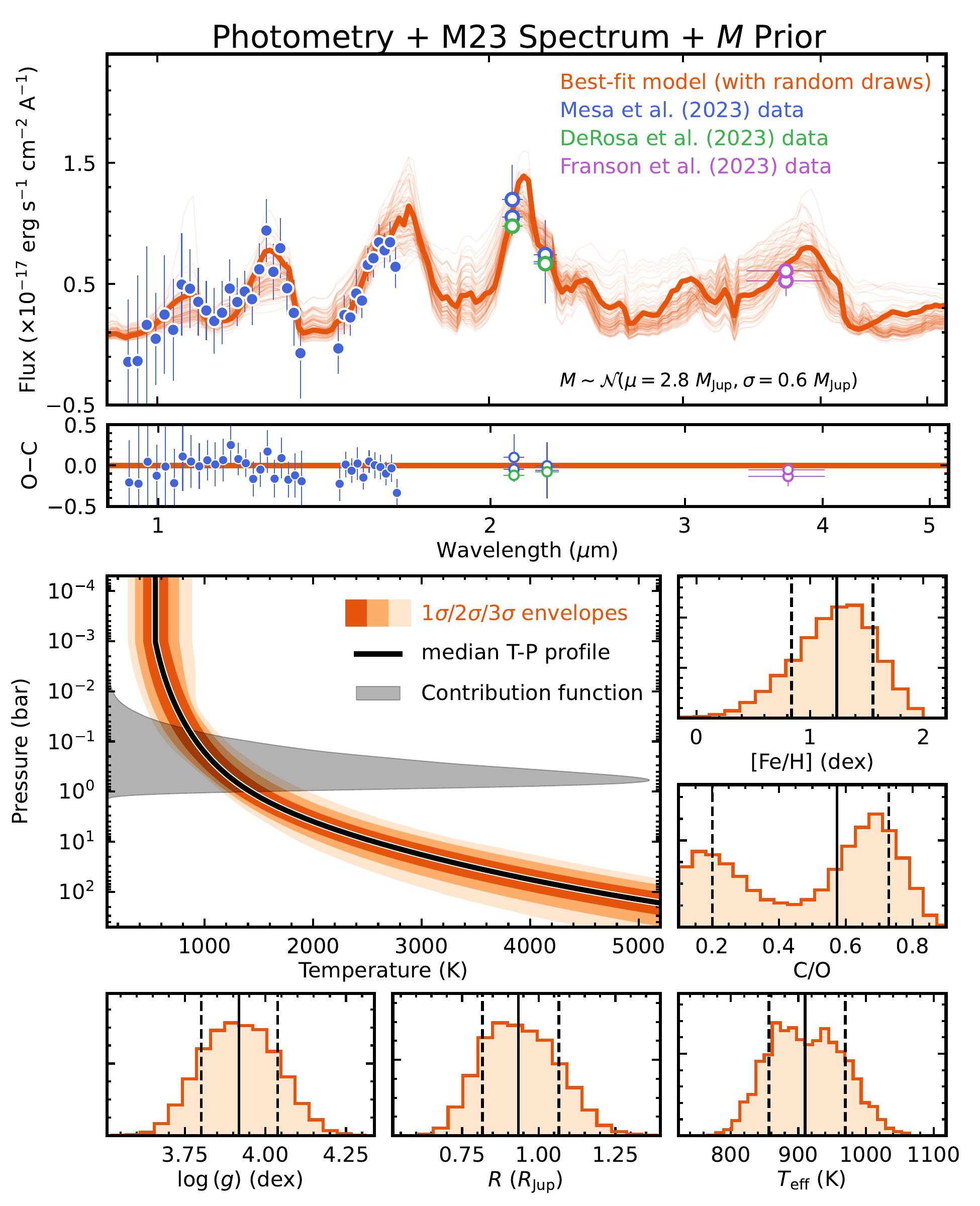}
\caption{Continued. Results of the retrieval analysis on $K1/K2/L'$ photometry and the \cite{2023AandA...672A..93M} spectrum of AF~Lep~b, with a constrained prior on $M$ and default priors (Table~\ref{tab:retrieval_params}) for the remaining parameters. The format is the same as Figure~\ref{fig:results_m23}.}
\end{center}
\end{figure*}

\addtocounter{figure}{-1}
\begin{figure*}[t]
\begin{center}
\includegraphics[height=7.5in]{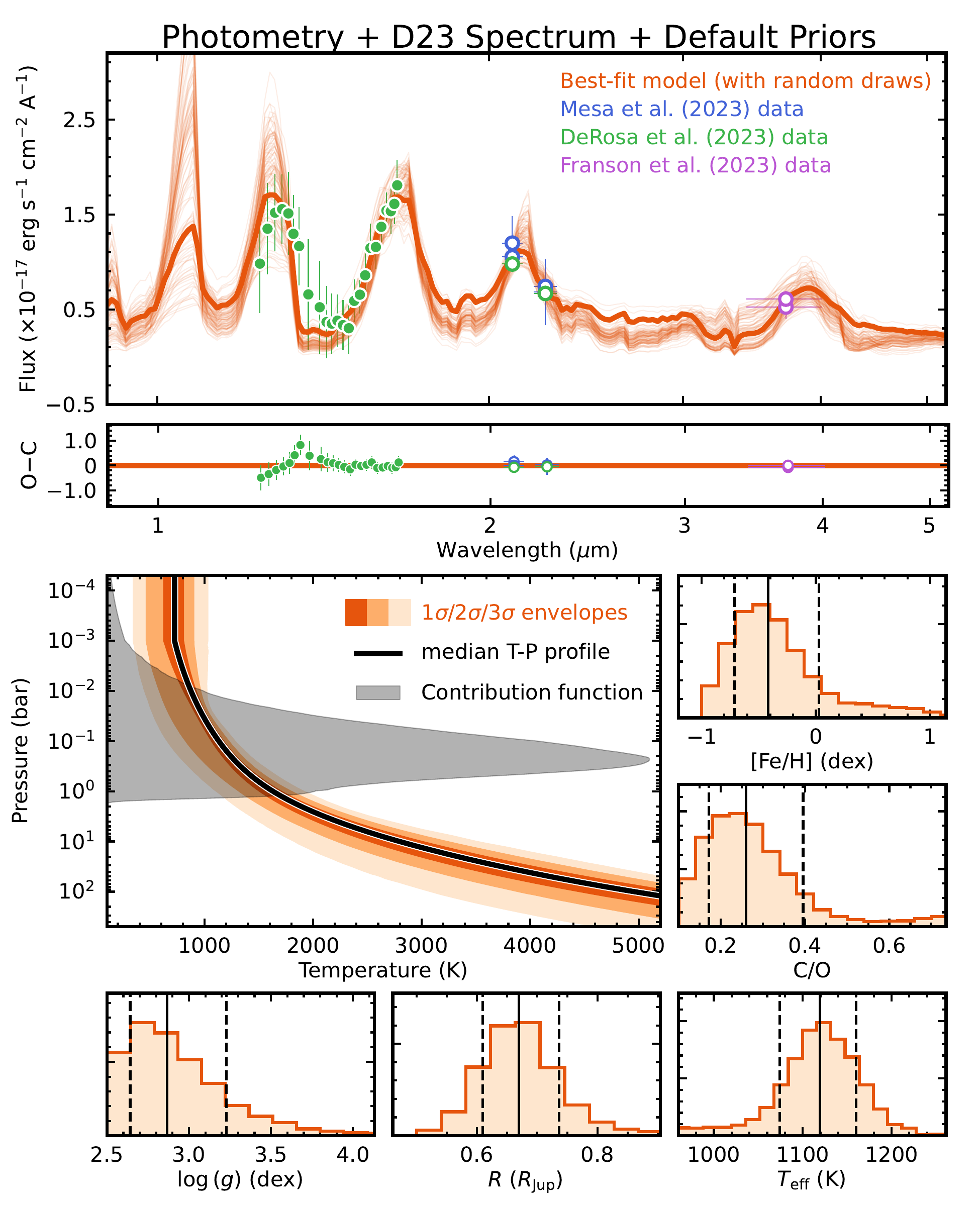}
\caption{Continued. Results of the retrieval analysis on $K1/K2/L'$ photometry and the \cite{2023AandA...672A..94D} spectrum of AF~Lep~b, with default parameter priors listed in Table~\ref{tab:retrieval_params}. The format is the same as Figure~\ref{fig:results_m23}.}
\end{center}
\end{figure*}

\addtocounter{figure}{-1}
\begin{figure*}[t]
\begin{center}
\includegraphics[height=7.5in]{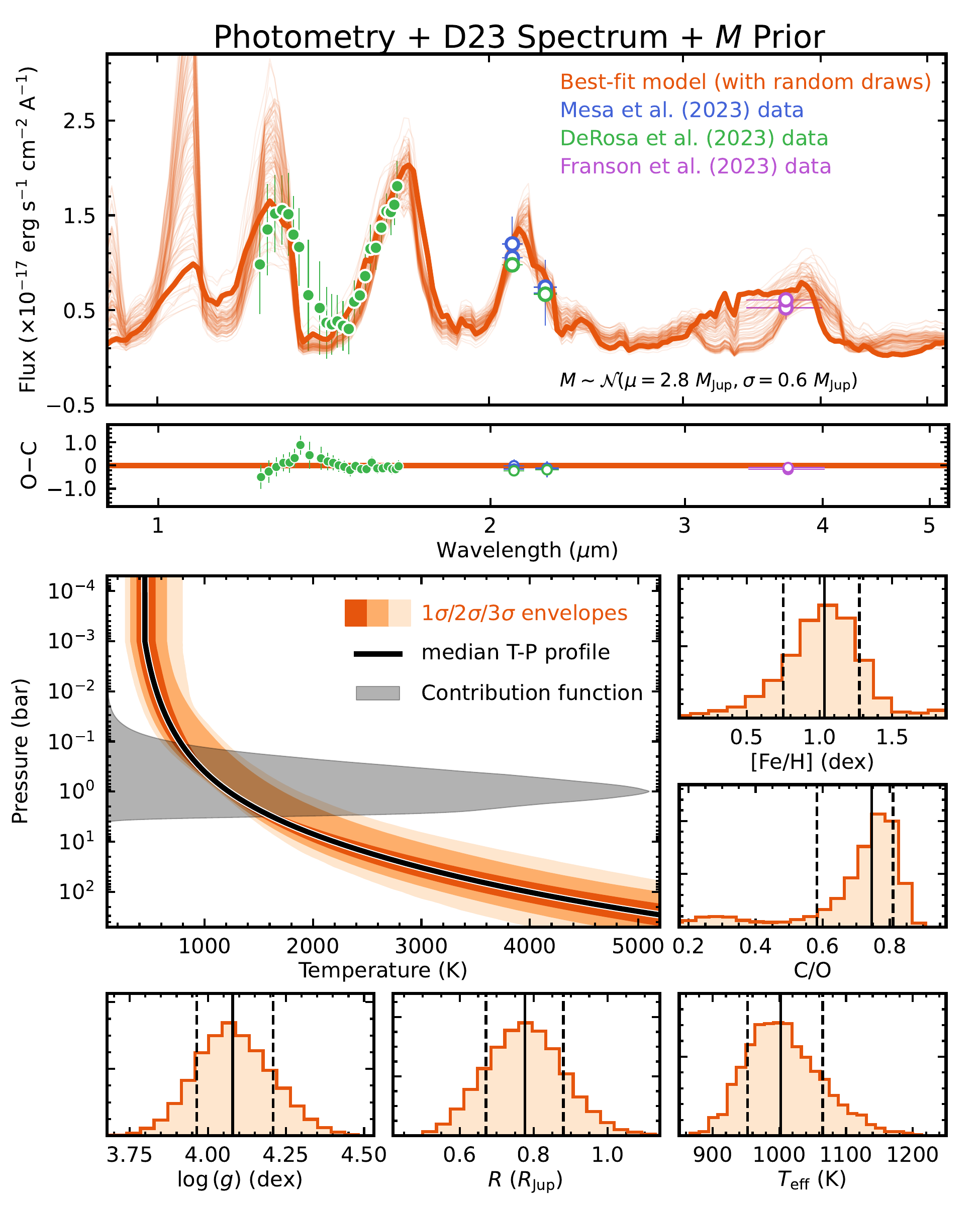}
\caption{Continued. Results of the retrieval analysis on $K1/K2/L'$ photometry and the \cite{2023AandA...672A..93M} spectrum of AF~Lep~b, with a constrained prior on $M$ and default priors (Table~\ref{tab:retrieval_params}) for the remaining parameters. The format is the same as Figure~\ref{fig:results_m23}.}
\end{center}
\end{figure*}

\addtocounter{figure}{-1}
\begin{figure*}[t]
\begin{center}
\includegraphics[height=7.5in]{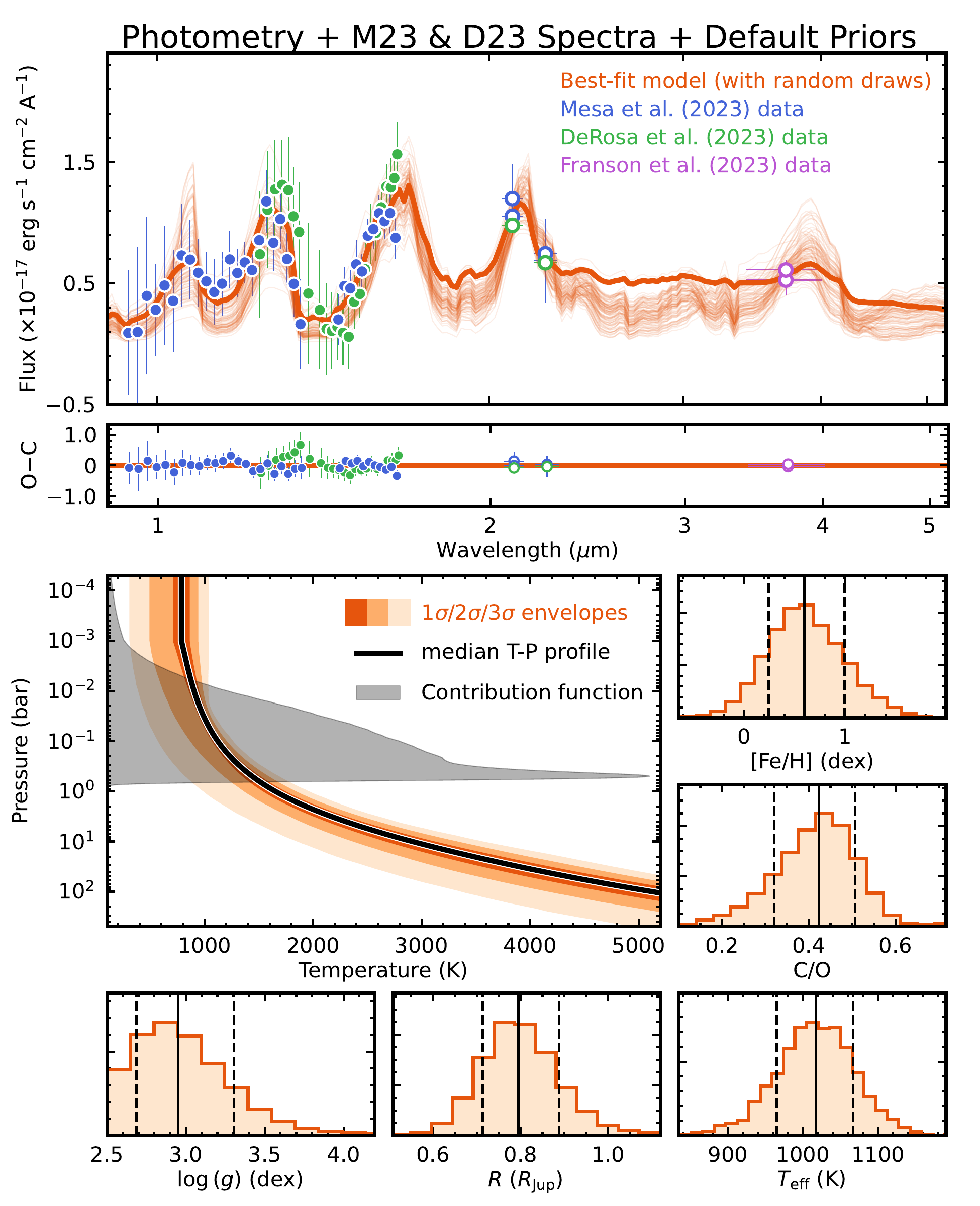}
\caption{Continued. Results of the retrieval analysis on $K1/K2/L'$ photometry and both spectra from \cite{2023AandA...672A..93M} and \cite{2023AandA...672A..94D}, with default parameter priors listed in Table~\ref{tab:retrieval_params}. The format is the same as Figure~\ref{fig:results_m23}.}
\end{center}
\end{figure*}

\addtocounter{figure}{-1}
\begin{figure*}[t]
\begin{center}
\includegraphics[height=7.5in]{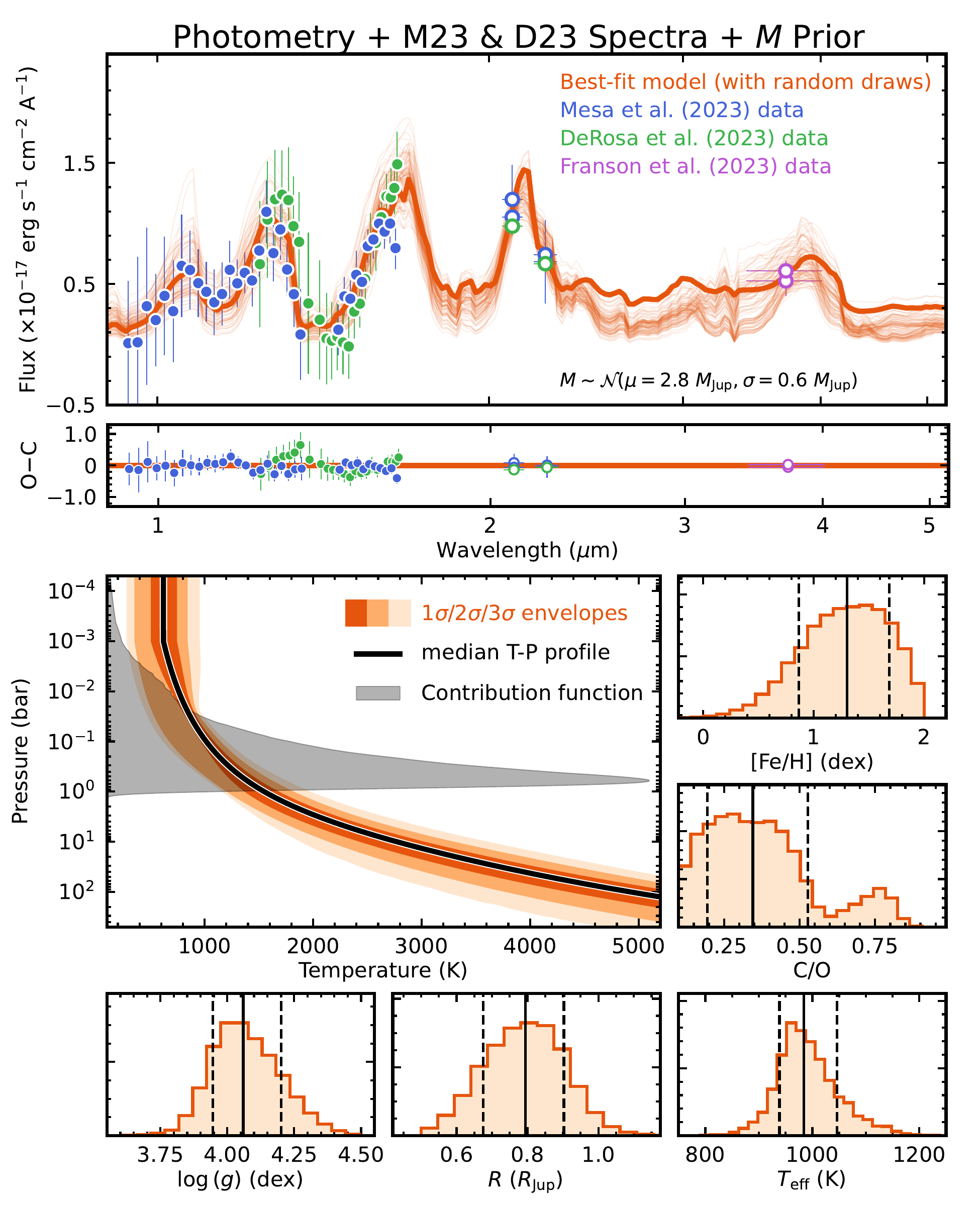}
\caption{Continued. Results of the retrieval analysis on $K1/K2/L'$ photometry and both spectra from \cite{2023AandA...672A..94D} and \cite{2023AandA...672A..94D}, with a constrained prior on $M$ and default priors (Table~\ref{tab:retrieval_params}) for the remaining parameters. The format is the same as Figure~\ref{fig:results_m23}.}
\end{center}
\end{figure*}

%%% spectra vs. models

\begin{figure*}[t]
\begin{center}
\includegraphics[height=7.5in]{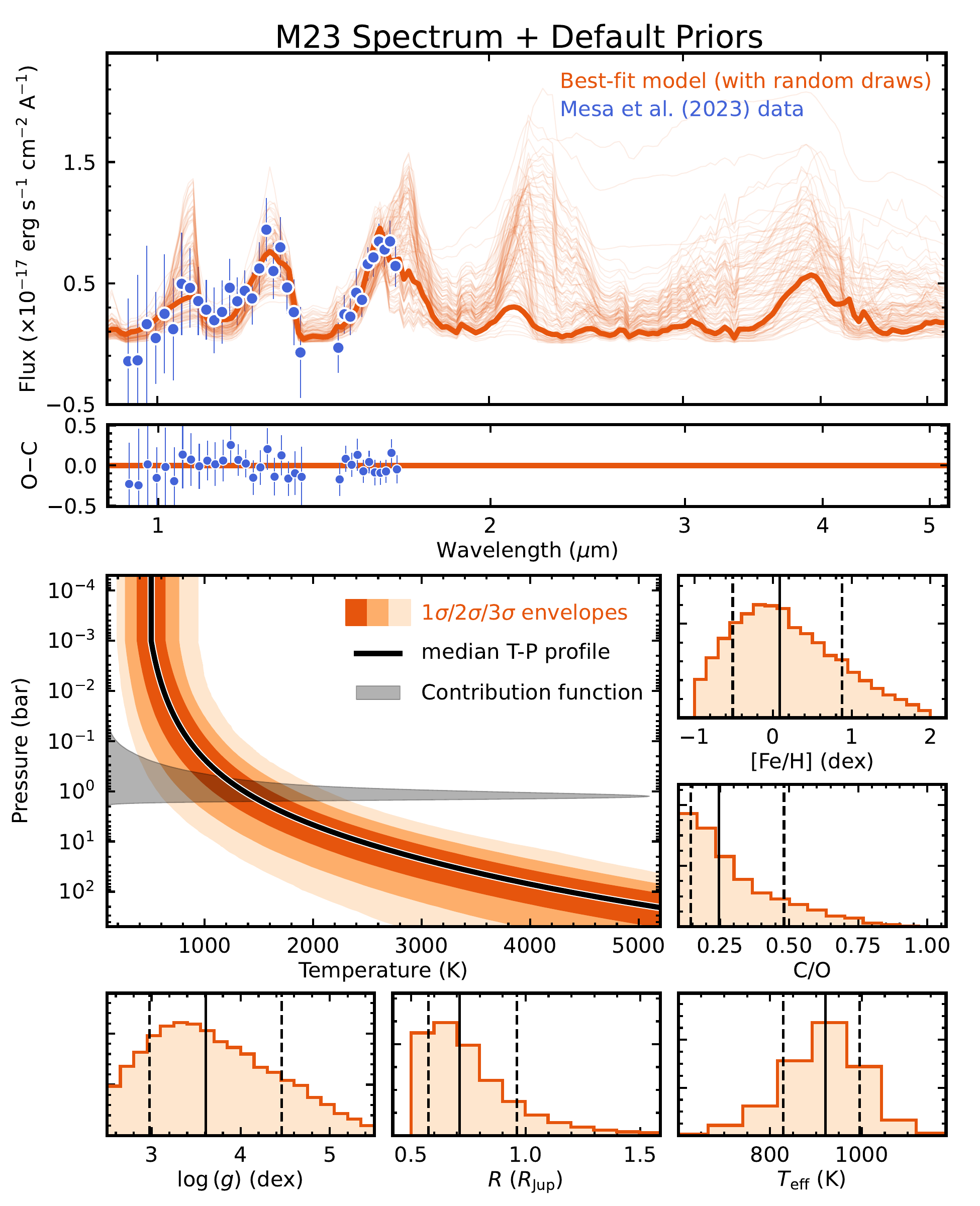}
\caption{Results of the retrieval analysis on the \cite{2023AandA...672A..93M} spectrum of AF~Lep~b, with default parameter priors listed in Table~\ref{tab:retrieval_params}. The format is the same as Figure~\ref{fig:results_m23}.}
\label{fig:retrieval_without_phot}
\end{center}
\end{figure*}

\renewcommand{\thefigure}{\arabic{figure}}
\addtocounter{figure}{-1}
\begin{figure*}[t]
\begin{center}
\includegraphics[height=7.5in]{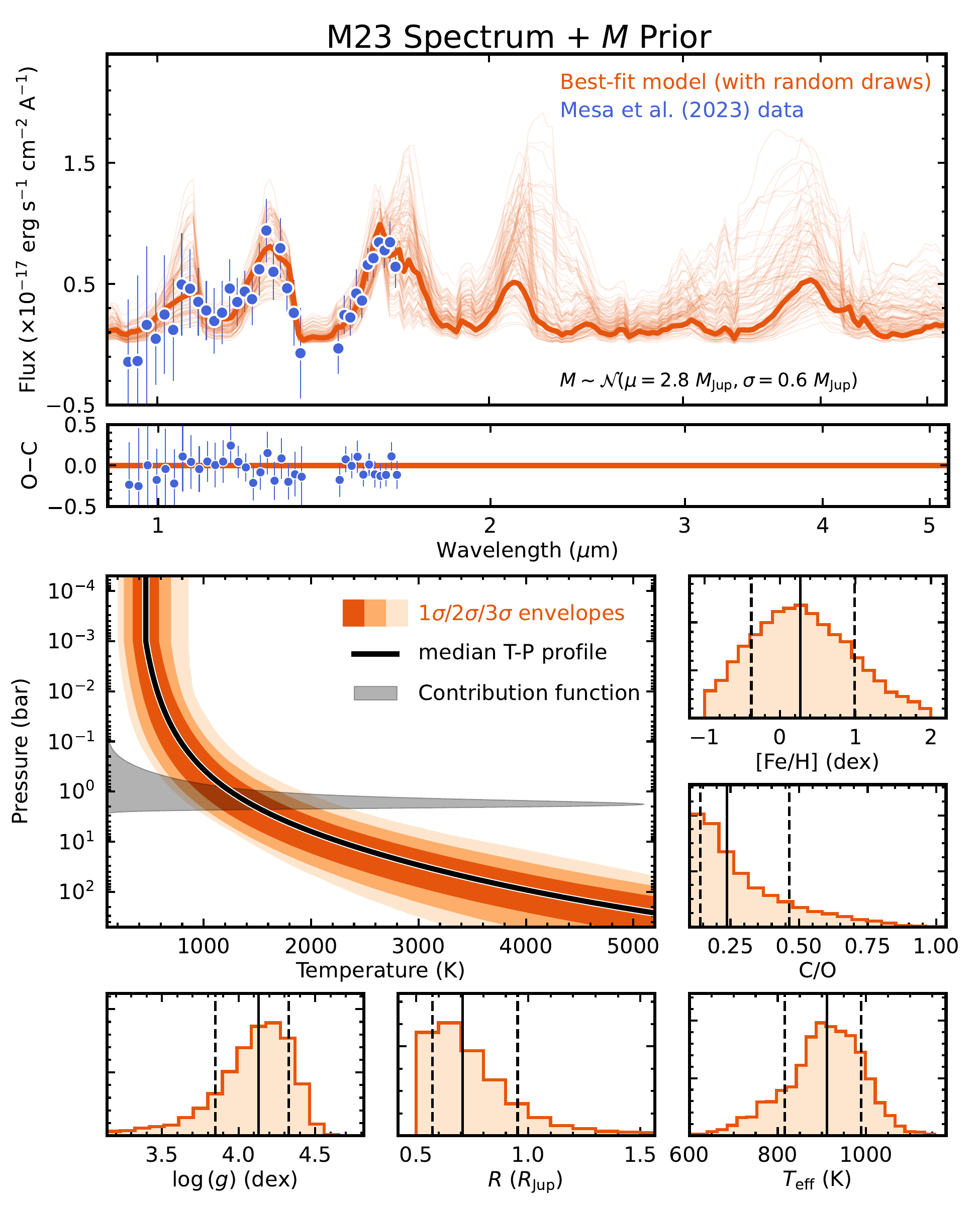}
\caption{Continued. Results of the retrieval analysis on the \cite{2023AandA...672A..93M} spectrum of AF~Lep~b, with a constrained prior on $M$ and default priors for the remaining parameters (Table~\ref{tab:retrieval_params}). The format is the same as Figure~\ref{fig:results_m23}.}
\end{center}
\end{figure*}

\addtocounter{figure}{-1}
\begin{figure*}[t]
\begin{center}
\includegraphics[height=7.5in]{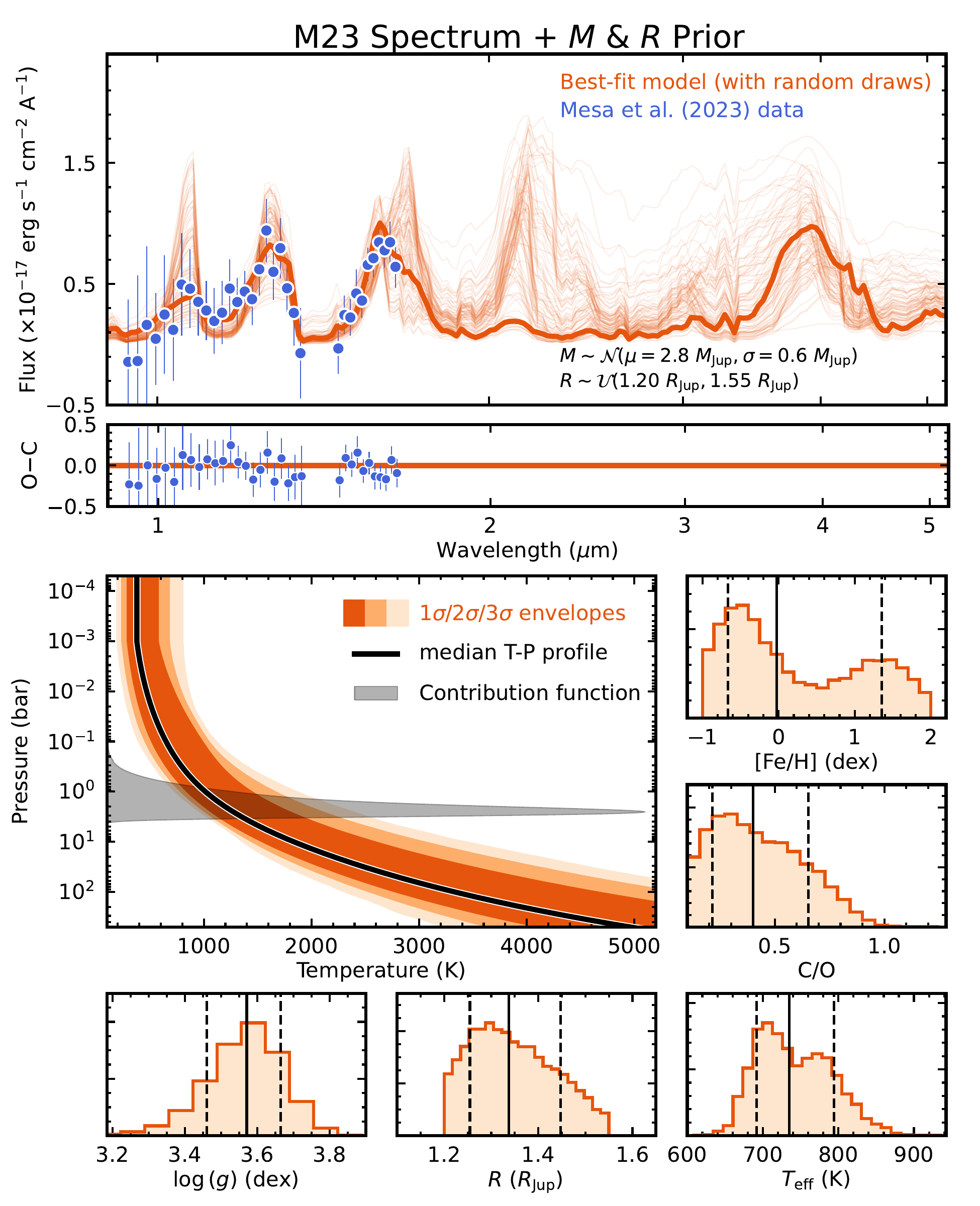}
\caption{Continued. Results of the retrieval analysis on the \cite{2023AandA...672A..93M} spectrum of AF~Lep~b, with constrained priors on both $M$ and $R$ and default priors for the remaining parameters (Table~\ref{tab:retrieval_params}). The format is the same as Figure~\ref{fig:results_m23}.}
\end{center}
\end{figure*}

\addtocounter{figure}{-1}
\begin{figure*}[t]
\begin{center}
\includegraphics[height=7.5in]{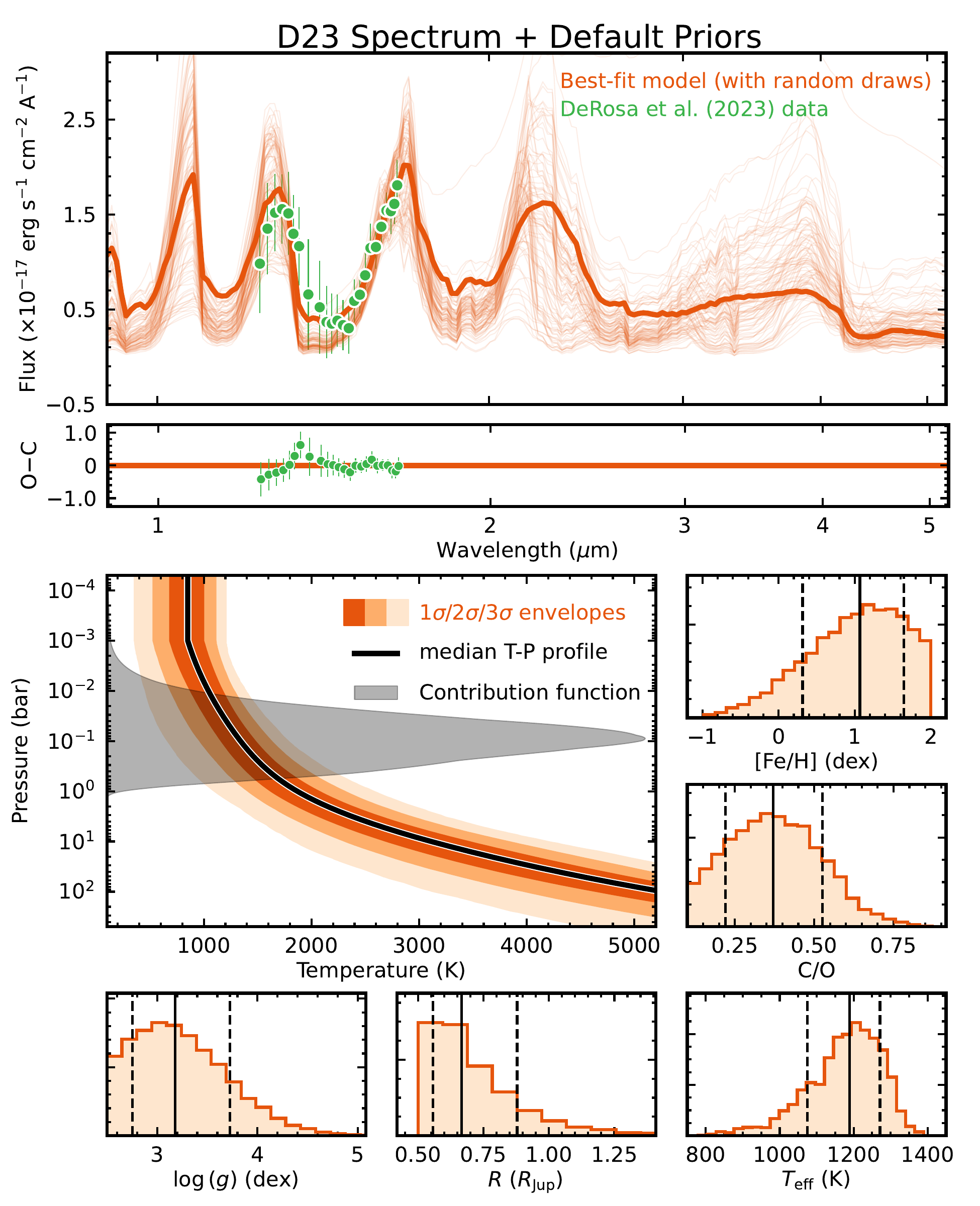}
\caption{Continued. Results of the retrieval analysis on the \cite{2023AandA...672A..94D} spectrum of AF~Lep~b, with default parameter priors listed in Table~\ref{tab:retrieval_params}. The format is the same as Figure~\ref{fig:results_m23}.}
\end{center}
\end{figure*}

\addtocounter{figure}{-1}
\begin{figure*}[t]
\begin{center}
\includegraphics[height=7.5in]{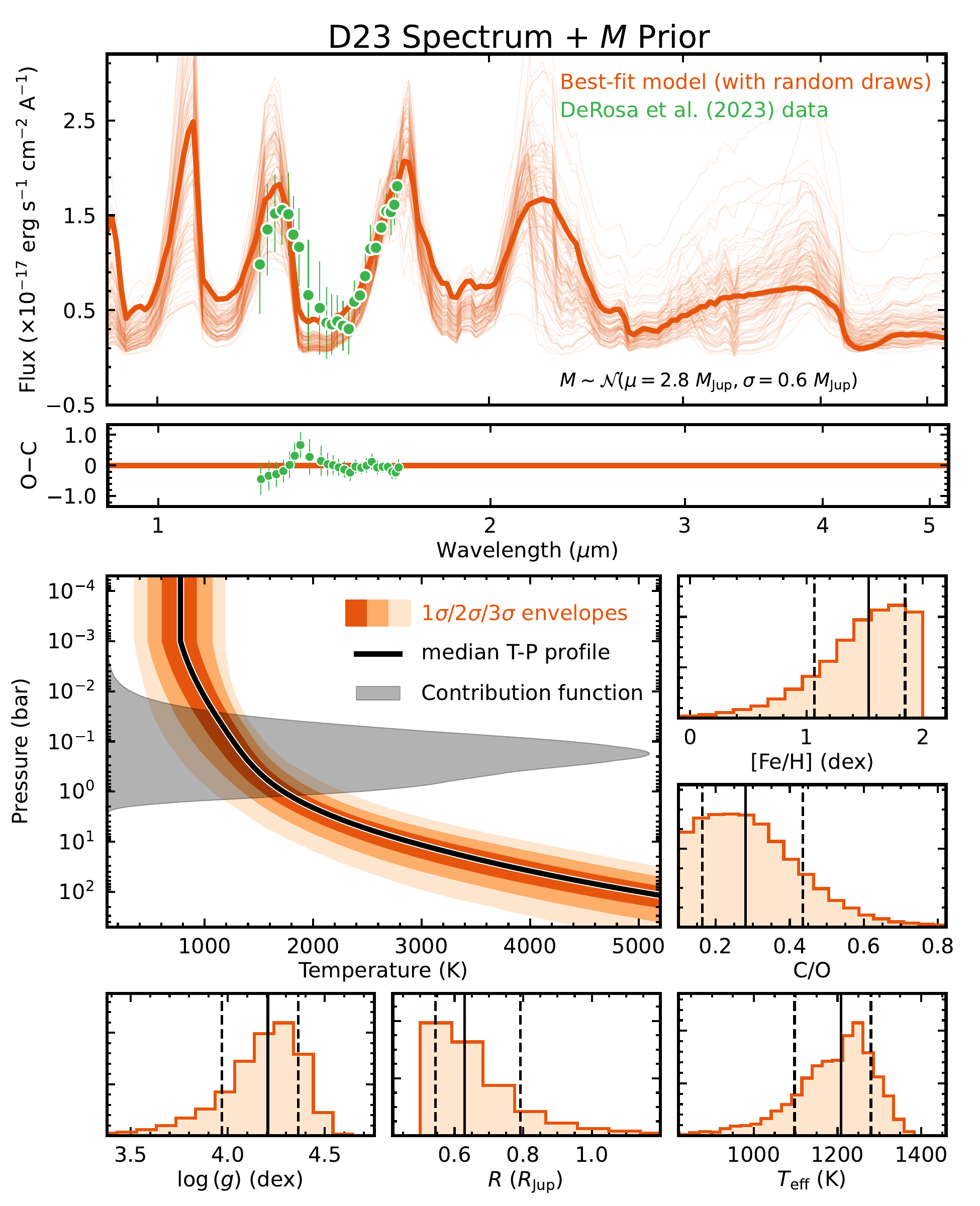}
\caption{Continued. Results of the retrieval analysis on the \cite{2023AandA...672A..94D} spectrum of AF~Lep~b, with a constrained prior on both $M$ and default priors for the remaining parameters (Table~\ref{tab:retrieval_params}). The format is the same as Figure~\ref{fig:results_m23}.}
\end{center}
\end{figure*}

\addtocounter{figure}{-1}
\begin{figure*}[t]
\begin{center}
\includegraphics[height=7.5in]{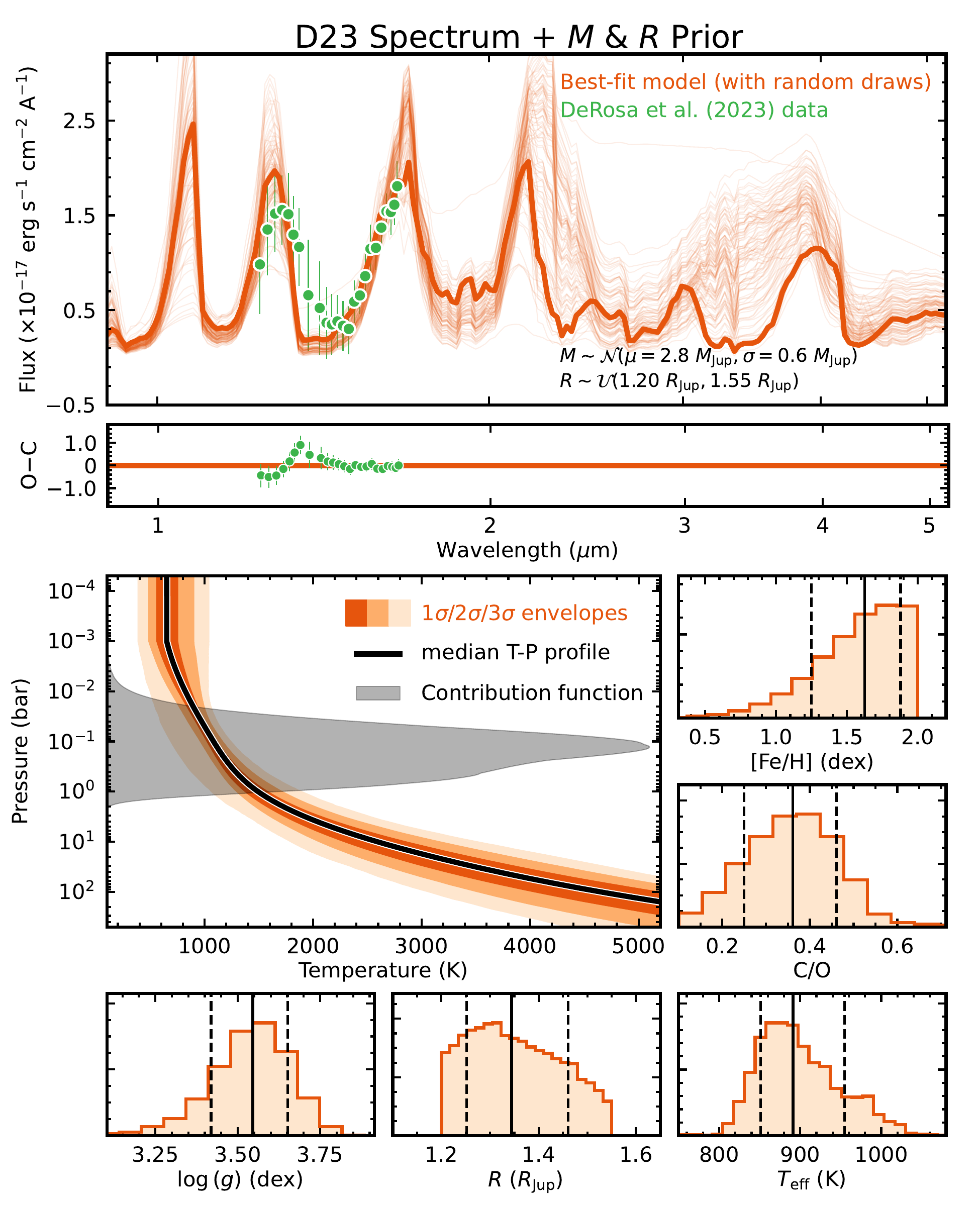}
\caption{Continued. Results of the retrieval analysis on the \cite{2023AandA...672A..94D} spectrum of AF~Lep~b, with constrained priors on both $M$ and $R$ and default priors for the remaining parameters (Table~\ref{tab:retrieval_params}). The format is the same as Figure~\ref{fig:results_m23}.}
\end{center}
\end{figure*}

\addtocounter{figure}{-1}
\begin{figure*}[t]
\begin{center}
\includegraphics[height=7.5in]{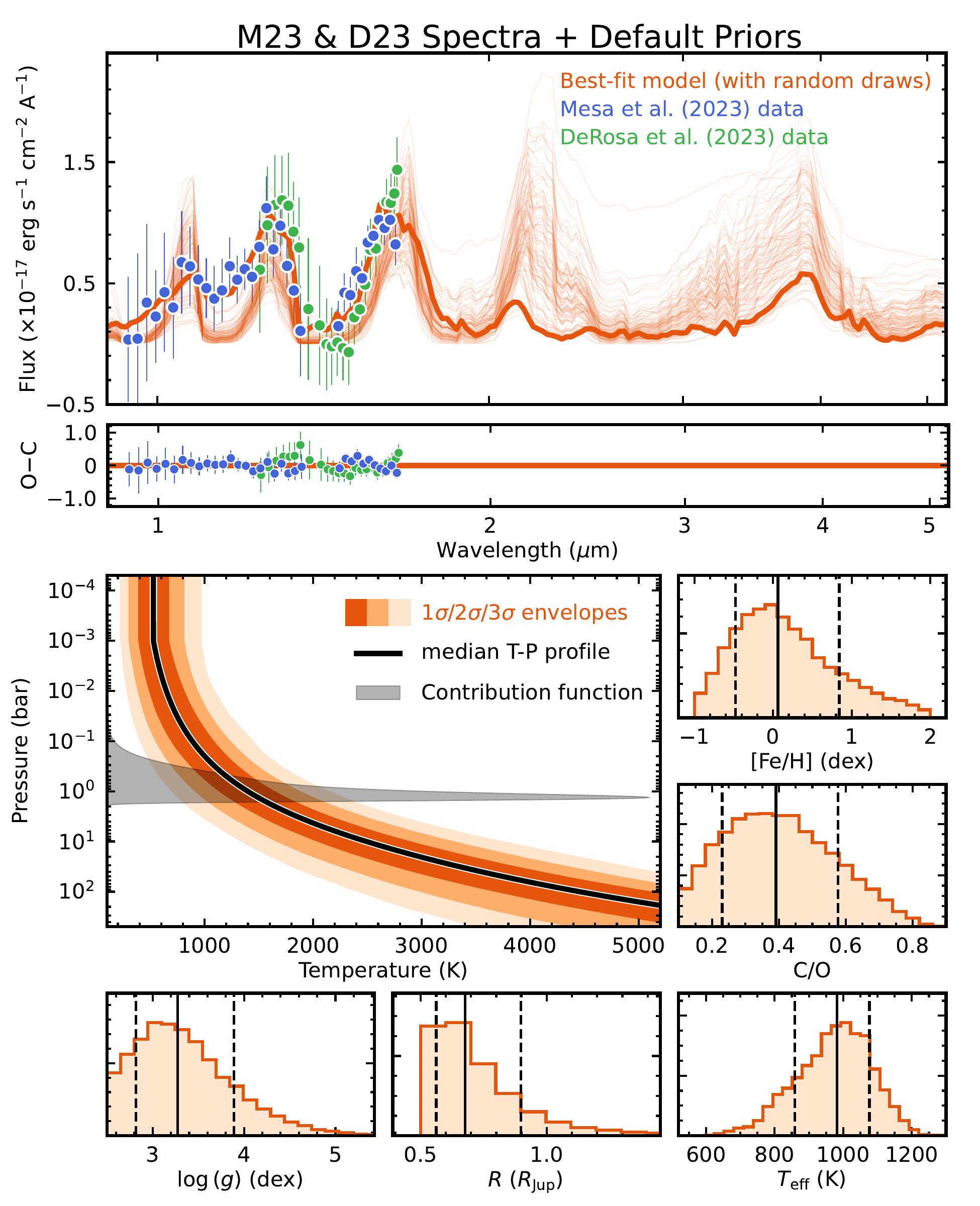}
\caption{Continued. Results of the retrieval analysis on the both spectra from \cite{2023AandA...672A..93M} and \cite{2023AandA...672A..94D}, with default parameter priors listed in Table~\ref{tab:retrieval_params}. The format is the same as Figure~\ref{fig:results_m23}.}
\end{center}
\end{figure*}

\addtocounter{figure}{-1}
\begin{figure*}[t]
\begin{center}
\includegraphics[height=7.5in]{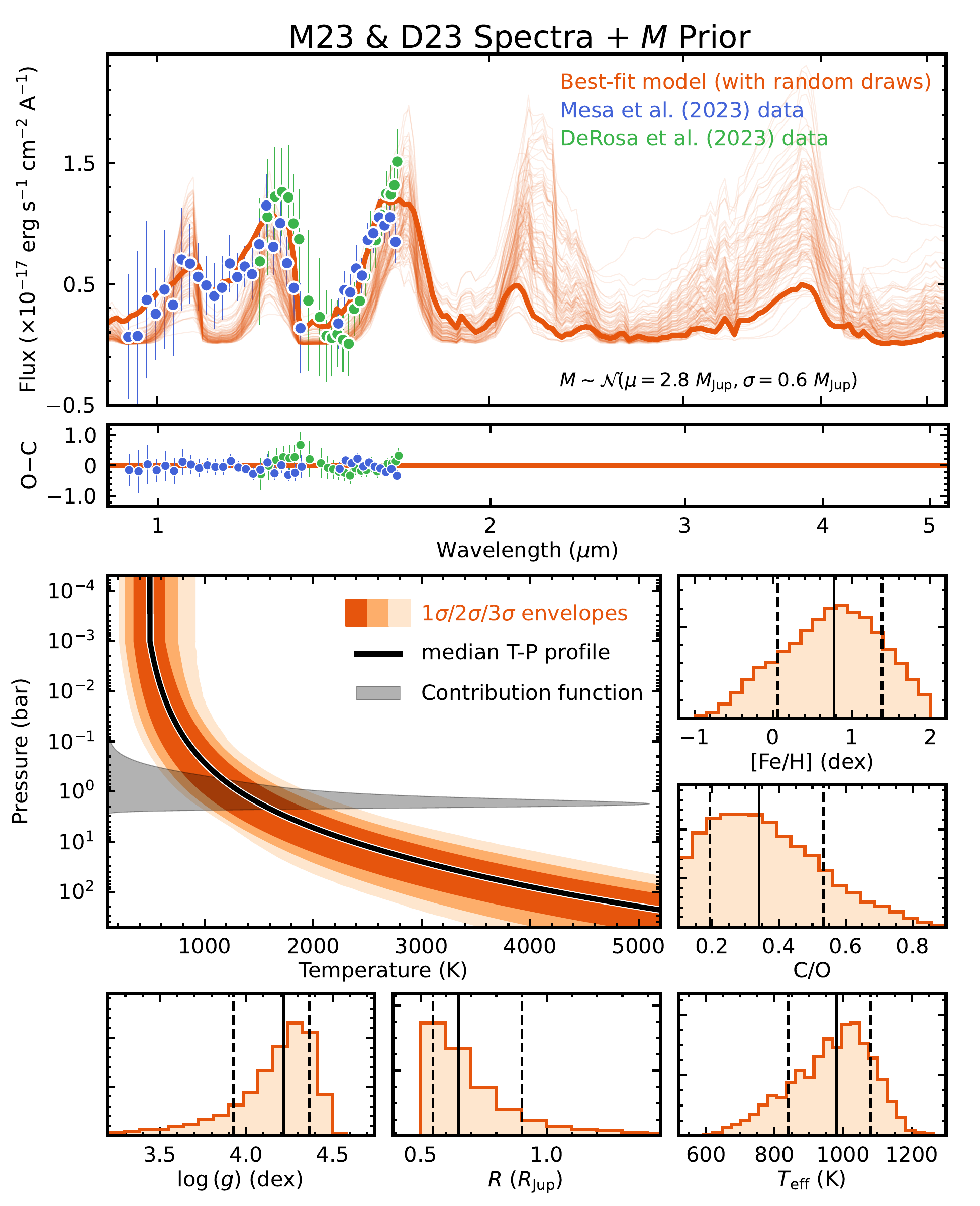}
\caption{Continued. Results of the retrieval analysis on $K1/K2/L'$ photometry and both spectra from \cite{2023AandA...672A..93M} and \cite{2023AandA...672A..94D}, with a constrained prior on $M$ and default priors for the remaining parameters (Table~\ref{tab:retrieval_params}). The format is the same as Figure~\ref{fig:results_m23}.}
\end{center}
\end{figure*}

\addtocounter{figure}{-1}
\begin{figure*}[t]
\begin{center}
\includegraphics[height=7.5in]{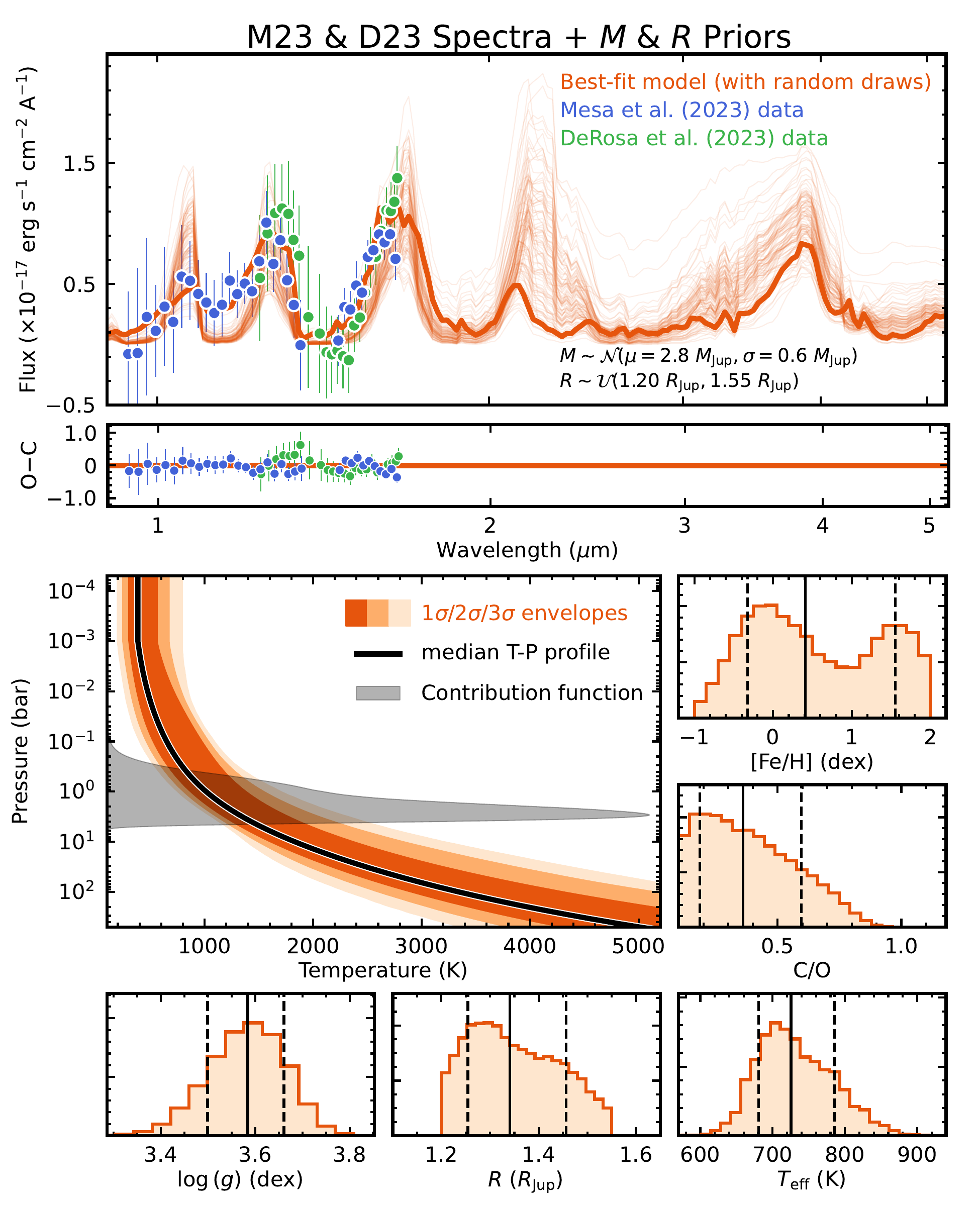}
\caption{Continued. Results of the retrieval analysis on $K1/K2/L'$ photometry and both spectra from \cite{2023AandA...672A..94D} and \cite{2023AandA...672A..94D}, with constrained priors on both $M$ and $R$ and default priors for the remaining parameters (Table~\ref{tab:retrieval_params}). The format is the same as Figure~\ref{fig:results_m23}.}
\end{center}
\end{figure*}

%%%%%%%%%%%%
%%%   TABLES   %%%
%%%%%%%%%%%%
\clearpage

{ 
\begin{deluxetable*}{lcccc}
\setlength{\tabcolsep}{12pt} 
\tablecaption{Properties of AF Lep b based on evolution models (assuming planet is 10 Myr younger than the host star)} \label{tab:younger_evoparams} 
\tablehead{ \multicolumn{1}{l}{Evolution Model} &  \multicolumn{1}{c}{$T_{\rm eff}$} &  \multicolumn{1}{c}{$\log{(L_{\rm bol}/L_{\odot})}$} &  \multicolumn{1}{c}{$\log{g}$} &  \multicolumn{1}{c}{$R$} \\ 
\multicolumn{1}{l}{} &  \multicolumn{1}{c}{(K)} &  \multicolumn{1}{c}{(dex)} &  \multicolumn{1}{c}{(dex)} &  \multicolumn{1}{c}{($R_{\rm Jup}$)} } 
\startdata 
\multicolumn{5}{c}{Hot-Start Models} \\ 
\hline 
\cite{2008ApJ...689.1327S}: cloudy ($f_{\rm sed} = 2$) and [Fe/H]$=0$ &  $754^{+72}_{-55}$  &  $-5.22^{+0.17}_{-0.15}$  &  $3.56^{+0.08}_{-0.07}$  &  $1.400^{+0.025}_{-0.022}$  \\ 
\cite{2008ApJ...689.1327S}: hybrid and [Fe/H]$=0$ &  $781^{+106}_{-82}$  &  $-5.20^{+0.24}_{-0.21}$  &  $3.60^{+0.08}_{-0.07}$  &  $1.328^{+0.024}_{-0.018}$  \\ 
\cite{2021ApJ...920...85M}: cloudless and [Fe/H]$=-0.5$ &  $744^{+99}_{-82}$  &  $-5.31^{+0.23}_{-0.21}$  &  $3.62^{+0.08}_{-0.09}$  &  $1.288^{+0.018}_{-0.014}$  \\ 
\cite{2021ApJ...920...85M}: cloudless and [Fe/H]$=0$ &  $756^{+99}_{-87}$  &  $-5.27^{+0.23}_{-0.22}$  &  $3.60^{+0.08}_{-0.09}$  &  $1.312^{+0.022}_{-0.016}$  \\ 
\cite{2021ApJ...920...85M}: cloudless and [Fe/H]$=+0.5$ &  $767^{+95}_{-85}$  &  $-5.23^{+0.22}_{-0.22}$  &  $3.58^{+0.08}_{-0.09}$  &  $1.339^{+0.026}_{-0.020}$  \\ 
\cite{2012ApJ...745..174S}: cloudless and [Fe/H]$=0$ &  $742^{+101}_{-89}$  &  $-5.29^{+0.23}_{-0.23}$  &  $3.59^{+0.09}_{-0.09}$  &  $1.329^{+0.023}_{-0.017}$  \\ 
\cite{2012ApJ...745..174S}: hybrid and [Fe/H]$=+0.5$ &  $723^{+98}_{-86}$  &  $-5.34^{+0.23}_{-0.23}$  &  $3.59^{+0.09}_{-0.09}$  &  $1.329^{+0.023}_{-0.017}$  \\ 
\cite{2020AandA...637A..38P}: cloudless and [Fe/H]$=0$ &  $760^{+99}_{-91}$  &  $-5.26^{+0.23}_{-0.23}$  &  $3.60^{+0.08}_{-0.09}$  &  $1.318^{+0.027}_{-0.020}$  \\ 
\cite{2003AandA...402..701B}: cloudless and [Fe/H]$=0$ &  $791^{+101}_{-90}$  &  $-5.20^{+0.22}_{-0.22}$  &  $3.60^{+0.08}_{-0.09}$  &  $1.317^{+0.023}_{-0.020}$  \\ 
\hline 
\multicolumn{5}{c}{Cold-Start Models} \\ 
\hline 
\cite{2021ApJ...920...85M}: cloudless and [Fe/H]$=-0.5$  &  $665^{+41}_{-44}$  &  $-5.52^{+0.11}_{-0.12}$  &  $3.62^{+0.09}_{-0.10}$  &  $1.278^{+0.013}_{-0.010}$  \\ 
\cite{2012ApJ...745..174S}: cloudless and [Fe/H]$=0$  &  $520^{+21}_{-22}$  &  $-5.96^{+0.07}_{-0.08}$  &  $3.64^{+0.10}_{-0.10}$  &  $1.253^{+0.012}_{-0.010}$  \\ 
\cite{2012ApJ...745..174S}: hybrid and [Fe/H]$=+0.5$  &  $528^{+24}_{-26}$  &  $-5.93^{+0.08}_{-0.09}$  &  $3.64^{+0.10}_{-0.10}$  &  $1.253^{+0.012}_{-0.010}$  \\ 
\enddata 
\end{deluxetable*} 
}

\end{document}